\documentclass[12pt, draftclsnofoot, onecolumn]{IEEEtran}
\IEEEoverridecommandlockouts
\usepackage{verbatim}
\usepackage[utf8x]{inputenc}
\usepackage{epstopdf}
\usepackage[pdftex]{graphicx}
\usepackage{tabularx}
\usepackage{booktabs}
\usepackage[table]{xcolor}
\usepackage{booktabs}
\usepackage{multirow}
\usepackage{ctable}
\usepackage[cmex10]{amsmath}
\usepackage{algorithm}
\usepackage{algpseudocode}
\usepackage{pifont}
\usepackage{amssymb}
\usepackage{cite}
\usepackage{subcaption}
\usepackage{textcomp}
\usepackage{array}
\graphicspath{{./Figures/}}

\newcommand*\xbar[1]{%
	\hbox{%
		\vbox{%
			\hrule height 0.5pt 
			\kern0.5ex
			\hbox{%
				\kern-0.1em
				\ensuremath{#1}%
				\kern-0.1em
			}%
		}%
	}%
}

\newcommand{\g}{\operatorname{g}}

\newcommand{\ue}{\operatorname{u}}
\newcommand{\f}{\operatorname{f}}

\begin{document}
	\bstctlcite{IEEEexample:BSTcontrol}
	\title{Uplink Interference Mitigation Techniques for Coexistence of 5G mmWave Users with Incumbents at 70 and 80 GHz} 
	\author{Ghaith Hattab,~\IEEEmembership{Student Member,~IEEE,} Eugene Visotsky,~\IEEEmembership{Member,~IEEE,} \\Mark Cudak,~\IEEEmembership{Member,~IEEE,} and Amitava Ghosh,~\IEEEmembership{Fellow,~IEEE}
		\thanks{This paper was presented in part at the IEEE Global communications Conference (Globecom), Singapore, December 2017 \cite{HattabGhosh2017a}.	G. Hattab was with Nokia Bell Labs. He is currently with the Department of Electrical	Engineering, University of California, CA, USA. E. Visotsky, M. Cudak, and A. Ghosh are with Nokia Bell Labs in Naperville, IL, USA. (email: ghattab@ucla.edu,\{eugene.visotsky, mark.cudak, amitava.ghosh\}@nokia-bell-labs.com).}}

	\maketitle
	
	\begin{abstract}
		The millimeter wave spectra at 71-76GHz (70GHz) and 81-86GHz (80GHz) have the potential to endow fifth-generation new radio (5G-NR) with mobile connectivity at gigabit rates. However, a pressing issue is the presence of incumbent systems in these bands, which are primarily point-to-point fixed stations (FSs). In this paper, we first identify the key properties of incumbents by parsing databases of existing stations in major cities to devise several modeling guidelines and characterize their deployment geometry and antenna specifications. Second, we develop a detailed uplink interference framework to compute the aggregate interference from outdoor 5G-NR users into FSs. We then present several case studies in dense populated areas, using actual incumbent databases and building layouts. Our simulation results demonstrate promising 5G coexistence at 70GHz and 80GHz as the majority of FSs experience interference well below the noise floor thanks to the propagation losses in these bands and the deployment geometry of the incumbent and 5G systems. For the few FSs that may incur higher interference, we propose several passive interference mitigation techniques such as angular-based exclusion zones and spatial power control. Simulation results show that the techniques can effectively protect FSs, without tangible degradation of the 5G coverage.   
	\end{abstract} 
	\vspace{0.5in}
	\newpage
	\begin{IEEEkeywords} 
		5G, coexistence, interference management, spectrum sharing, mmWave, wireless backhaul.
	\end{IEEEkeywords}

	\section{Introduction}
	Fifth-generation new radio (5G-NR) is envisioned to be the first cellular standard with millimeter wave (mmWave) spectrum access \cite{Xiao2017,ITU2015}. Such paradigm shift towards mmWave access is necessary to scale with the explosive growth of mobile traffic and to provide unparalleled network capacity, with peak data rates reaching tens of Gbps \cite{AndrewsZhang2014a}. Indeed, the mmWave spectrum has attracted significant attention from standard bodies, industry, and the academic community, culminating recently when the Federal Communications Commission (FCC) has opened up 3.85GHz of licensed spectrum for cellular services, and specifically at 28GHz (27.5-28.35GHz) and 39GHz (37-40GHz) \cite{FCC2016b}. Nevertheless, there is still an additional 10GHz of licensed spectra at 70GHz (71-76GHz) and 80GHz (81-86GHz) that are left for future consideration as candidate bands for mmWave mobile networks \cite{FCC2016b,ITU2015b}. 
	
	The advantages of using 70GHz and 80GHz bands, also known as the \emph{e-band}, are twofold. First, each band can easily provide a contiguous high bandwidth, e.g., 2GHz, in contrast to 28GHz and 39GHz, where each provides a maximum of 850MHz and 1.6GHz, respectively. Second, the e-band is available worldwide, enabling economies of scale through universal adoption of common mmWave devices. Equally important, operating at the higher end of the mmWave spectrum is not significantly different from operating at 28GHz as the channel models are the same \cite{RanganErkip2014}, and the increase in path loss can be compensated by using an array with a larger number of antenna elements. In addition, several prototypes have shown the feasibility of mmWave systems over 70GHz. For instance, Nokia and Huawei have already demonstrated experimental 5G systems designed to operate at 73.5 GHz \cite{Cudak2014,Inoue2017,Huawei2016}.  
	
	One key challenge of using the 70GHz and 80GHz bands is the presence of existing incumbents, which are primarily fixed stations (FSs) that provide point-to-point services such as wireless backhaul. Per FCC regulations, these incumbents must be protected from harmful interference. Thus, our objective is to study the feasibility of the coexistence of 5G systems with existing FSs and to develop interference mitigation techniques that ensure harmonious spectrum sharing. 
	
	\subsection{Related work}
	Several works have studied spectrum sharing paradigms for mmWave networks \cite{Gupta2016a,Rebato2017,Boccardi2016}. However, these works have solely focused on sharing among different mobile operators, e.g., sharing frequency channels, infrastructure, etc. Spectrum sharing of 5G systems and other services has recently attracted attention. For instance,  the work in \cite{Ghorbanzadeh2015,Ghorbanzadeh2016,Khawar2016} focus on the 5G coexistence with radar systems, whereas the work in \cite{Beltran2016} studies the coexistence with WiFi. While these aforementioned works are limited to sub-6GHz, the mmWave access paradigm has also spurred interest in coexistence studies. For example, the work in \cite{Kim2015} and \cite{Kim2016} study the feasibility of 5G coexistence with incumbents at 28GHz, which are satellite systems, and the coexistence with fixed service at 39GHz. In \cite{Guidolin2015}, the impact of FSs interference on the throughput of UEs operating at 28GHz is studied. A more relevant work to this paper is the one in \cite{Kim2017}, which studies the coexistence of 5G with FSs at 70GHz. However, the work makes several modeling assumptions, e.g., only a single FS is assumed to exist at a fixed distance from the 5G system and only a fixed portion of links are assumed to be non-line-of-sight (NLOS). In addition, the work in \cite{Kim2017} focuses on the 5G downlink (DL) interference. To mitigate the uplink (UL) interference, a probing device is proposed to be installed on the FS to report excessive interference to the 5G system. In this work, however, we focus on UL passive interference mitigation techniques, i.e., we propose techniques that do no require any coordination between the 5G system and the incumbents or require probing devices.

	\subsection{Contributions}
	The main contributions of this paper are summarized as follows.
	
	\begin{itemize}
		\item \textbf{Characterizing incumbents:} We analyze databases of existing FSs in four major areas in the United States, characterizing their deployment geometry and key antenna specifications. The analysis provides insights on the feasibility of 5G coexistence and gives benchmarks for accurate modeling of FSs, which can be of interest to the academic community. 
		\item \textbf{5G uplink interference analysis:} We present a detailed interference analysis framework to compute the aggregate uplink (UL) interference from 5G users into FSs. We also present random models for user's azimuth and elevation antenna directions to help reduce the simulator complexity without degrading the simulation's accuracy. 
		\item  \textbf{Passive interference mitigation:} We propose several passive interference techniques that do not require any coordination between the 5G system and the incumbent systems. Specifically, we propose sector-based and beam-based exclusion zones where 5G base stations (gNBs) switch off certain beams to protect victim FS receivers. While these techniques are shown to be effective, they can affect the 5G downlink (DL) coverage. Thus, we propose spatial power control, defining quiet beams where associated users transmit at lower power. We discuss the implementation of such techniques for 5G-NR. 
	\end{itemize}
	The coexistence feasibility and the effectiveness of the proposed mitigation techniques are validated via three case studies, where we deploy 5G systems in dense urban and suburban areas. The studies use the databases of existing FSs and actual building layouts for accurate interference analysis. Our results have shown that the majority of FSs are protected from harmful interference due to the high propagation losses at 70GHz and 80GHz, the high attenuation due to the misalignment between the user and the FS's antenna boresight, and the deployment geometry of FSs and 5G systems. For the few FSs that experience higher interference, the proposed mitigation techniques provide significant protection, and they are more effective than switching off gNBs that are in vicinity of FSs. Finally, as a by-product of the simulation set-up, we validate the performance of 5G networks in 70GHz and 80GHz and show the distribution of the beams used by the gNB and the user, making design insights for mobile network operators and vendors. 
	
	\subsection{Paper Organization} 
	The rest of the paper is organized as follows. The system model is presented in Section \ref{sec:model}. The study of FSs' deployment and the interference analysis framework are presented in Section \ref{sec:FSdatabase} and Section \ref{sec:interferenceAnalysis}, respectively. The proposed mitigation techniques are discussed in Section \ref{sec:mitigation}. Simulation results are presented in Section \ref{sec:simulations}, and the conclusions are drawn in Section \ref{sec:conclusion}.

	\section{System Model}\label{sec:model}
	
	\subsection{5G base stations (gNBs)}
	We consider a street-level deployment of gNBs such that each one is deployed at a street corner at height $h_{\g}$ and the inter-site distance (ISD) between every site is approximately $d_{\operatorname{ISD}}$.\footnote{In the simulation set-up, we first deploy gNBs in a grid with a fixed ISD of $d_{\operatorname{ISD}}$, covering the entire simulated area. Then, we look at the location of each dropped gNB to check if it lies at a street corner. If the gNB does not lie at a corner, we move it from its initial location to the nearest street corner, given that there are no gNBs located there. Such deployment strategy is followed by mobile operates, where gNBs (or small cells) are deployed  at street corners every few blocks.} Each site consists of four sectors, i.e., each sector covers an area of 90$^\circ$. 
	
\begin{figure}[t!]
		\center
		\includegraphics[width=2.25in]{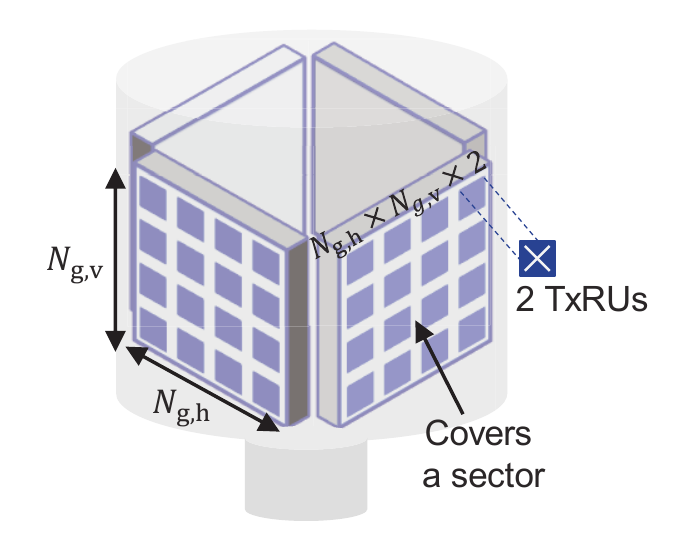} 
		\caption{An illustrative example of a 5G gNB site.}
		\label{fig:gNBsite}
	\end{figure}

	Each sector is equipped with a large-scale cross-polarized antenna array of size $N_{\g,\operatorname{h}}\times N_{\g,\operatorname{v}}\times2$. The antenna array is assumed to be mechanically tilted downward at angle $\phi_{\g}$ (few degrees) as the majority of outdoor UEs are at a ground-level, whereas the gNB is few meters above the ground. Each antenna element has a gain of $G_{\g}$ and a transmit power of $P_{\g}$ and is half-wavelength apart from the nearest antenna element. An illustrative example of a gNB site is given in Fig. \ref{fig:gNBsite} \cite{Cudak2014}.
	
	\subsection{5G Users}
	We only consider outdoor user equipment  terminals (UEs), that are randomly deployed over space, as FSs are outdoors and the attenuation due to penetration losses for indoor UEs is very high at 70GHz and 80GHz. Each UE is equipped with a cross-polarized antenna array of size $N_{\ue,\operatorname{h}}\times N_{\ue,\operatorname{v}}\times2$, where each antenna element has a gain and a transmit power of $G_{\ue}$ and $P_{\ue}$, respectively. The UE array height is assumed to be $h_{\ue}$, and it is titled upward at angle $\phi_{\ue}$\footnote{The actual mechanical tilt will depend on the UE, yet assuming an upward one can be considered as a worst case scenario. We note that we also consider a randomized tilt in Section \ref{sec:interferenceAnalysis}.C}. The UE is also assumed to have two panels, i.e., two sectors, with each one covering 180$^\circ$. Thus, the user can sense beams in all directions, but only one panel will be active after user and beam association. 
	During cell selection and association, the UE measures the received power of reference signals sent over different beams from gNBs in vicinity of the UE. Then, the UE connects to the beam with the highest received power (other beam association algorithms or criteria can be considered \cite{Liu2018,Cui2018,Alkhateeb2017,Giordani2016,Barati2016}).
	
	\subsection{Incumbent Fixed Stations}
	We consider FSs that operate in the 71-76GHz and 81-86GHz bands, and they are currently registered in the FCC's database as incumbents are required to be in the database for operating in these bands \cite{Comsearch2018}. Thus, their exact three-dimensional locations are used. Similarly, we extract their antenna specifications, e.g., beamwidth, gain, azimuth orientation, and tilt. While different FSs may operate at different center frequencies in the aforementioned bands, we assume in this paper that all of them share the same spectrum with the 5G system, as a worst case scenario. 
	
	\subsection{Antenna Patterns} 
\begin{figure}[t!]
		\centering
		\begin{subfigure}[t]{.3\textwidth}
			\centering
			\includegraphics[width=2.25in]{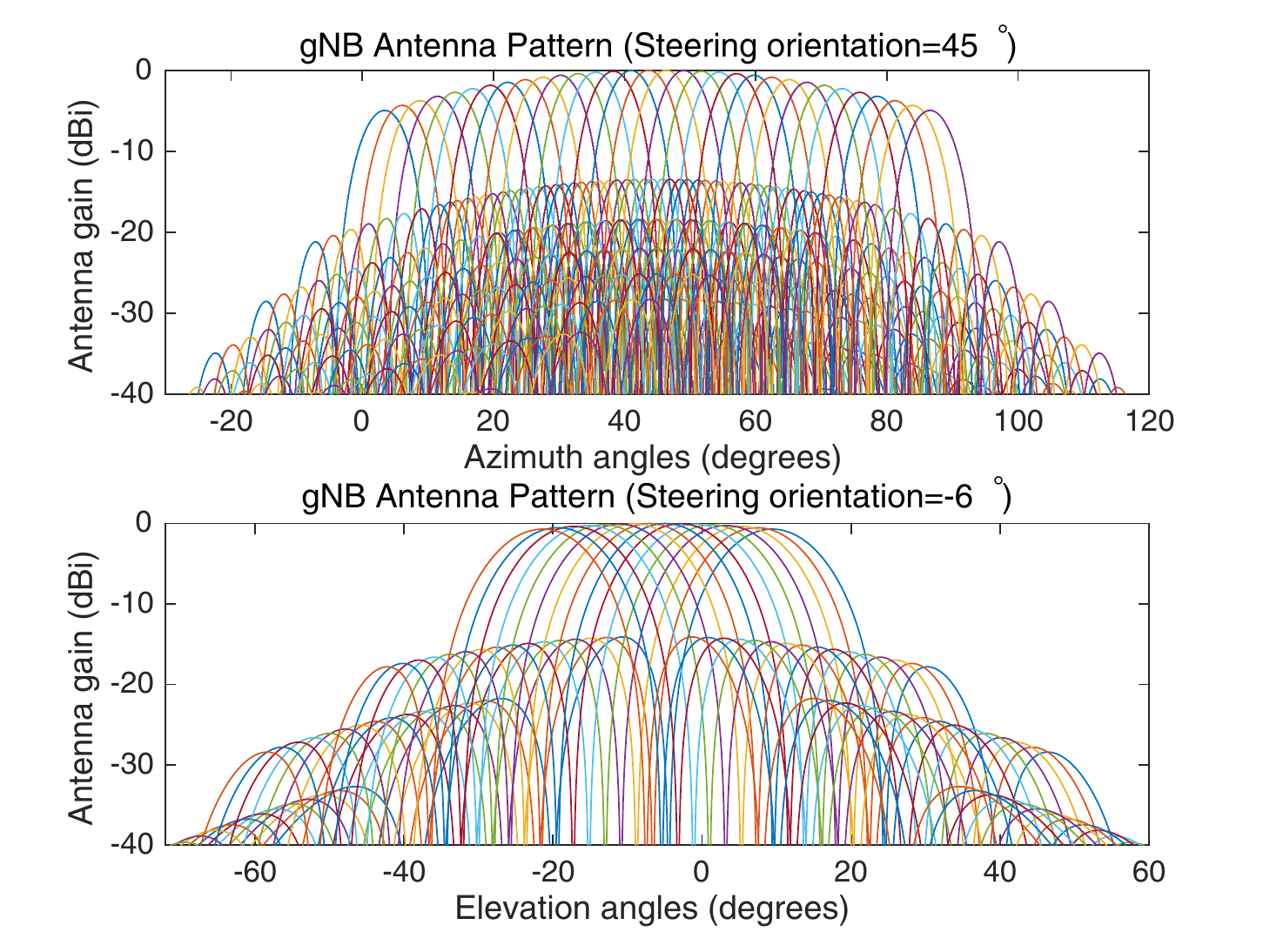}
			\caption{gNB beams}
			\label{fig:gNBAntennaPatterns}
		\end{subfigure}~
		\begin{subfigure}[t]{.3\textwidth}
			\centering
			\includegraphics[width=2.25in]{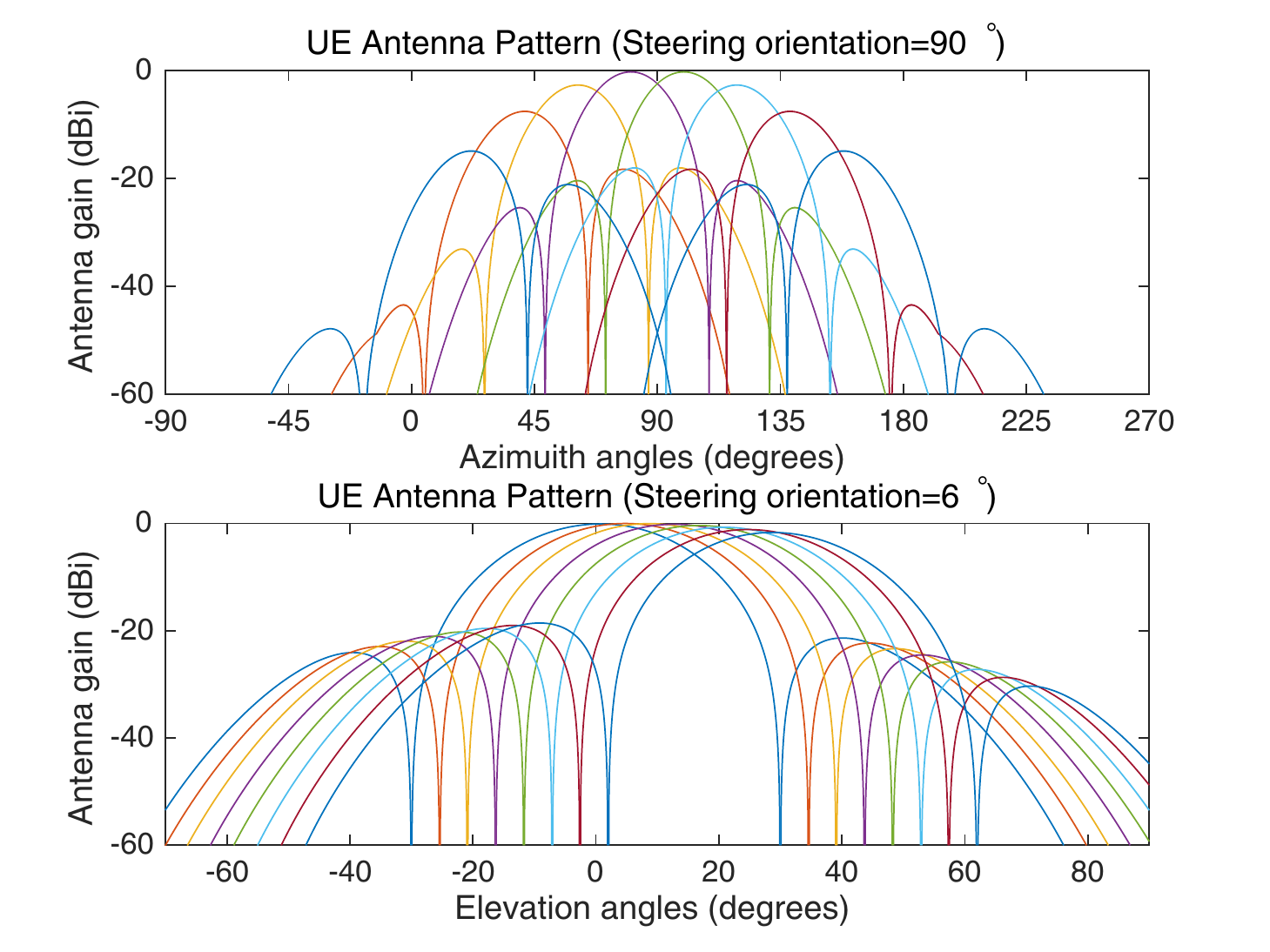}
			\caption{UE beams}
			\label{fig:UEAntennaPatterns}
		\end{subfigure}~
		\begin{subfigure}[t]{.3\textwidth}
			\centering
			\includegraphics[width=2.25in]{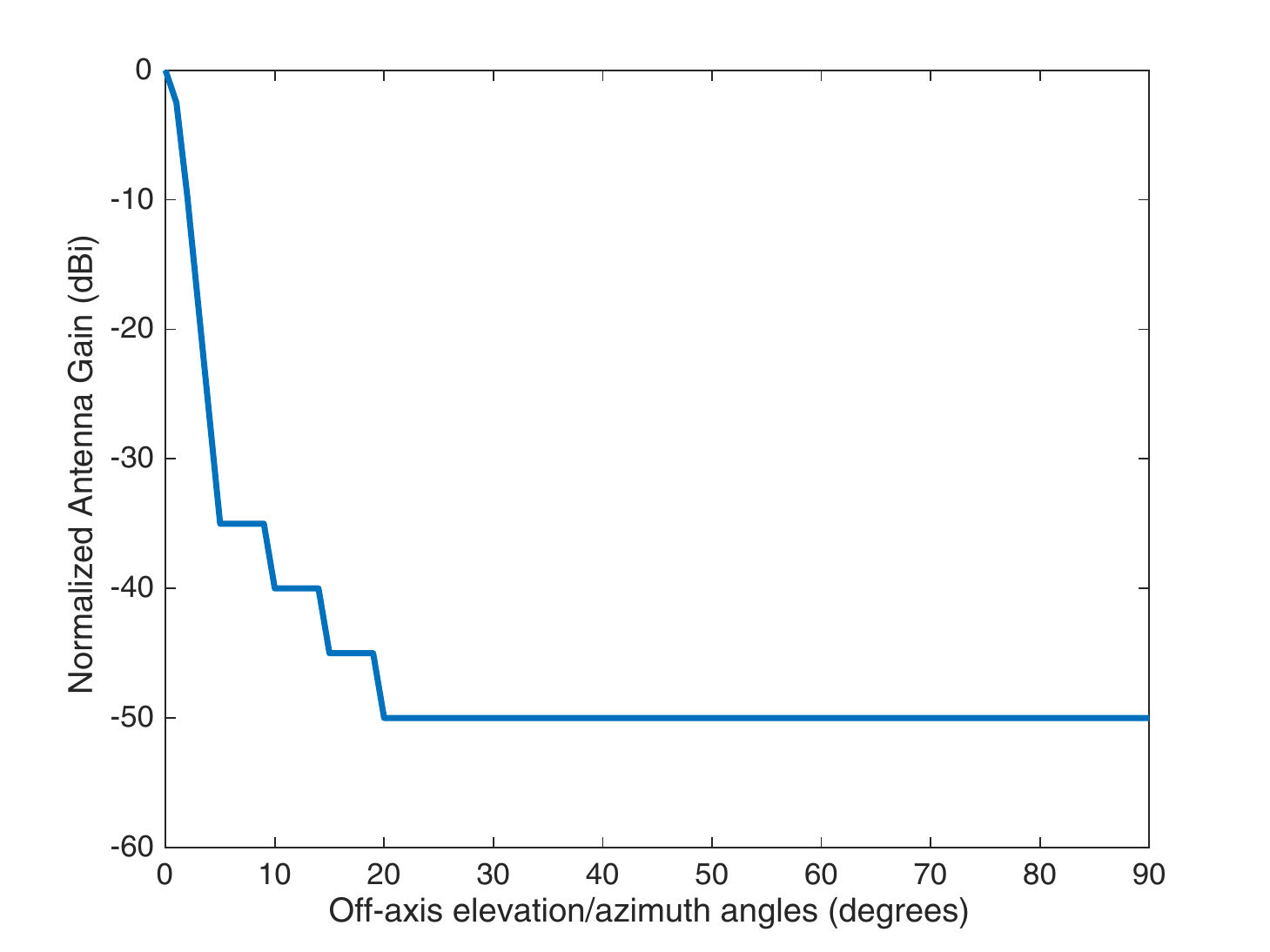}
			\caption{FS beams}
			\label{fig:FSAntennaPattern}
		\end{subfigure}
		\caption{Antenna beam patterns of: (a) the gNB; (b) the UE, and (c) the FS }
		\label{fig:AntennaPattern}
	\end{figure}

	For beam association and data communications, the gNB can use one of the $4N_{\g,\operatorname{h}}N_{\g,\operatorname{v}}$ available beams, where we assume the number of beams per dimension is twice the number of antennas in that dimension.\footnote{The number of beams, or directions to sweep, is a design parameter that also depends on the type of antenna used. Sweeping the angular domain with more beams improves the coverage of the 5G system as narrower beams, with higher gain, are used. However, finer sweeping typically increases the search space, increasing the complexity and delay of initial access.} The azimuth (or elevation) beam pattern beamwidth is approximately $\theta_{\g,\operatorname{BP}}^{\operatorname{BW}}\approx 102/N_{\g,\operatorname{h}}$ (or $\phi_{\g,\operatorname{BP}}^{\operatorname{BW}}\approx 102/N_{\g,\operatorname{v}}$) \cite{Skolnik2001}. We further assume a parabolic element pattern such that the normalized azimuth and elevation attenuations are, in dB, \cite{3GPP2017}
	\begin{equation}
	A_{\g,\operatorname{EP}}(\theta)= 12 \left(\frac{\theta}{\theta_{\g,\operatorname{EP}}^{\operatorname{BW}}}\right)^2~~\text{and}~~  A_{\g,\operatorname{EP}}(\phi) = 12 \left(\frac{\phi}{\phi_{\g,\operatorname{EP}}^{\operatorname{BW}}}\right)^2, 
	\end{equation}
	where $\theta_{\g,\operatorname{EP}}^{\operatorname{BW}}$ and $\phi_{\g,\operatorname{EP}}^{\operatorname{BW}}$ are the element pattern 3dB beamwidths in azimuth and elevation, respectively. The same definitions are applied for the UE side, replacing the subscript $\g$ with $\ue$. Fig. \ref{fig:gNBAntennaPatterns} and Fig. \ref{fig:UEAntennaPatterns} show the antenna patterns of 5G gNBs and UEs, respectively, where it is assumed that the gNB and UE arrays are, respectively, of size $16\times8\times2$ and $4\times4\times2$. 
	
	For the incumbent system, we assume all FSs have antenna patterns that, at least, meet the FCC's regulation as specified in \cite{FCC2017}. Essentially, the regulation specifies the minimum radiation suppression for a given angle from the centerline of the main beam. Fig. \ref{fig:FSAntennaPattern} shows the normalized antenna gain for a given off-axis angle. Due to the high directivity of the FS's antenna, it is shown that a slight misalignment with the main boresight is enough to incur significant signal attenuation. A summary of the main parameters used is provided in Table \ref{tab:parameters}.

	\begin{table}[!t]
		\caption{Main parameters and their values if applicable}
		\label{tab:parameters}
		\scriptsize
		\centering
		\def\arraystretch{0.7}
		\begin{tabular}{|l|l|l|}
			\hline
			Symbol   									&Description 						& Value(s) if applicable 				\\\hline
			$h_{(\cdot)}$						        &Height of gNB or UE		        & $h_{\g}=6$m; $h_{\ue}=1.5$m \\\hline
			$d_{\operatorname{ISD}}$					&Inter-site distance 			& $d_{\operatorname{ISD}}=200$m\\\hline
			$N_{(\cdot),\operatorname{h}}$				&Number of columns in an array 	    &$N_{\g,\operatorname{h}}=16$; $N_{\ue,\operatorname{h}}=4$ \\\hline
			$N_{(\cdot),\operatorname{v}}$				&Number of rows in an array 	    &$N_{\g,\operatorname{v}}=8$; $N_{\ue,\operatorname{v}}=4$ \\\hline			
			$\phi_{(\cdot)}$						    &Antenna tilt 	        			& $\phi_{\g}=-6^\circ$; $\phi_{\ue}=6^\circ$ \\\hline			
			$G_{(\cdot)}$						    	&Antenna gain	        	&$G_{\g}=G_{\ue}=5$dBi \\\hline		
			$P_{(\cdot)}$						    	&Antenna transmit power 		&$P_{\g}=7$dBm; $P_{\ue}=1$dBm \\\hline	
			$\theta_{(\cdot),\operatorname{(\cdot)}}^{\operatorname{BW}}$ &3dB beamwidth of beam/element patterns in azimuth  & 
			$\theta_{\g,\operatorname{BP}}^{\operatorname{BW}}=6^\circ$; $\theta_{\ue,\operatorname{BP}}^{\operatorname{BW}}=25^\circ$; $\theta_{\g,\operatorname{EP}}^{\operatorname{BW}}=\theta_{\ue,\operatorname{EP}}^{\operatorname{BW}}=65^\circ$\\\hline
			$\phi_{(\cdot),\operatorname{(\cdot)}}^{\operatorname{BW}}$ &3dB beamwidth of beam/element patterns in elevation & 
			$\phi_{\g,\operatorname{BP}}^{\operatorname{BW}}=12^\circ$; $\phi_{\ue,\operatorname{BP}}^{\operatorname{BW}}=65^\circ$; $\phi_{\g,\operatorname{EP}}^{\operatorname{BW}}=\phi_{\ue,\operatorname{EP}}^{\operatorname{BW}}=65^\circ$\\\hline
			$A_{(\cdot),\operatorname{FTBR}}$			&Front-to-back ratio loss  & $A_{\f,\operatorname{FTBR}}=55$dB; 
			$A_{\g,\operatorname{FTBR}}=A_{\ue,\operatorname{FTBR}}=30$dB\\\hline
			$F_{(\cdot)}$						        &Noise figure		    & $F_{\ue}=9$dB \\\hline		
			$B$						        			&Channel bandwidth	& $B=1$GHz \\\hline
			$f_c$										&Carrier frequency    & $f_c=\{73.5,83.5\}$GHz\\\hline
			$\mathbf{x}_{\operatorname{a}}$					    &$(x,y)$-coordinates of $\operatorname{a}$&\\\hline
			$d_{\operatorname{a}\rightarrow\operatorname{b}}$   &2D distance from  $\operatorname{a}$ to  $\operatorname{b}$ (m)&\\\hline
			$\operatorname{PL}_{\operatorname{a}\rightarrow\operatorname{b}}$ &Path loss from  $\operatorname{a}$ to  $\operatorname{b}$ (dB)&\\\hline
			$\mathcal{X}(\cdot)$						&Log-normal shadowing with standard deviation of $\sigma_{(\cdot)}$& $\sigma_{\operatorname{LOS}}=4$dB; $\sigma_{\operatorname{NLOS}}=7.82$dB \\\hline
			$\beta$										&Indicator variable that denotes a blockage event& Blockage: $\beta=1$; No blockage: $\beta=0$\\\hline
			$G_{(\cdot),\operatorname{max}}$		    &Maximum antenna gain	(dBi)&	\\\hline
		\end{tabular}
	\end{table}
	
	\section{Analysis of FSs Deployment}\label{sec:FSdatabase}
	In this section, we study the deployment of FSs to get some guidelines on their deployment geometry and features. The insights help understand how the deployment of FSs affects the coexistence with 5G systems. Equally important, they can be also used as a benchmark for modeling FSs using stochastic-based approaches \cite{Haenggi2012}. 
	
	We parse the databases of FSs deployed in four major metropolitan areas: Chicago, New York, Los Angeles, and San Fransisco \cite{Comsearch2018}. Each database covers an area of radius  300km. Table \ref{tab:FSdatabase} summarizes the analyzed databases. A link is defined as a two-way communication between two FSs, whereas a pair is defined as a link with unique spatial coordinates of the FSs. Thus, the same pair could have multiple links, each over a different channel in 70GHz and/or 80GHz.
	
	\begin{table}[!t]
		\caption{Current number of links and pairs in each database}
		\label{tab:FSdatabase}
		\centering
		\scriptsize
		\def\arraystretch{0.8}
		\begin{tabular}{|l|c|c|}
			\hline
			Database   		&  No. of links & No. of pairs \\\hline
			Chicago			&  1743			& 512\\
			New York		&  5303			& 1685\\
			Los Angeles		&  1013			& 911\\
			San Francisco	&  1892			& 1801\\	
			\hline
		\end{tabular}
	\end{table}
	
	\subsection{Spatial Distribution}
\begin{figure}[t!]
		\centering
		\begin{subfigure}[t]{0.4\textwidth}
			\centering
			\includegraphics[width=2.25in]{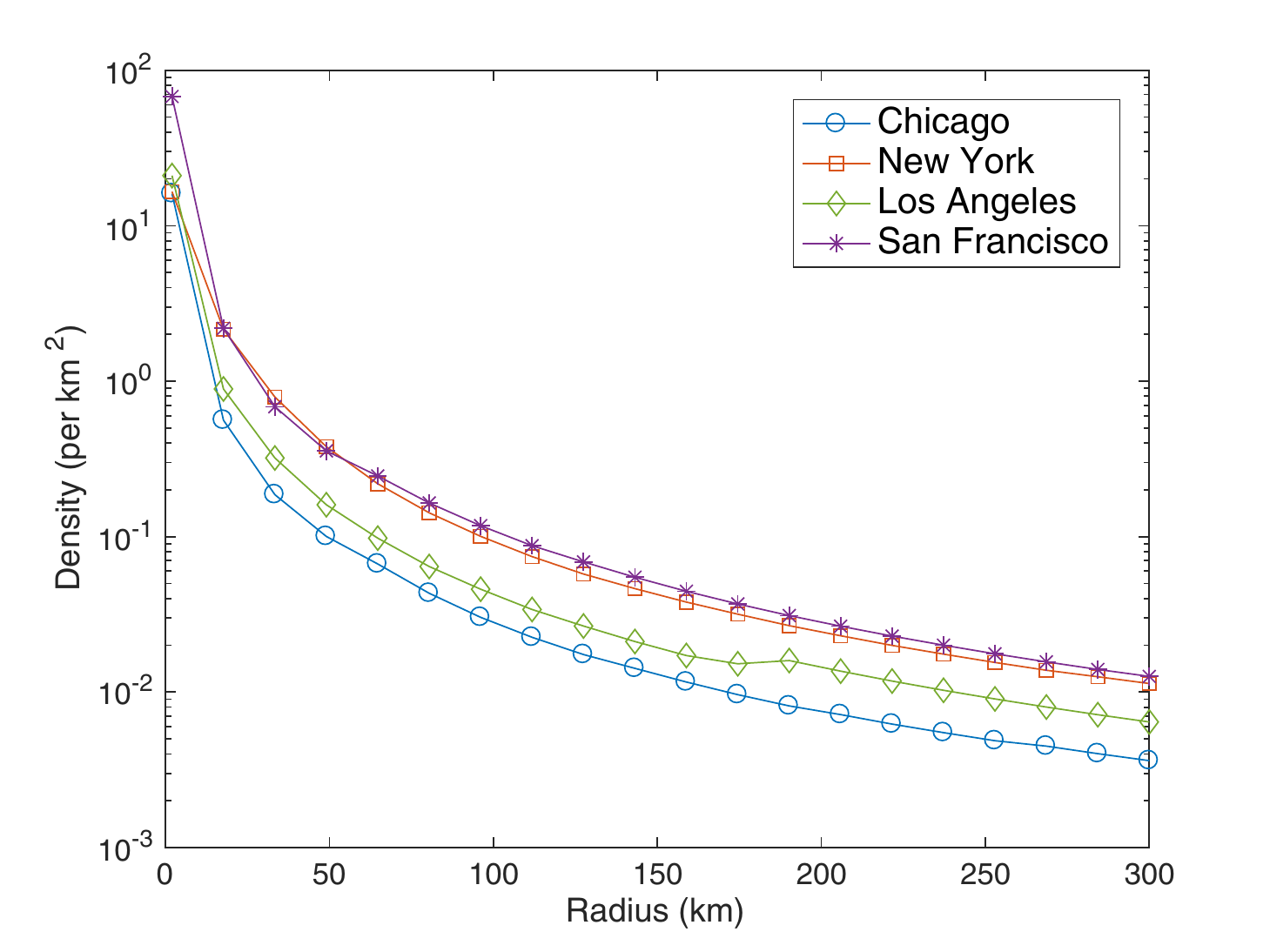}
			\caption{Density with variations of region's radius}
			\label{fig:FSDensityVsRadius}
		\end{subfigure}~~
		\begin{subfigure}[t]{.4\textwidth}
			\centering
			\includegraphics[width=2.25in]{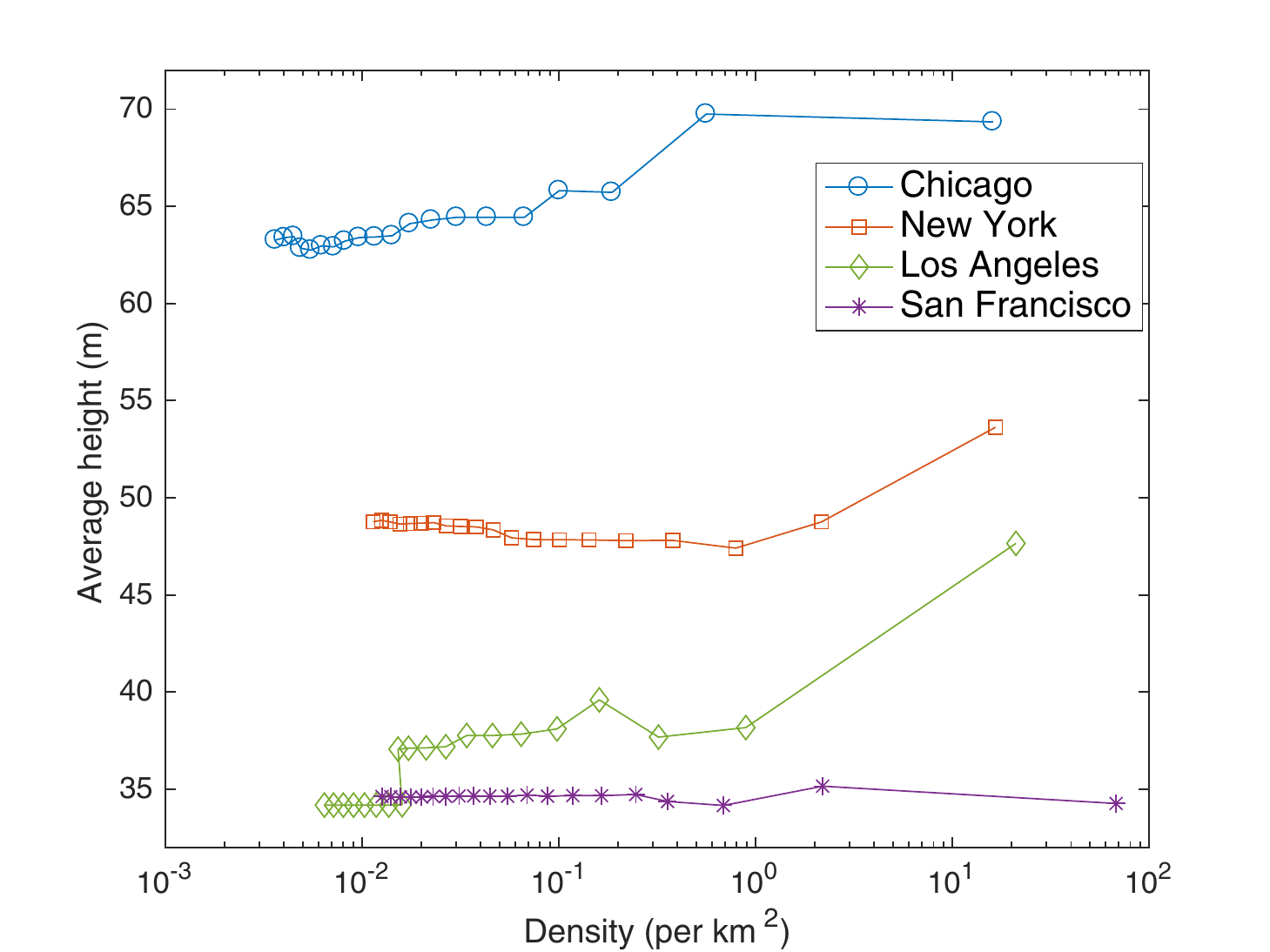}
			\caption{Average height for a given density}
			\label{fig:FSHeightVsDensity}
		\end{subfigure}
		\\
		\begin{subfigure}[t]{0.4\textwidth}
			\centering
			\includegraphics[width=2.25in]{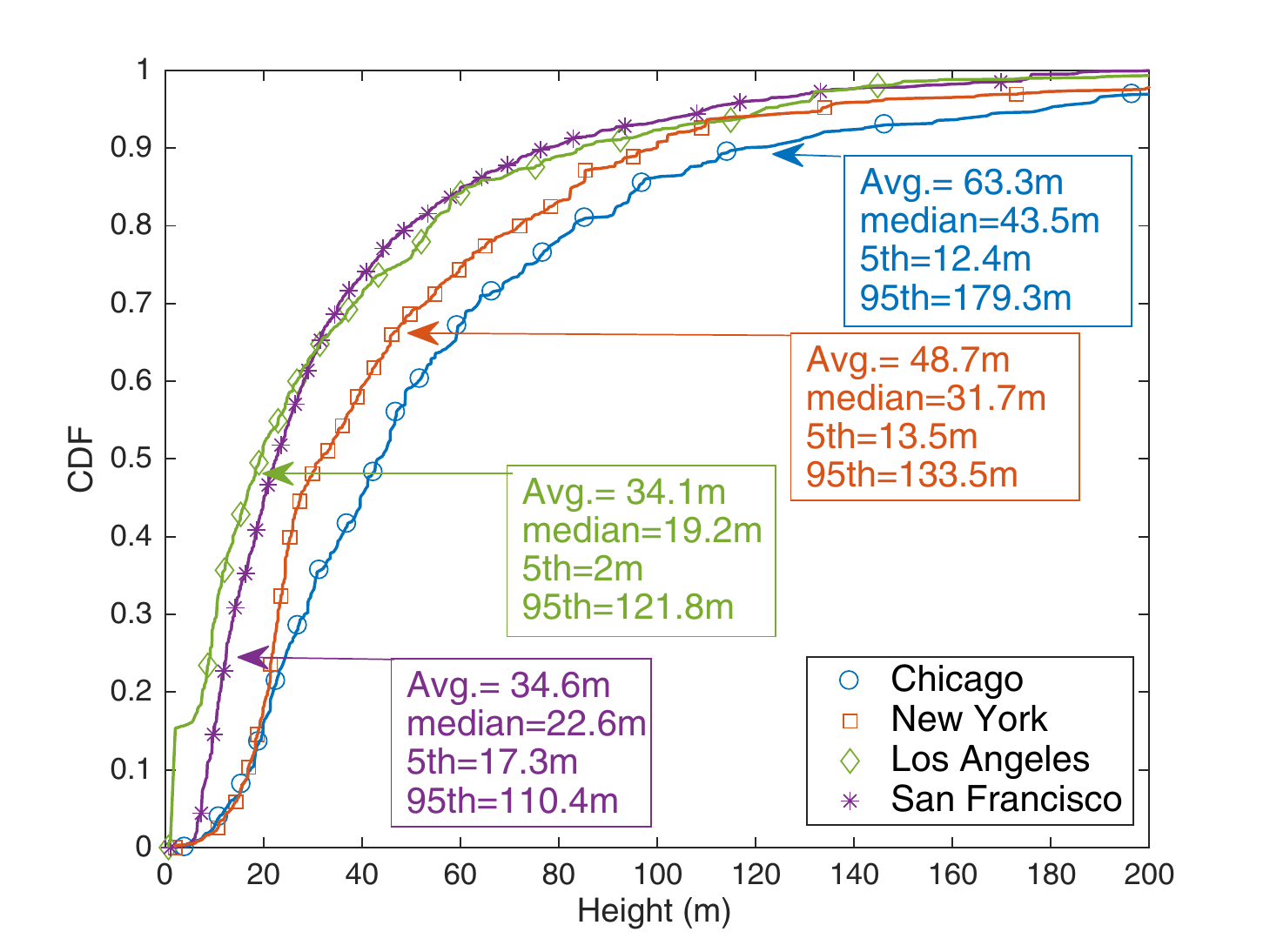}
			\caption{CDF of FSs' height}
			\label{fig:FSHeightCDF}
		\end{subfigure}~~
		\begin{subfigure}[t]{.4\textwidth}
			\centering
			\includegraphics[width=2.25in]{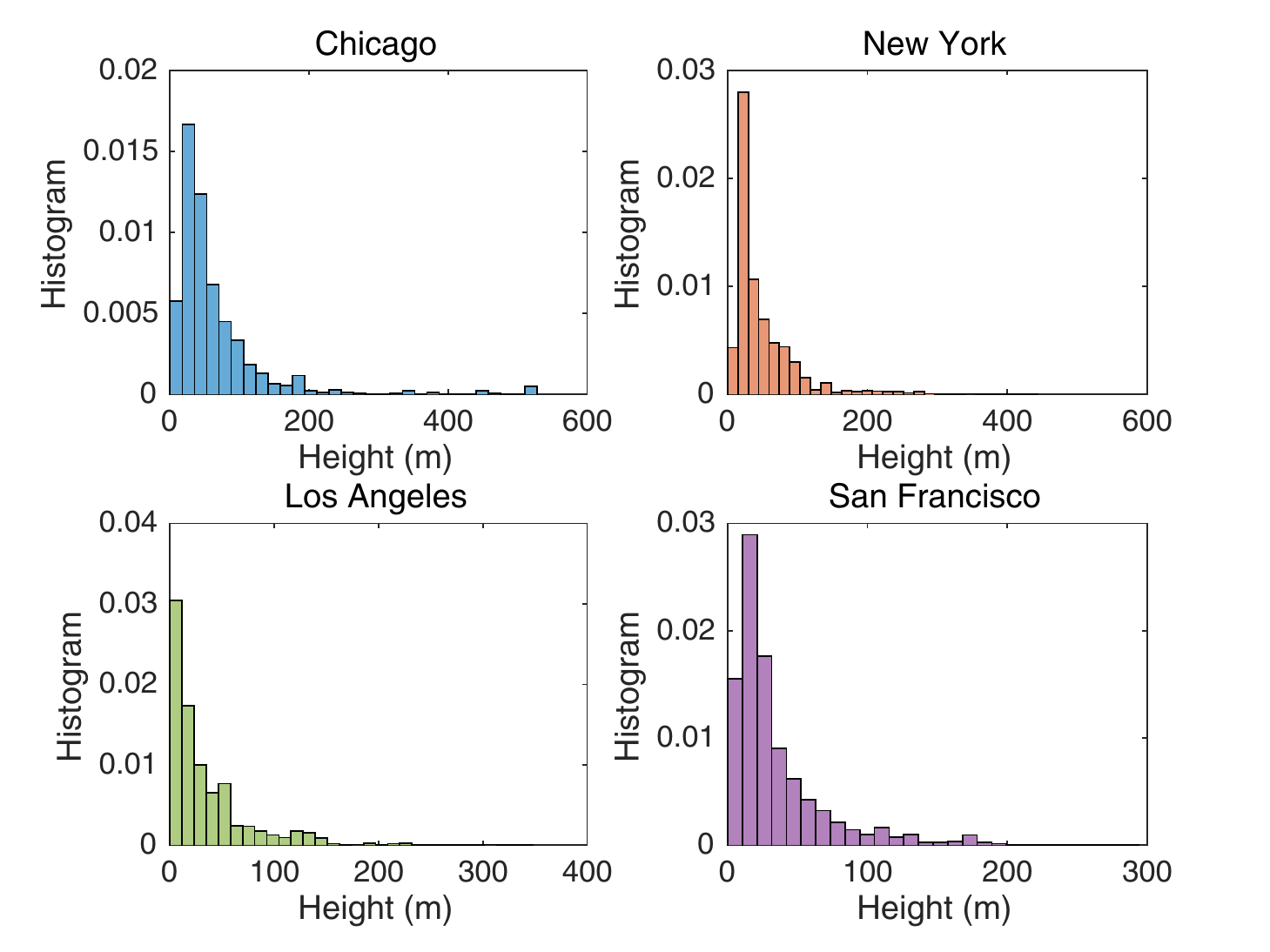}
			\caption{PDF of FSs' height}
			\label{fig:FSHeightPDF}
		\end{subfigure}
		\caption{FSs' spatial deployment.}
		\label{fig:FSspatial}
	\end{figure}

	We first analyze the spatial distribution of these FSs.In Fig. \ref{fig:FSDensityVsRadius}, we show the density of FSs around a city center with variations of the region radius, i.e., we compute the number of FSs in an area of a given radius, where the area is centered around one of the city's main hubs (e.g., Willis Tower for Chicago, the Empire State Building for New York, and the financial districts of Los Angeles and San Fransisco). It is evident that FSs are non-uniformly distributed over space, and specifically they tend to have higher density near city centers while they become very sparsely deployed in suburban areas as city centers have higher density of people, buildings, and attractions, elevating the need for denser fixed backhaul links. Overall, FSs have low density relative to existing cellular networks. 
	
	For each FS density shown in Fig. \ref{fig:FSDensityVsRadius}, we also compute the average height among those FSs deployed in a given area, and show in Fig. \ref{fig:FSHeightVsDensity} these heights for the different densities of FSs. It is shown that, except for San Francisco, the average height generally increases in denser areas compared to lightly dense areas, showing that the deployment height appears to be correlated with the average building heights in these areas. From the 5G coexistence perspective, this implies that the density of FSs in urban areas should not be worrisome as these stations tend to be deployed at altitudes that are above  5G cell sites. In contrast, FSs are likely to be deployed at relatively low heights in suburban areas, yet their density is very low in such regions. 
	
	Fig. \ref{fig:FSHeightCDF} and Fig. \ref{fig:FSHeightPDF} show the cumulative density function (CDF) and the probability density function (PDF) of the FSs' deployment height. The average and median heights are at least 34m and 19m, respectively. More importantly, 95\% of FSs are deployed above 12m for most metropolitan areas. Note that for LA, the fifth percentile is 2m, but this is relative to ground, i.e., many of FSs in LA are actually deployed on hills. Since 5G sites are expected to be deployed at heights of four to six meters, gNBs will be below the majority of FSs, limiting the 5G interference on FSs and vice versa. 
	
	\subsection{Antenna Specifications}
\begin{figure}[t!]
		\centering
		\begin{subfigure}[t]{0.4\textwidth}
			\centering
			\includegraphics[width=2.25in]{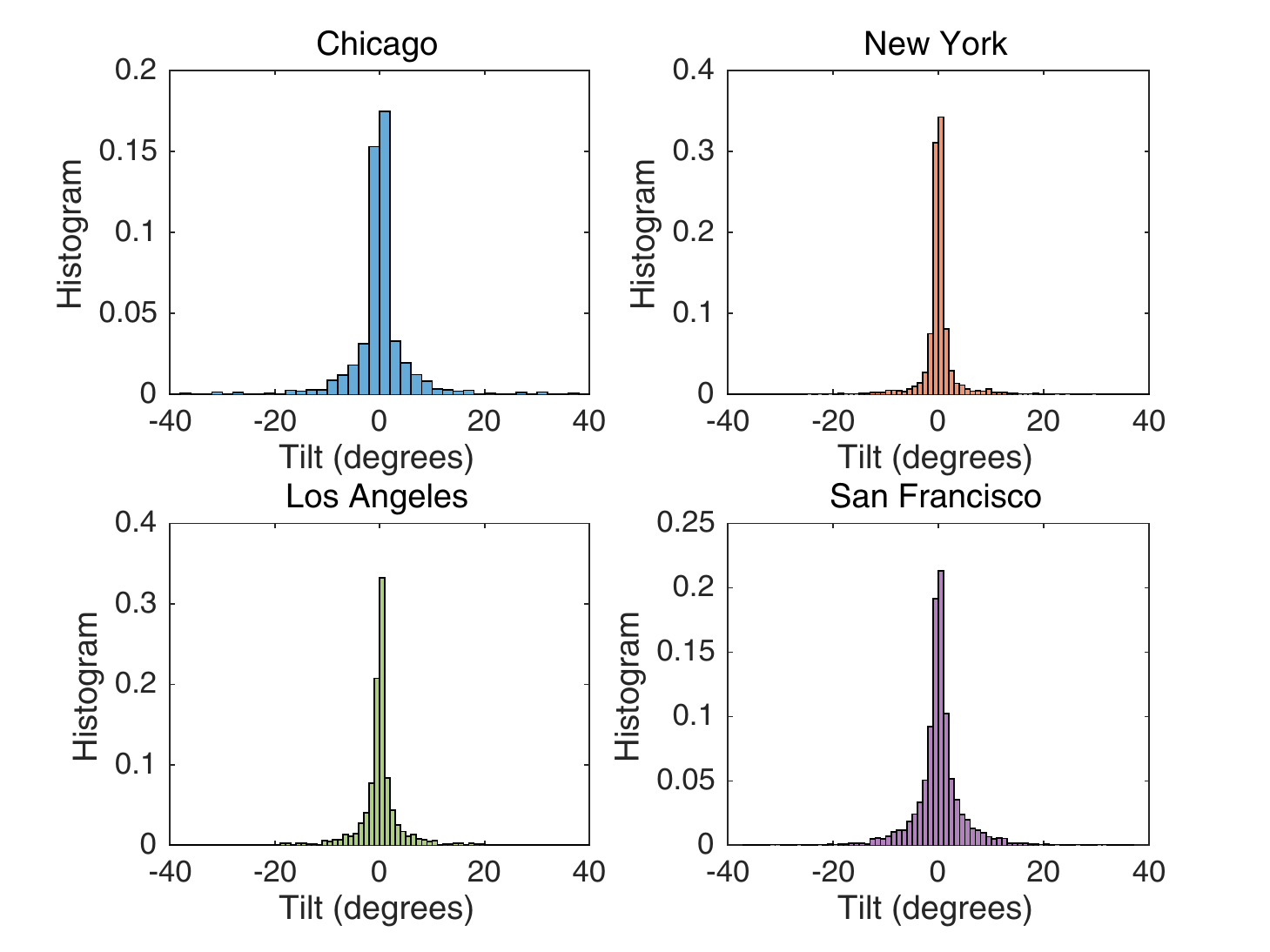}
			\caption{Tilt histograms}
			\label{fig:FSTilt}
		\end{subfigure}~~
		\begin{subfigure}[t]{.4\textwidth}
			\centering
			\includegraphics[width=2.25in]{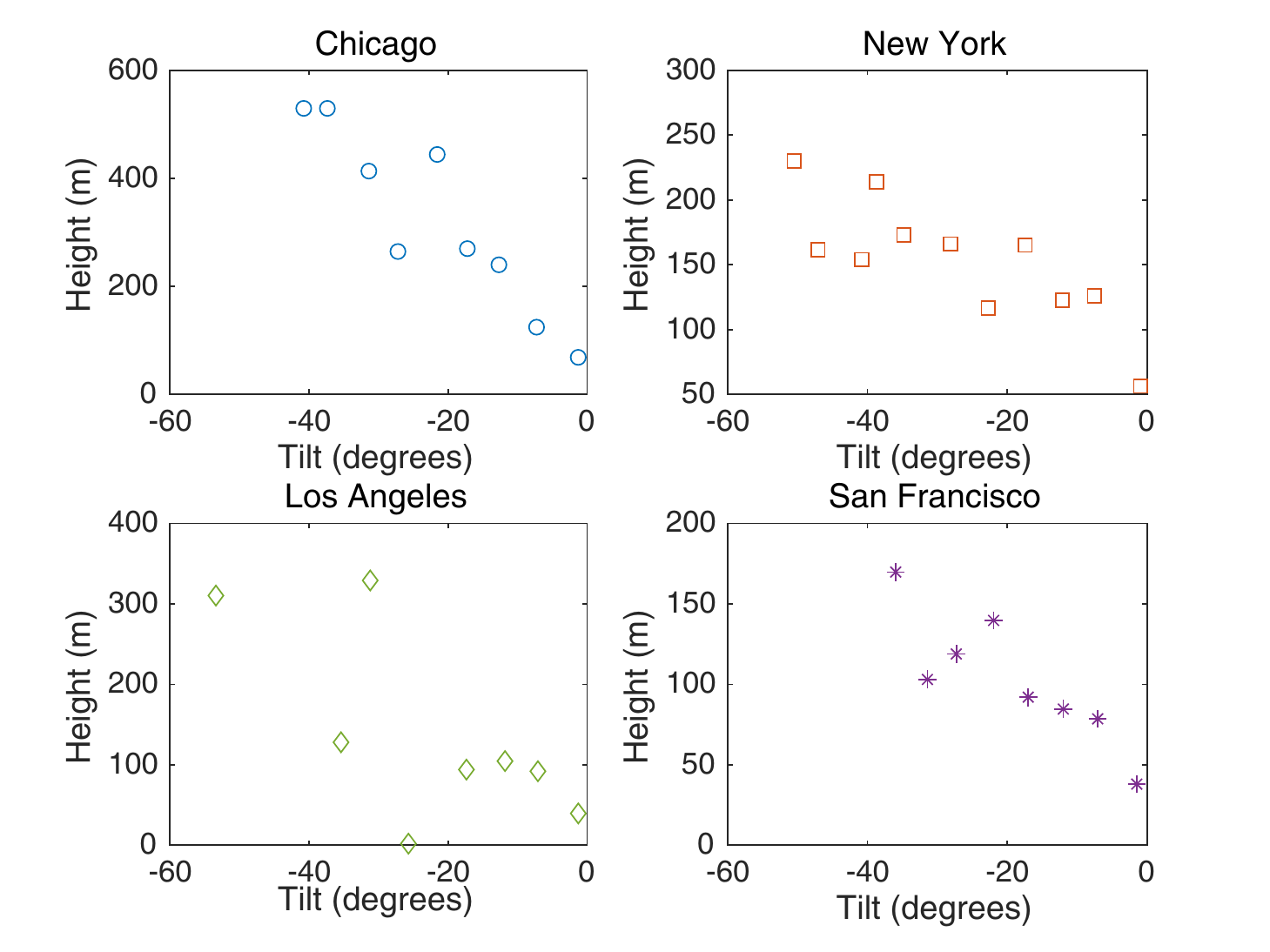}
			\caption{Average height for a given tilt}
			\label{fig:FSHeightTilt}
		\end{subfigure}
		\\
		\begin{subfigure}[t]{0.4\textwidth}
			\centering
			\includegraphics[width=2.25in]{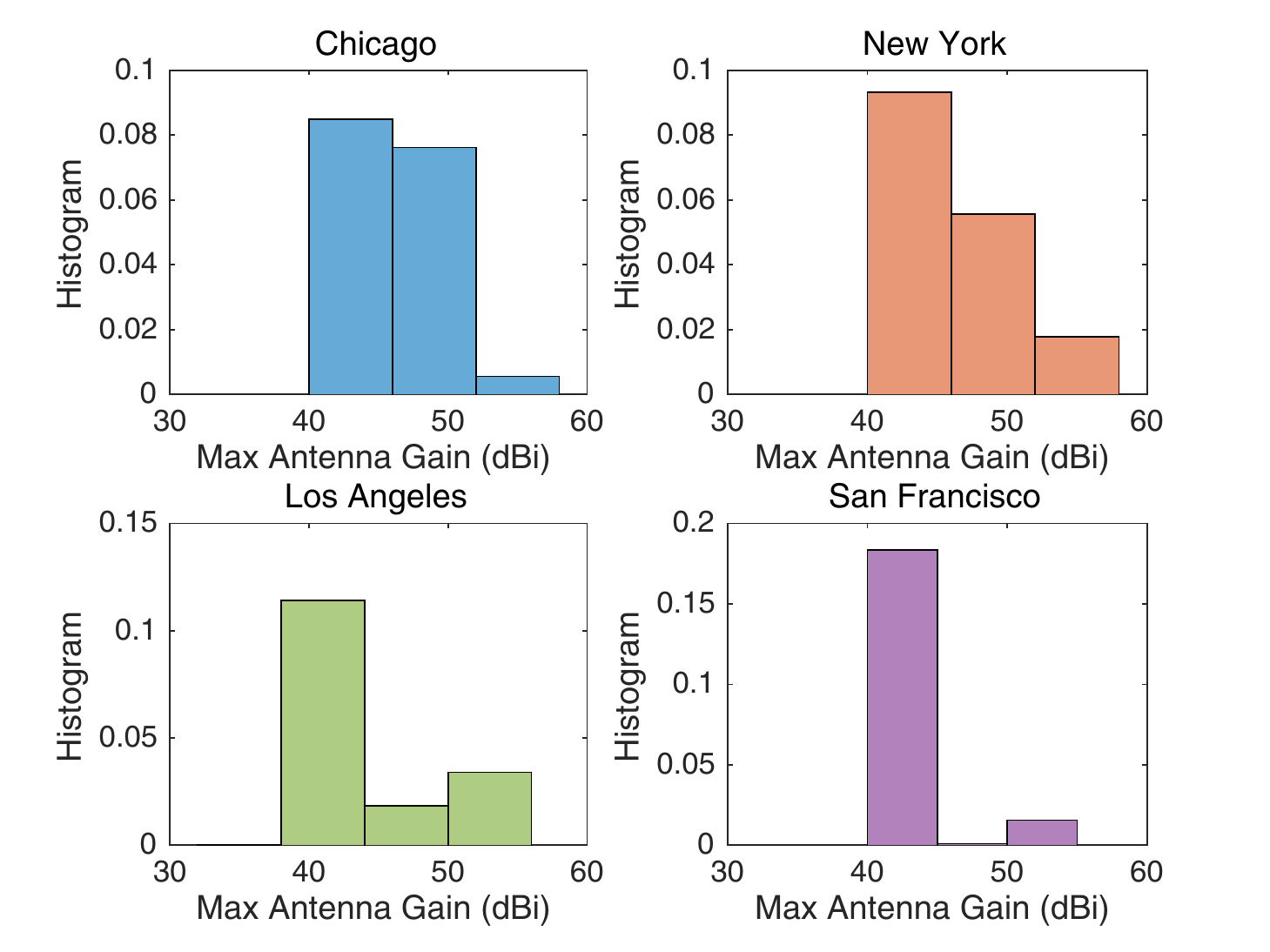}
			\caption{Antenna gain histograms}
			\label{fig:FSGain}
		\end{subfigure}~~
		\begin{subfigure}[t]{.4\textwidth}
			\centering
			\includegraphics[width=2.25in]{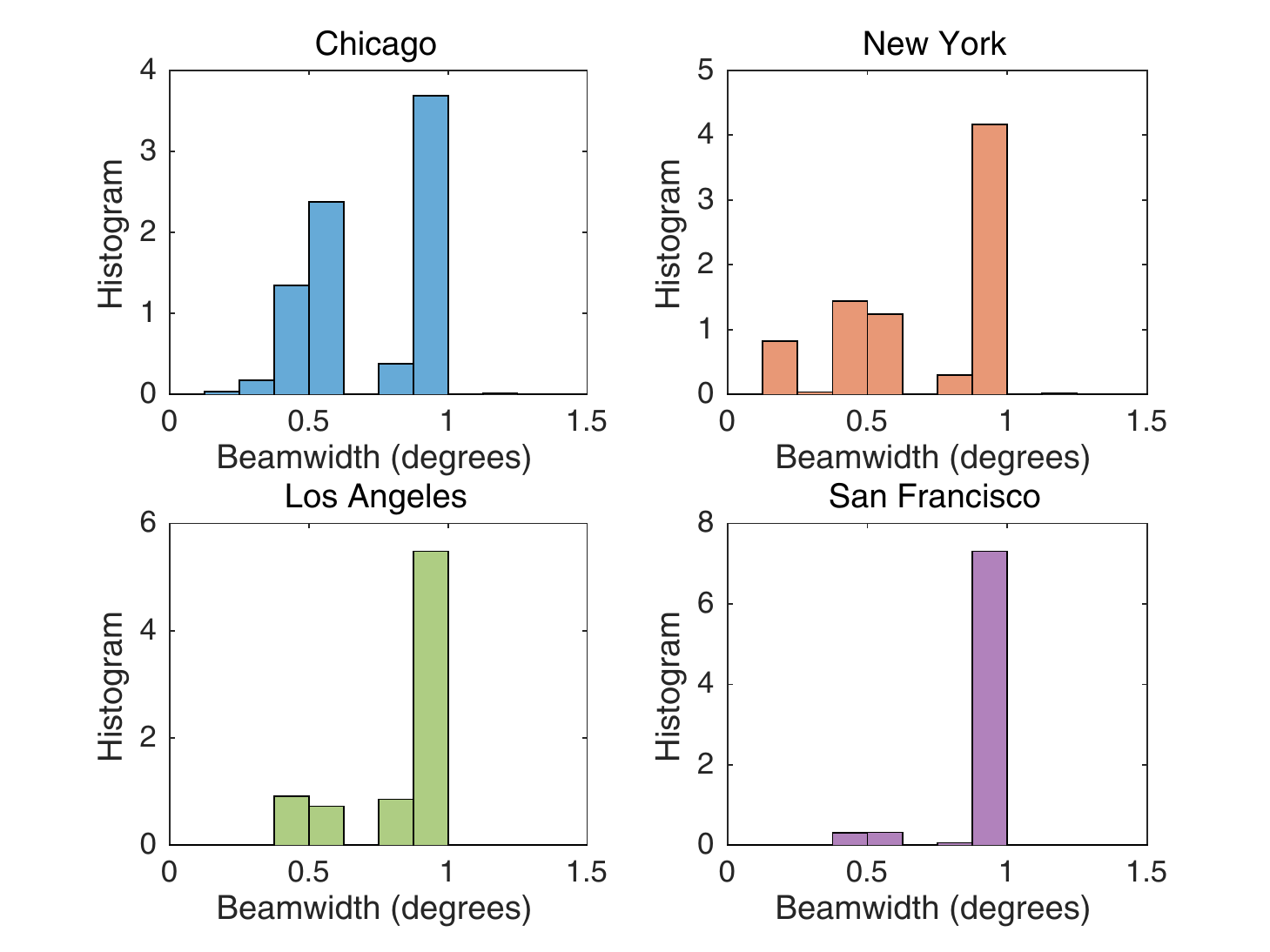}
			\caption{Beamwidth histograms}
			\label{fig:FSBeamwidth}
		\end{subfigure}
		\caption{FSs' antenna information}
		\label{fig:FSdeployment2}
	\end{figure}

	Another critical aspect of FSs' deployment is their physical antenna orientation. Fig. \ref{fig:FSHeightTilt} shows the histogram of the antenna's tilt, verifying that the vast majority of FSs have their tilt angles pointing horizontally. For instance, more than 93\% of FSs have their tilt angles within $[-10,10]$ degrees. There are only few FSs with high negative tilts, i.e., they point to the street level. These FSs, however, are typically deployed on top of high-rise buildings as verified in Fig. \ref{fig:FSHeightTilt}. In other words, there is a correlation between the deployment height and the negative tilt. Thus, although these FSs will have a higher chance to experience UE interference, as they point to the ground, 5G signals will typically experience a larger path loss given the height of these FSs. 
	
	Another key feature of FSs is their high antenna gain. Indeed, as shown in Fig. \ref{fig:FSGain}, the antenna gain is typically from 40dBi to 55dBi. Such high gains are necessary for long-range coverage at millimeter wave frequencies, but can be troublesome for other transmitter-receiver pairs in vicinity. For this reason, the maximum 3dB beamwidth, per FCC regulations, should be less than or equal to $1.2^{\circ}$ \cite{FCC2017}. This is verified in Fig. \ref{fig:FSBeamwidth}, where the vast majority of FSs have beamwidths at $1^{\circ}$. From a 5G coexistence perspective, the UE must be tightly aligned with the FS for it to cause tangible interference. Otherwise, most 5G signals will be highly attenuated, falling outside the FS receiver's beam (cf. Fig. \ref{fig:FSAntennaPattern}).

	\subsection{Comments on incumbent modeling and 5G coexistence}
	The aforementioned analysis of the different incumbents' databases helps provide several modeling guidelines of incumbent FSs. For instance, using the popular homogeneous Poisson Point Process (HPPP) \cite{Haenggi2012} to model the locations of FSs may not be practical if the region of interest is large, as FSs tend to be non-uniformly distributed over space. In addition, due to the disparities between the height of FS deployment and the 5G mmWave deployment, it is more meaningful to consider three-dimensional stochastic processes (or two-dimensional processes with the third dimension being a constant that reflects the mean height of the buildings in a given area). For antenna parameters, it is observed the majority of FSs have similar characteristics, and thus it suffices to assume all of them have the same antenna gain and beamwidth, and further assume they point horizontally in elevation.
	
	From a coexistence perspective, the deployment strategy of FSs is favorable for future 5G deployment over 70GHz and 80GHz for the following reasons:
	\begin{itemize}
		\item FSs are generally deployed above 12m, whereas 5G cell sites will be only at 4 to 6 meters above the ground for street-level deployment, and hence they will be well below FSs.
		\item The vast majority of FSs are oriented horizontally, i.e., they are directed above 5G deployments. For the few FSs that point to the street level, these are typically at high altitudes, increasing the path loss between the UE and the FS. 
		\item The ultra-narrow beamwidths of FSs can help significantly attenuate UE interfering signals when they fall outside the main lobe. 
	\end{itemize}
	
	\section{Analysis of UE Interference on FSs}\label{sec:interferenceAnalysis}
	In this section, we present our framework to compute the aggregate interference from the 5G system into incumbent systems. The approach used is applicable to the coexistence of any two wireless communication systems that rely on directional beams.
	
	We focus on the 5G system operating in the uplink mode, i.e., we study the UE interference into FSs, for the following reasons. First, UEs typically have positive tilt angles compared to 5G gNBs, and thus the former are more likely to interfere with FSs. Second, the mobility of UEs makes their locations appear random, while gNBs' deployment can be optimized to ensure minimal interference on FSs. We note that in the Simulations Section, we show that although gNBs have higher transmit power and antenna gains, the 5G DL aggregate interference is not higher than 	that of the UL, primarily because gNBs tend to have beams pointing to the ground.
	
	The interference seen at a victim FS is an aggregation of all UEs transmitting in the UL to their respective gNBs. This interference depends mainly on three components: (i) The path loss between the UE and the FS, (ii) the attenuation due to the FS's antenna pattern, and (iii) the attenuation due to the UE's antenna pattern. We describe each one in details next. In what follows, $\mathbf{x}_{\ue}$, $\mathbf{x}_{\f}$, and $\mathbf{x}_{\f,{\operatorname{tx}}}$ denote the $(x,y)$-coordinates of the interfering UE, the victim FS receiver, and the corresponding FS transmitter, respectively. In addition, $d_{\operatorname{a}\rightarrow\operatorname{b}}$ denotes the 2D distance between $\operatorname{a}$ and $\operatorname{b}$ while $\mathbf{a}\bullet\mathbf{b}$ denotes the dot product between two vectors $\mathbf{a}$ and $\mathbf{b}$, i.e., $\mathbf{a}^{\textsf{T}}\mathbf{b}$.
	
	
	\subsection{Path Loss between a User and a Fixed Station}
	Signals can be significantly attenuated if they are blocked by objects at mmWave frequencies, i.e., it is critical to consider whether the link is LOS or NLOS for path loss computations at such high frequencies. To this end, we use the 3GPP path loss model \cite{3GPP2017}, which is expressed, in dB, as\footnote{The model can be generalized to include indoor losses and indoor-to-outdoor penetration losses, when UEs are located indoors, as shown in \cite{3GPP2017}.}
	\begin{equation}
	\operatorname{PL}_{\ue\rightarrow\f} = \mathbf{1}_{(\beta=0)} \operatorname{PL}_{\operatorname{LOS}}(\mathbf{x}_{\ue},\mathbf{x}_{\f},h_{\ue},h_{\f},f_c)+\mathbf{1}_{(\beta=1)} \operatorname{PL}_{\operatorname{NLOS}}(\mathbf{x}_{\ue},\mathbf{x}_{\f},h_{\ue},h_{\f},f_c)+\mathcal{X}(\beta),
	\end{equation}
	where $\operatorname{PL}_{\operatorname{LOS}}(\cdot)$ is the line-of-sight (LOS) path loss, $\operatorname{PL}_{\operatorname{NLOS}}(\cdot)$ is the non-LOS (NLOS) path loss, $\mathcal{X}$ is the log-normal shadow fading, $\beta\in\{0,1\}$ is a binary variable that indicates whether the UE-FS is blocked by a building or not, and $\mathbf{1}(\cdot)$ is the indicator function. We note that $\operatorname{PL}_{\operatorname{LOS}}$ and $\operatorname{PL}_{\operatorname{NLOS}}$ are functions of the distance between the UE and the FS, their heights, and the center frequency $f_c$, as given in \cite{3GPP2017}. Essentially, the path loss is a multi-slope model with different path loss exponents depending on the distance between the UE and the FS. Also, the standard deviation of the log-normal shadow fading depends on whether the link is LOS or NLOS \cite{3GPP2017}. 
	
	In this work, we rely on actual building layouts to determine whether the link is LOS or NLOS. We intentionally ignore blockage by other objects, e.g., foliage \cite{Thomas2014}, cars, etc., to emulate a worst case scenario as additional blockage should reduce the interference. We note that in the Simulations Section, we compare the LOS probability using the actual building layouts with the theoretical LOS probability used by the 3GPP model, which is expressed as \cite{3GPP2017}
	\begin{equation}
	\label{eq:3GPPLOS}
	\mathbb{P}_{\operatorname{LOS}}(d_{\ue\rightarrow\f})= \min\left\{\frac{18}{d_{\ue\rightarrow\f}},1\right\}\times\left(1-\exp\left(-\frac{d_{\ue\rightarrow\f}}{36}\right)\right) + \exp\left(-\frac{d_{\ue\rightarrow\f}}{36}\right).
	\end{equation}

\begin{figure}[t!]
		\center
		\includegraphics[width=3in]{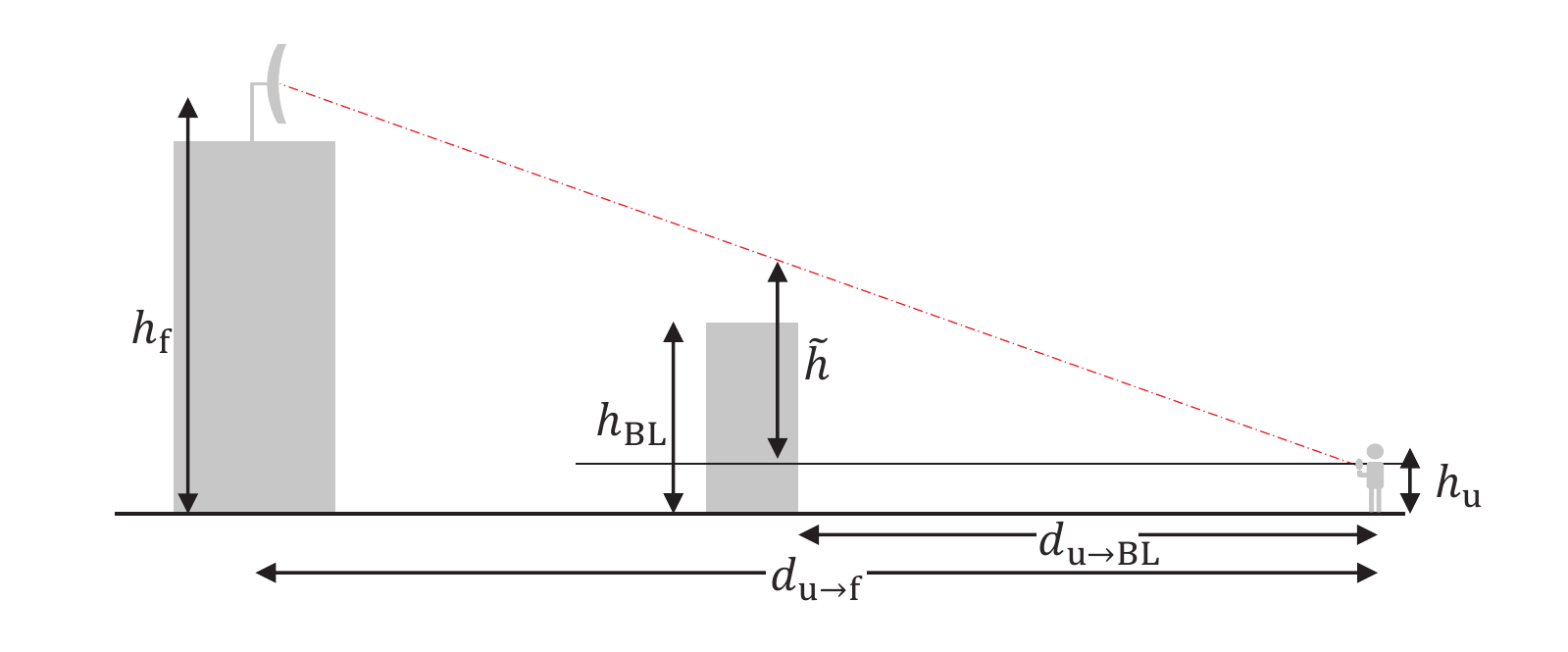} 
		\caption{A blockage event in 3D occurs when $\tilde h+h_{\ue}\leq h_{\operatorname{BL}}$.}
		\label{fig:blockageCheck}
	\end{figure}

	As stated earlier, we define a blockage event as having the UE-FS blocked by a building. This is computed as follows. Assuming the $xy$-plan represents the ground, we first check whether the line that connects between the UE and the FS is blocked by a building, which is defined as a 2D polygon. If the polygon does intersect with the line, we then check whether it blocks the line with the 3D version of the polygon, where the third dimension is the building's height,  $h_{\operatorname{BL}}$. Specifically, let $d_{\ue\rightarrow\operatorname{BL}}$ be the distance between the UE and the building and $d_{\ue\rightarrow\f}$ be the distance between the UE and the FS.  Then, a blockage event occurs if $\tilde h + h_{\ue}\leq h_{\operatorname{BL}}$, where
	\begin{equation}
	\tilde h = d_{\ue\rightarrow\operatorname{BL}} \times \tan \left(\tan^{-1}\left(\frac{h_{\f}-h_{\ue}}{d_{\ue\rightarrow\f}}\right)\right).
	\end{equation}
	This is visualized in Fig. \ref{fig:blockageCheck}.
	
	\subsection{Attenuation due to FS Antenna Pattern}
	As illustrated in Fig. \ref{fig:FSAntennaPattern}, a small misalignment between the received signal and the FS's antenna boresight results in significant attenuation. Thus, it is critical to accurately compute the interfering signal angle-of-arrival at the FS antenna. Define the line connecting the UE to the FS as the \emph{interference axis}. Let the off-axis azimuth angle $\theta_{\f\rightarrow\ue}^{\operatorname{off}}$ be the angle between the FS's antenna boresight and the interference axis, then we have 
	\begin{equation}
	\theta_{\f\rightarrow\ue}^{\operatorname{off}} = \cos^{-1}\left(\mathbf{u}_{\f\rightarrow\f,{\operatorname{tx}}}\bullet\mathbf{u}_{\f\rightarrow\ue}\right),
	\end{equation}
	where $\mathbf{u}_{\f\rightarrow\f,{\operatorname{tx}}}= \frac{\mathbf{x}_{\f,{\operatorname{tx}}}-\mathbf{x}_{\f}}{\|\mathbf{x}_{\f,{\operatorname{tx}}}-\mathbf{x}_{\f}\|}$ is the unit vector in the azimuth direction of the FS's antenna boresight, and $\mathbf{u}_{\f\rightarrow\ue}=\frac{\mathbf{x}_{\ue}-\mathbf{x}_{\f}}{\|\mathbf{x}_{\ue}-\mathbf{x}_{\f}\|}$ is the unit vector from the FS's antenna towards the UE.  Similarly, let $\phi_{\f\rightarrow\ue}^{\operatorname{off}}$ be the off-axis elevation angle, then it can be shown that
	\begin{equation}
	\phi_{\f\rightarrow\ue}^{\operatorname{off}}= \tan^{-1}\left(\frac{h_{\f}-h_{\ue}}{d_{\f\rightarrow\ue}}\right)+\phi_{\f},
	\end{equation}
	where $ \phi_{\f}$ is the FS's antenna tilt. All these vectors and  off-axis angles are shown in Fig. \ref{fig:FSangles}. Finally, the combined azimuth and elevation attenuation at the FS victim receiver is expressed as 
	\begin{equation}
	\label{eq:FSpattern}
	G_{\f\rightarrow\ue}=
	G_{\f,\operatorname{max}} - \min\left\{A_{\f}(\theta_{\f\rightarrow\ue}^{\operatorname{off}})+A_{\f}(\phi_{\f\rightarrow\ue}^{\operatorname{off}}),A_{\f,\operatorname{FTBR}}\right\},
	\end{equation}
	where $G_{\f,\operatorname{max}}$ is the maximum antenna gain in dBi, $A_{\f,\operatorname{FTBR}}$ is the front-to-back ratio loss (FTBR) in dB, and $A_{\f}(\cdot)$ is the attenuation for a given off-axis angle, and it corresponds to the antenna pattern that matches the FCC regulations (cf. Fig. \ref{fig:FSAntennaPattern}) \cite{FCC2017}.

	\subsection{UE Radiated Power (EIRP) Into FS Antenna}
	
	\subsubsection{Actual directions}
\begin{figure}[t!]
		\centering
		\begin{subfigure}[t]{.45\textwidth}
			\centering
			\includegraphics[width=3.25in]{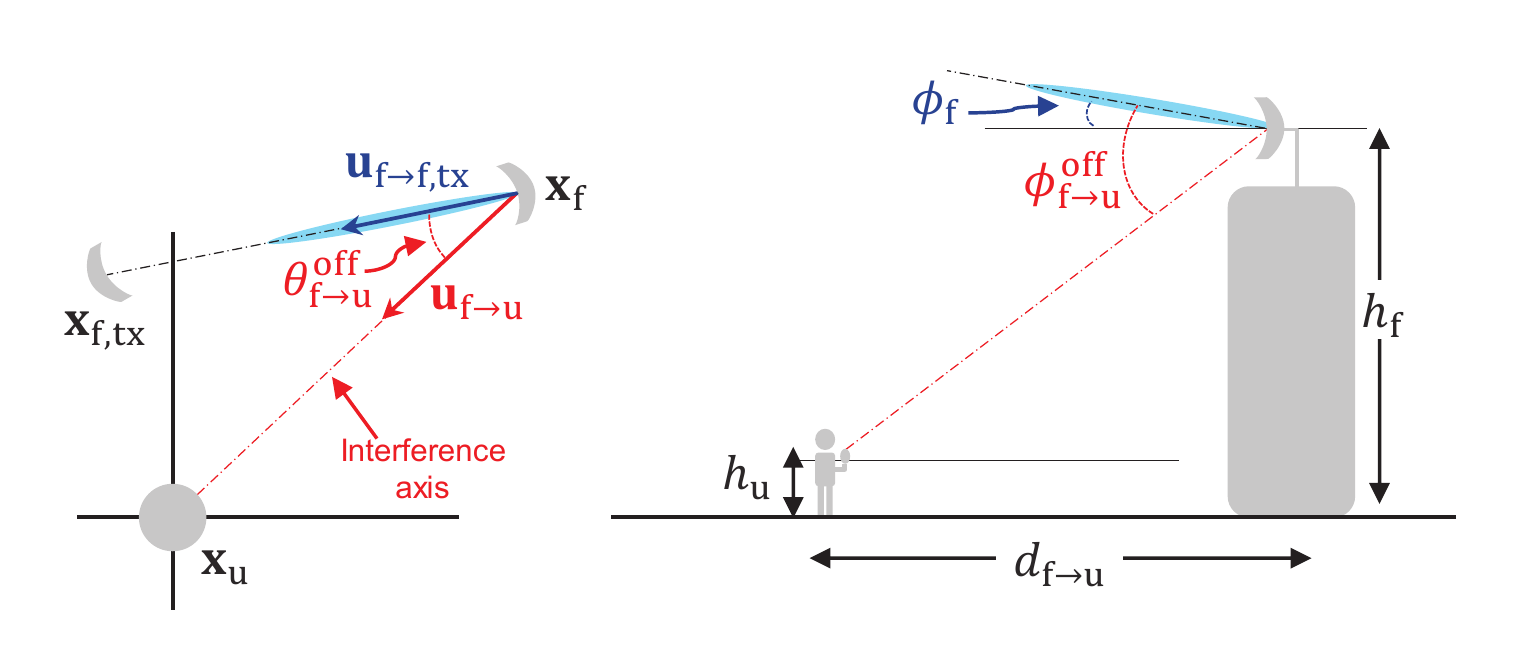}
			\caption{With respect to the FS}
			\label{fig:FSangles}
		\end{subfigure}~~
		\begin{subfigure}[t]{.45\textwidth}
			\centering
			\includegraphics[width=3.25in]{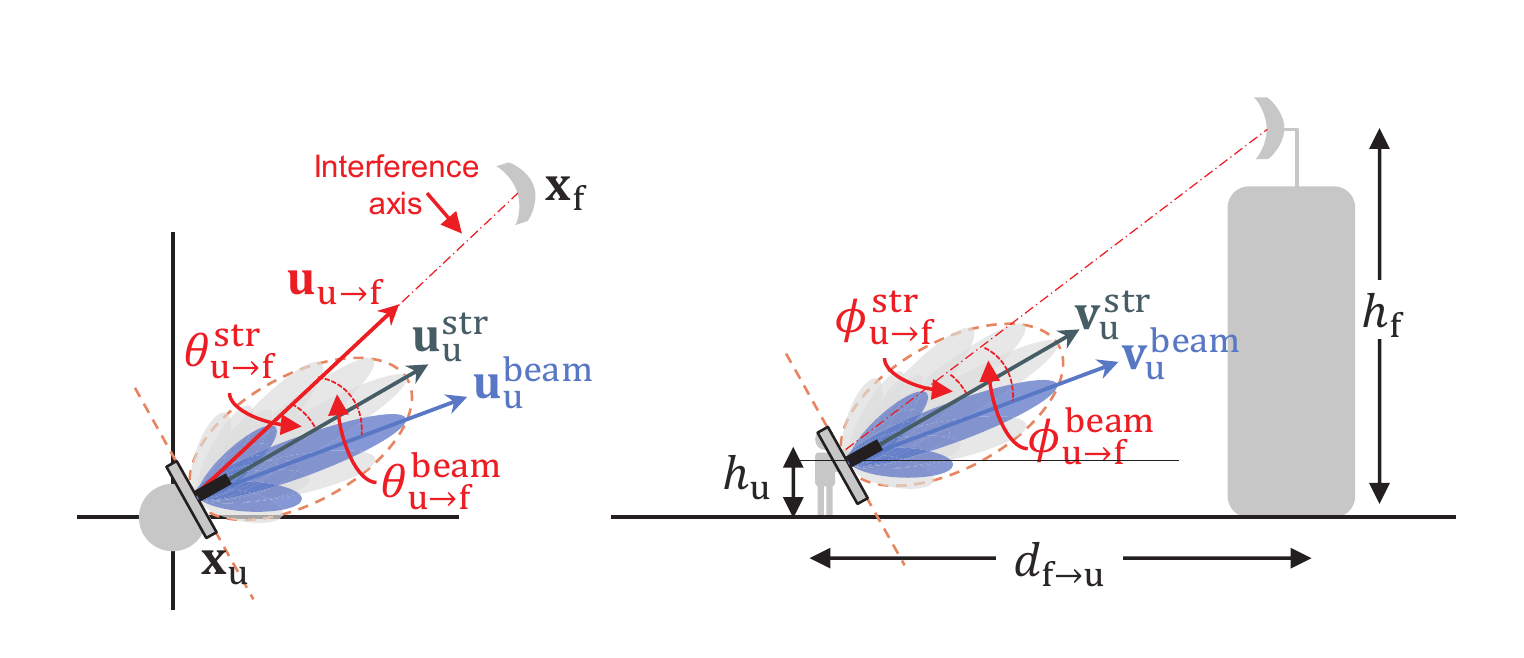}
			\caption{With respect to the UE}
			\label{fig:UEangles}
		\end{subfigure}
		\caption{Off-axis azimuth and elevation angles}
		\label{fig:AntennaPatterns}
	\end{figure}

	The directions of the UE's two opposite panels are defined by unit vectors in the direction of the panels' boresight. We assume that these directions are random in azimuth such that the boresight of the first one is distributed uniformly as $\mathcal{U}(0,180)$ while the other one is pointing in the opposite direction, i.e., 180$^\circ$ from the first one. Only one of the UE antenna panels is active during data communications. Let $\mathbf{u}_{\ue}^{\operatorname{str}}$ denote the unit vector in the azimuth direction of the UE's panel that is active. Also, let $\mathbf{u}_{\ue}^{\operatorname{beam}}$ denote the unit vector in the azimuth direction of the main lobe of the UE's beam used in the UL, which corresponds to the beam with the maximum received power during user and beam association. We similarly define $\mathbf{v}_{\ue}^{\operatorname{str}}$ and $\mathbf{v}_{\ue}^{\operatorname{beam}}$ for the elevation directions. Then, the total radiated power from the UE into the direction of the victim FS is expressed, in dBm, as 
	\begin{equation}
	\begin{aligned}
	E_{\ue\rightarrow\f} &= P_{\ue} + 10\log_{10}(2N_{\ue,\operatorname{h}}N_{\ue,\operatorname{v}})+G_{\ue,\operatorname{max}} - (A_{\ue,\operatorname{BP}}(\theta_{\ue\rightarrow\f}^{\operatorname{beam}})+A_{\ue,\operatorname{BP}}(\phi_{\ue\rightarrow\f}^{\operatorname{beam}}))\\ &-\min\{A_{\ue,\operatorname{EP}}(\theta_{\ue\rightarrow\f}^{\operatorname{str}})+A_{\ue,\operatorname{EP}}(\phi_{\ue\rightarrow\f}^{\operatorname{str}}),A_{\ue,\operatorname{FTBR}}\},
	\end{aligned}
	\end{equation}
	where $G_{\ue,\operatorname{max}}$ is the maximum antenna gain and $A_{\ue,\operatorname{FTBR}}$ is the FTBR loss. The azimuth off-axis angles are computed as
	\begin{equation}
	\theta_{\ue\rightarrow\f}^{\operatorname{beam}}=\cos^{-1}\left(\mathbf{u}_{\ue\rightarrow\f}\bullet\mathbf{u}_{\ue}^{\operatorname{beam}}\right),
	\end{equation}
	and 
	\begin{equation}
	\theta_{\ue\rightarrow\f}^{\operatorname{str}}=\cos^{-1}\left(\mathbf{u}_{\ue\rightarrow\f}\bullet\mathbf{u}_{\ue}^{\operatorname{str}}\right),
	\end{equation}
	where $\mathbf{u}_{\ue\rightarrow\f}=-\mathbf{u}_{\f\rightarrow\ue}$. The elevation off-axis angles are computed as 
	\begin{equation}
	\phi_{\ue\rightarrow\f}^{\operatorname{beam}}= \tan^{-1}\left(\frac{h_{\f}-h_{\ue}}{d_{\f\rightarrow\ue}}\right)- \angle{\mathbf{v}_{\ue}^{\operatorname{beam}}},
	\end{equation}
	and 
	\begin{equation}
	\phi_{\ue\rightarrow\f}^{\operatorname{str}}=\tan^{-1}\left(\frac{h_{\f}-h_{\ue}}{d_{\f\rightarrow\ue}}\right)- \angle{\mathbf{v}_{\ue}^{\operatorname{str}}},
	\end{equation}
	where $\angle{\cdot}$ denotes the angle of the vector. All of the relevant vectors and off-axis angles are illustrated in Fig. \ref{fig:UEangles}.

	\subsubsection{Random directions}
	We also present a random model for the 	UE's azimuth and elevation directions. This model does not require the deployment of gNBs, and hence ignores the computational complexity in simulating user and beam association. The model assumes that the UE uses the beam in the direction of the antenna's main boresight, i.e.,  $\mathbf{u}_{\ue}^{\operatorname{beam}}=\mathbf{u}_{\ue}^{\operatorname{str}}$ and $\mathbf{v}_{\ue}^{\operatorname{beam}}=\mathbf{v}_{\ue}^{\operatorname{str}}$. To this end, we model the azimuth direction as a uniform random variable $\tilde \theta_{\ue}\sim\mathcal{U}(0,360)$, whereas the elevation direction is modeled as
	\begin{equation}
	\label{eq:randomEl}
	\tilde \phi_{\ue} = \tan^{-1}\left(\frac{h_{\g}-h_{\ue}}{d_{\g\rightarrow\ue}}\right),
	\end{equation} 
	where $d_{\g\rightarrow\ue}\sim\mathcal{U}(d_0,d_{\operatorname{ISD}}/2)$, where $d_0>0$ is some constant, e.g., in this work we consider $d_0=10$m. These angles are used to compute the unit vectors needed for azimuth and elevation off-axis angles. 
	
	\subsection{UE Aggregate Interference}
	The interference caused by the $i$-th UE on the FS is given as 
	\begin{equation}
	I_{i,\operatorname{dBm}} = E_{i\rightarrow\f}+ G_{\f\rightarrow i} - \operatorname{PL_{i\rightarrow\f}}.
	\end{equation}
	We use the interference-to-noise ratio (INR) to determine the effect of 5G interference on the incumbent, which is expressed as
	\begin{equation}
	\operatorname{INR}_{\operatorname{dB}}= 10\log_{10}(I_{\operatorname{agg}}) - \left(10\log_{10}(N_0 B)+F_{\f}\right),
	\end{equation}
	where $I_{\operatorname{agg}}=\sum_{i} 10^{I_{i,\operatorname{dBm}}/10}$, $N_0$ is the noise power spectral density (mW/Hz), $B$ is the bandwidth (Hz), and $F_{\f}$ is the noise figure of the FS (dB).
	
	\section{Passive Interference Mitigation Techniques}\label{sec:mitigation}
	In this section, we propose several interference mitigation techniques to protect the incumbent FSs. We focus on two critical aspects. First, the techniques should be passive, i.e., they do not require any coordination with FSs, and second they should be practical to implement to appeal for mobile operators and vendors.
	
	\subsection{Sector-based Mitigation}
	In this technique, we propose to switch off sectors, creating sector-based exclusion zones. The key idea is that the 5G UE beam directions are typically reciprocal to those of 5G gNBs. Thus, if such reciprocal directions point to FSs, then the UE must be discouraged from using them, i.e., the sector with a reciprocal direction pointing towards the FSs should be switched off. More formally, let $\mathbf{u}_{\g}^{\operatorname{str},i}$ be the unit vector in the direction of the $i$-th sector boresight and $-\mathbf{u}_{\g}^{\operatorname{str},i}$ is its reciprocal direction. Then, the $i$-th sector is switched off if
	\begin{equation}
	\label{eq:secLoc}
	s_{l,i} =\left\{
	\begin{array}{ll}
	1,	& \cos^{-1}\left(-\mathbf{u}_{\g}^{\operatorname{str},i}\bullet \mathbf{u}_{\g\rightarrow\f}\right) \leq \psi_s\\
	0,	& \text{otherwise}
	\end{array}\right.,
	\end{equation}
	where $\psi_s$ is a predetermined decision threshold. 
	A more relaxed sector exclusion criterion is to switch sectors off if they are not only aligned with the FS's location but also its antenna orientation. Such criterion can still reduce the interference experienced at FSs as a slight mis-alignment with FS's antenna incurs significant signal attenuation. More formally, the $i$-th sector can be switched off if
	\begin{equation}
	\label{eq:secOr}
	s_{o,i} = \boldsymbol{1}_{(s_{l,i}=1)}\times \boldsymbol{1}_{(\cos^{-1}(\mathbf{u}_{\g}^{\operatorname{str},i}\bullet \mathbf{u}_{\f\rightarrow\operatorname{f,tx}}) \leq \psi_s)}.
	\end{equation}
	We refer to (\ref{eq:secLoc}) as location-based mitigation and (\ref{eq:secOr}) as orientation-based mitigation. Both techniques are demonstrated in Fig. \ref{fig:sectorBased}. 
	
	\subsection{Beam-based Mitigation}
\begin{figure}[t!]
		\centering
		\begin{subfigure}[t]{1\textwidth}
			\centering
			\includegraphics[width=4in]{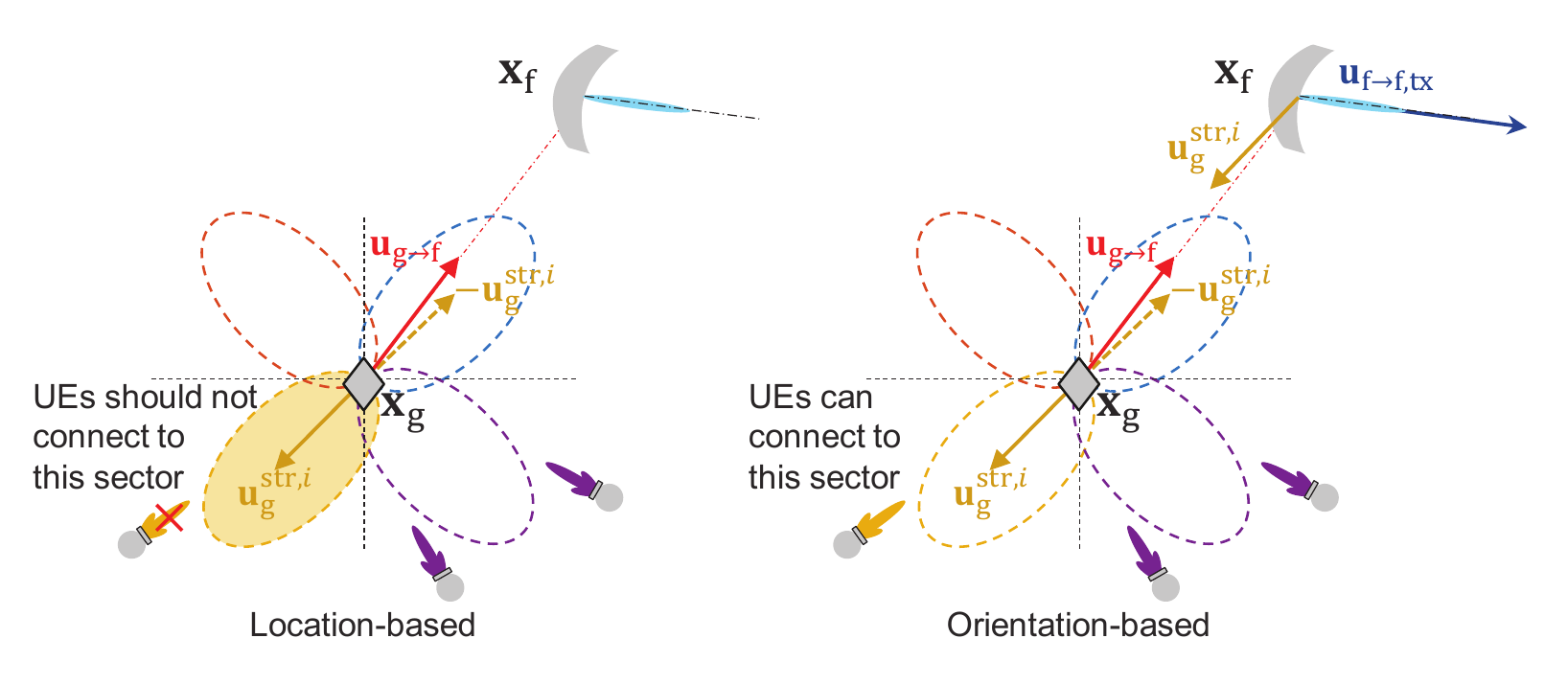}
			\caption{Sector-based mitigation}
			\label{fig:sectorBased}
		\end{subfigure}\\
		\begin{subfigure}[t]{1\textwidth}
			\centering
			\includegraphics[width=4in]{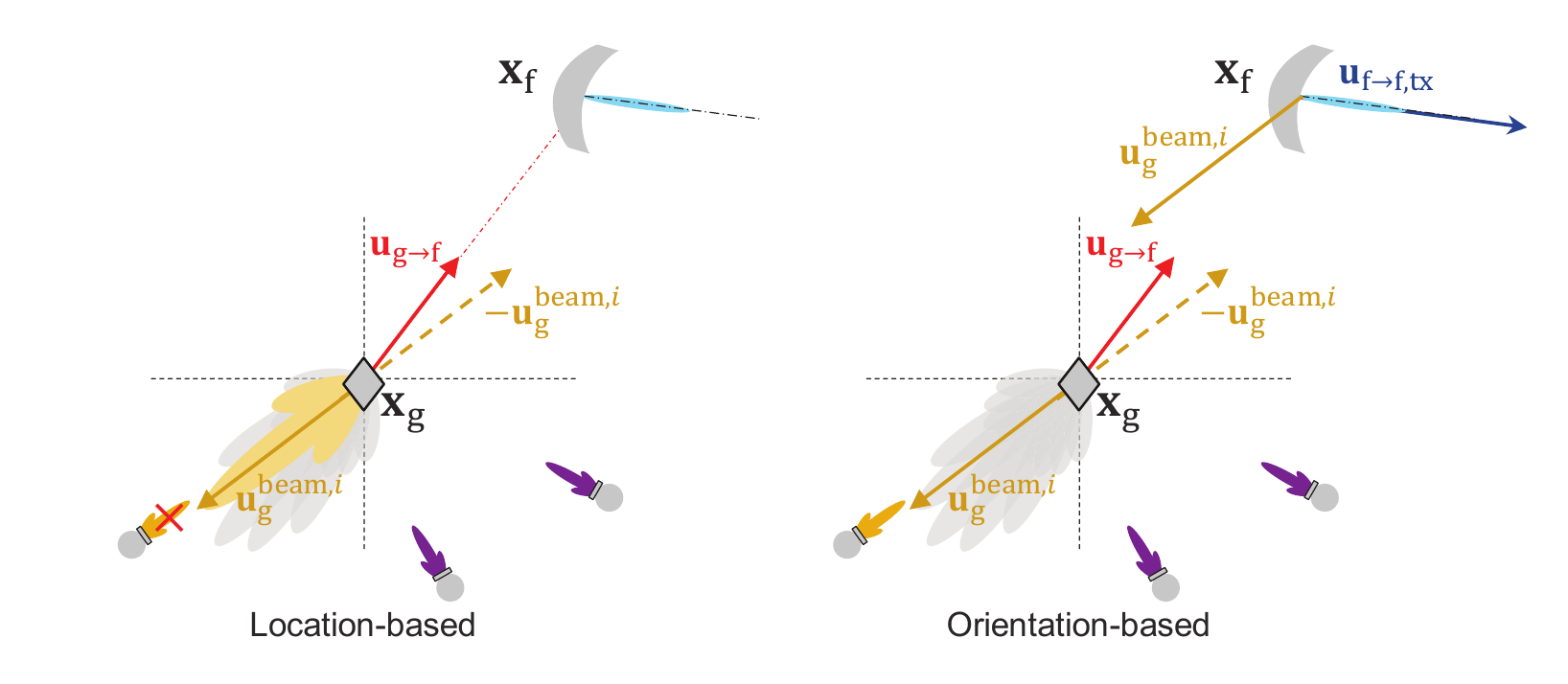}
			\caption{Beam-based mitigation}
			\label{fig:beamBased}
		\end{subfigure}
		\caption{Illustration of passive mitigation techniques}
		\label{fig:ExclusionZones}
	\end{figure}

	In the sector-based mitigation, only four decisions need to be made a priori for each gNB, making the approach simple to implement. This, however, may result in tangible coverage holes, affecting the performance of the 5G system. To this end, we can make exclusion zones at a finer scale, where decisions are made on a beam-by-beam basis instead. Specifically, the $i$-th beam is switched off if 
	\begin{equation}
	\label{eq:beamLoc}
	b_{l,i} =\left\{
	\begin{array}{ll}
	1,	& \cos^{-1}\left(-\mathbf{u}_{\g}^{\operatorname{beam},i}\bullet \mathbf{u}_{\g\rightarrow\f}\right) \leq \psi_b\\
	0,	& \text{otherwise}
	\end{array}\right.,
	\end{equation}
	where $\psi_b$ is a predetermined beam decision threshold and $\mathbf{u}_{\g}^{\operatorname{beam},i}$ is a unit vector in the direction of the $i$-th beam. In other words, the same sector could have beams switched on and beams switched off, depending on whether the beam meets the criterion in (\ref{eq:beamLoc}) or not. We can also make decisions based on the orientation of the FS's along with its location, i.e., 
	\begin{equation}
	\label{eq:beamOr}
	b_{o,i} = \boldsymbol{1}_{(b_{l,i}=1)}\times \boldsymbol{1}_{(\cos^{-1}(\mathbf{u}_{\g}^{\operatorname{beam},i}\bullet \mathbf{u}_{\f\rightarrow\operatorname{f,tx}}) \leq \psi_b)}.
	\end{equation}
	Beam-based exclusion zone is shown in Fig. \ref{fig:beamBased}. 
	
	\subsection{Spatial Power Control}
	%

	The aforementioned techniques can be classified as \emph{angular} exclusion zones, leading inevitably to lower downlink coverage with higher degradation if sector-based zones are used instead of beam-based. Alternative to switching beams (or sectors) off, we can implement power control, where the key idea is to transmit at lower power for beams that have higher alignment with the incumbent receiver. In this paper, we seek a simple \emph{binary} power control algorithm, where two power levels can be used depending on whether the beam is classified as a \emph{regular} beam or  as a \emph{quiet} one. Specifically, if the off-angle between the beam reciprocal direction and the incumbent receiver is below a predetermined threshold, then the beam is classified as \emph{quiet} or \emph{almost blank}, and thus the UE will transmit at low power. If the beam is not aligned with the incumbent, then the UE transmits at the maximum allowable power. To summarize, we have the following UL power control
	\begin{equation}
	\label{eq:PC}
	P_{\operatorname{UL},\ue,i^\star} = \left\{
	\begin{array}{ll}
	P_{\operatorname{lo}},  &\cos^{-1}\left(-\mathbf{u}_{\g}^{\operatorname{beam},i^\star}\bullet \mathbf{u}_{\g\rightarrow\f}\right) \leq \psi_b\\
	P_{\operatorname{up}},  &\text{otherwise}
	\end{array}\right.,
	\end{equation} 
	where $i^\star$ is the index of the gNB beam that the UE connects to, and $P_{\operatorname{lo}}$ and $P_{\operatorname{up}}$ represent the low and high transmit powers, respectively.  Note that it is natural to extend this approach to sectors or make it with respect to the orientation of the FS instead of its location.
	
	\emph{Remark:} More sophisticated power control algorithms can be considered, particularly when a multi-user access scheme is used. For example, one candidate formulation is to optimize the power allocated over each beam direction such that the aggregate interference on the incumbent receiver is minimized.
	
	
	\subsection{Implementation}
	Implementation of angular exclusion zones should be straightforward. Indeed, upon the deployment of the gNBs in a given region, the mobile operator must identify the FSs in vicinity using the FCC's database, where the operator can extract their locations and azimuth directions, which will be used to compute the necessary unit vectors. The operator then switch sectors (or beams) depending on the protection criterion used. Thus, UEs cannot find any reference signals from those sectors (or beams), and hence they do not connect to them during user and beam association. Clearly, the operator may need to update the sector-based (or beam-based) decisions if the FS's databased is changed, e.g., switch back sectors if an incumbent license is expiring, etc., which typically happens at a long-time scale.  
	
	To implement spatial binary power control, the operator must tag each beam, from the possible DL gNB beams, with an indicator variable denoting whether the beam is a regular beam or a quiet one, which is determined by computing $\cos^{-1}\left(-\mathbf{u}_{\g}^{\operatorname{beam},i^\star}\bullet \mathbf{u}_{\g\rightarrow\f}\right)$. The indicator value and the allowable transmit of the beam are then embedded in the reference signal sent over the beam during user association. This is done over the physical broadcast channel (xPBCH or ePBCH), and thus during synchronization, the UE can decode the master and system information blocks (MIB and SIB), identifying the UL transmit power limit over that beam. 
	
	We remark that the passive mitigation techniques primarily require the design of the angular protection thresholds, e.g., $\psi_s$ and $\psi_b$. For instance, using a higher threshold value provides more protection to incumbents, yet this may come at the expense of the 5G system coverage. Since we use the 3GPP channel model, it is difficult to analytically determine the threshold that strikes a good balance between the 5G coverage performance and the INR at the incumbent receiver. For this reason, a mobile network operator may run preliminary computer simulations, e.g., using the interference framework used in this paper, to determine $\psi_s$ or $\psi_b$.

	\section{Simulation Results}\label{sec:simulations}
\begin{figure}[t!]
		\centering
		\begin{subfigure}[t]{.3\textwidth}
			\centering
			\includegraphics[width=1.85in]{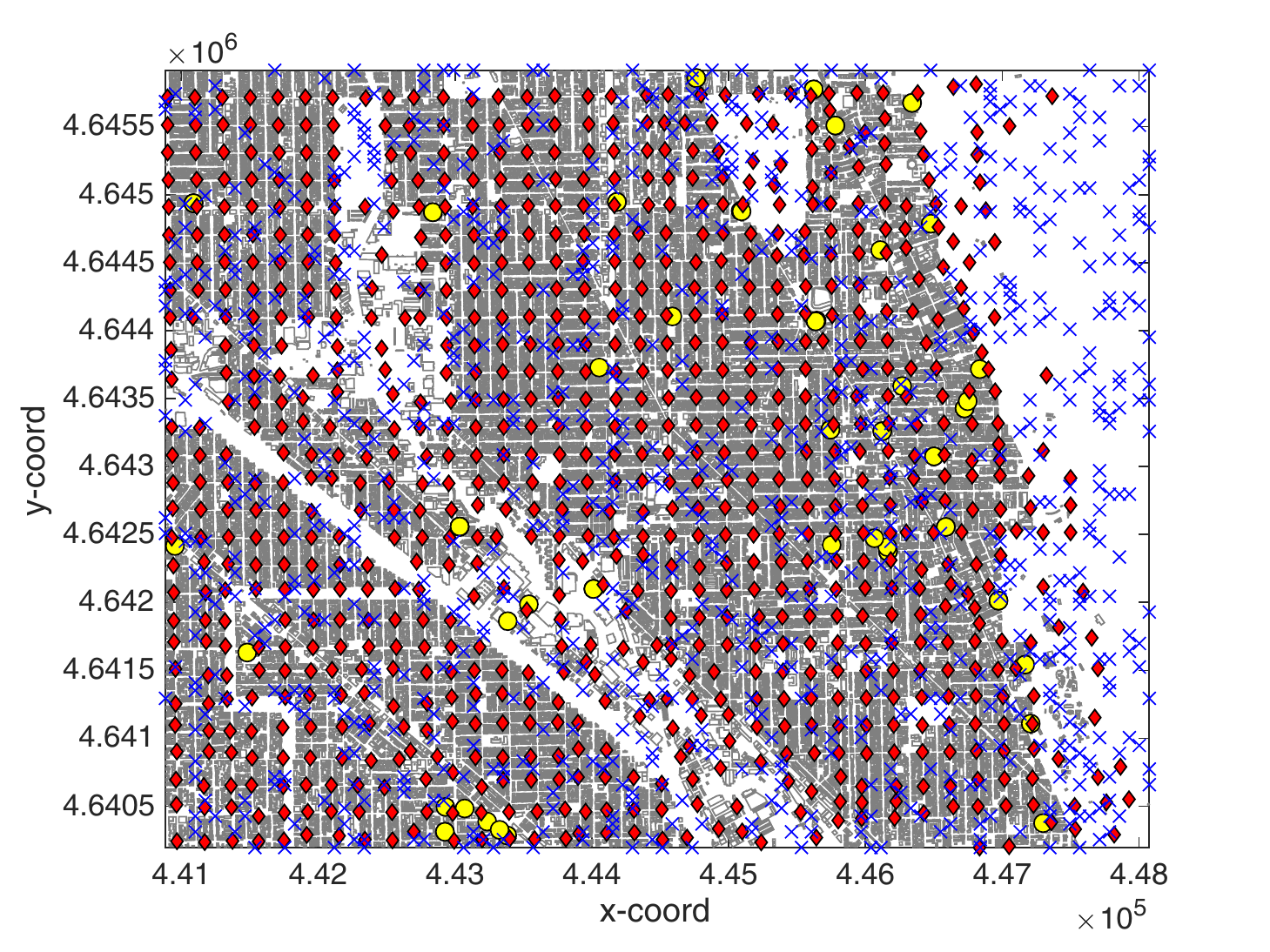}
			\caption{Lincoln Park (852 gNBs)}
		\end{subfigure}~~
		\begin{subfigure}[t]{.3\textwidth}
			\centering
			\includegraphics[width=1.85in]{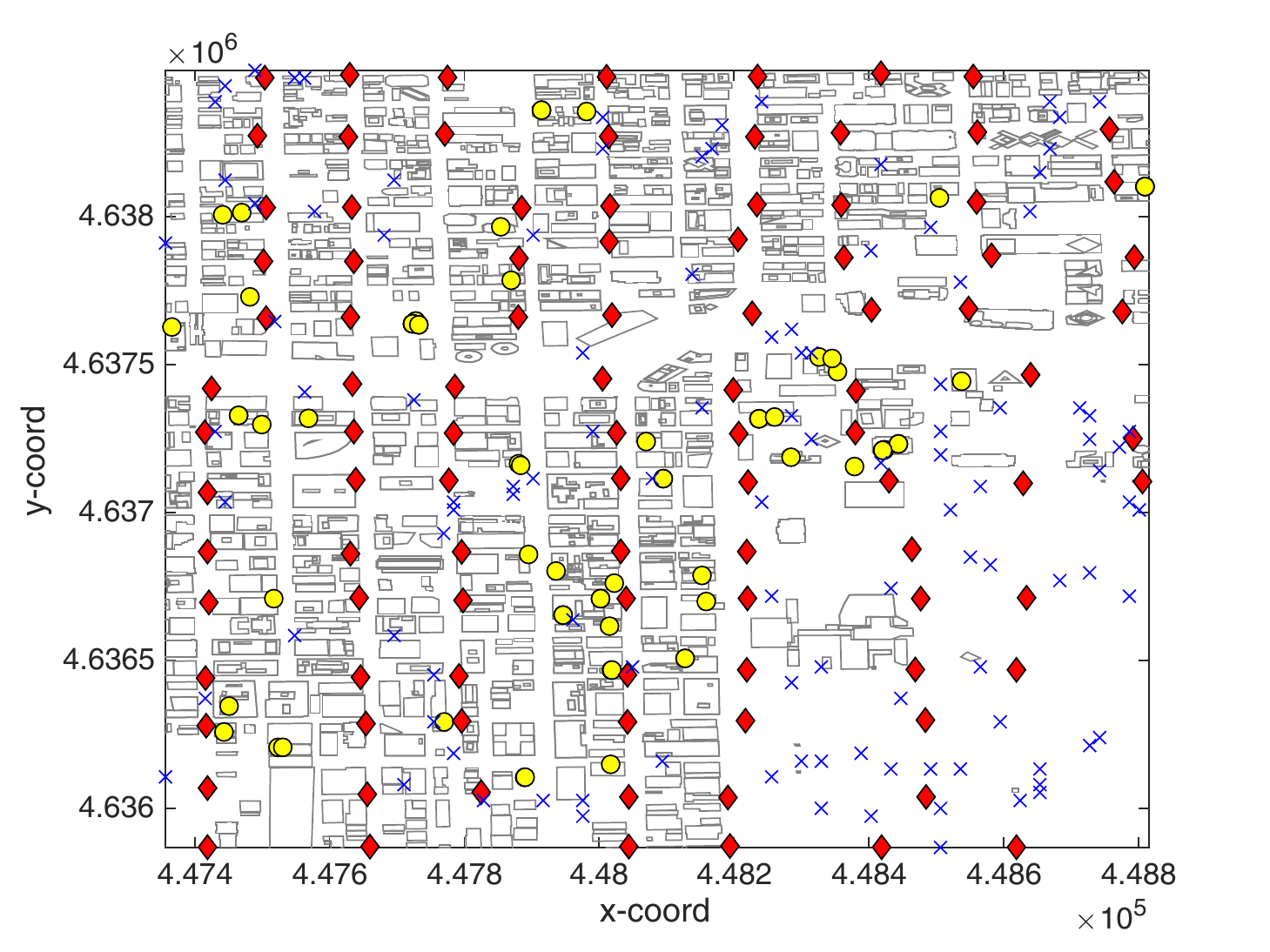}
			\caption{Chicago Loop (108 gNBs)}
		\end{subfigure}~~
		\begin{subfigure}[t]{.3\textwidth}
			\centering
			\includegraphics[width=1.85in]{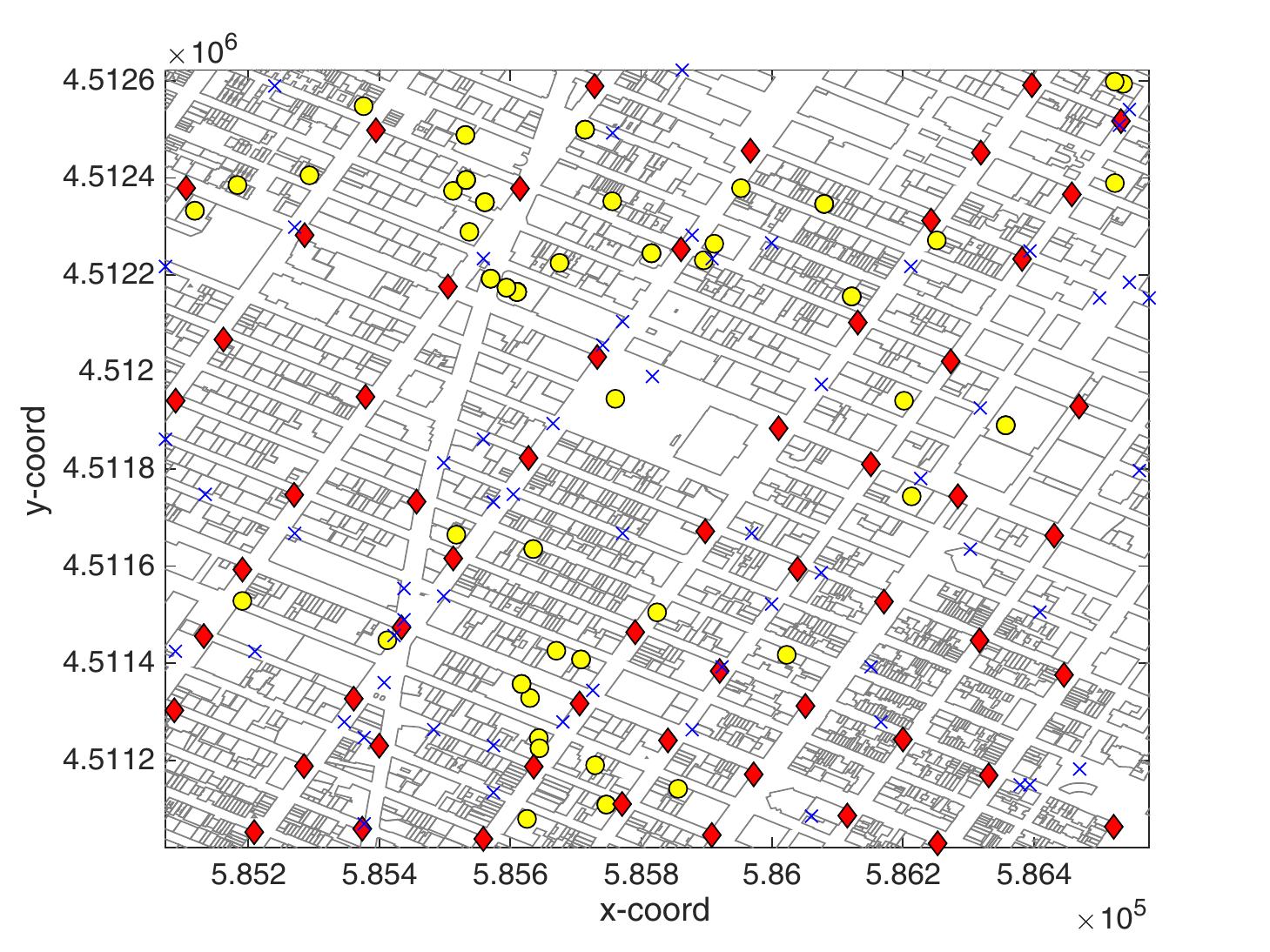}
			\caption{Lower Manhattan (58 gNBs)}
		\end{subfigure}
		\caption{Simulation scenarios. Here `$\circ$' denotes the FS, `$\Diamond$' denotes the gNB, and `$\times$' denotes the UE.}
		\label{fig:scenarios}
	\end{figure}

	We study the aggregate UE interference on FSs deployed in Lincoln Park, Chicago Loop, and Lower Manhattan. We deploy gNBs on a grid in each of the aforementioned cities, where $d_{\operatorname{ISD}}=200$m. Further, we randomly deploy outdoor UEs in each city, and assume an UL instantaneous traffic load of 25\%, i.e., each gNB site, which consists of four sectors, serves one UE in a given time slot. Fig. \ref{fig:scenarios} shows one spatial realization of the three deployment scenarios.
	
	Fo the channel model, we use the 3GPP NR-UMi model \cite{3GPP2017}. All other important simulation parameters are given in Table \ref{tab:parameters}. We consider the center frequencies: 73.5GHz and 83.5GHz, and assume that the UE maximum radiated power, without any attenuation, is 33dBm or 43dBm \cite{Huo2017,FCC2016}. Per FCC regulations, we consider $A_{\f,\operatorname{FTBR}}=55$dB \cite{FCC2017}. For noise power, we assume $B=1$GHz and $N_0$ is computed at temperature 290K. Finally, the FS's location, height, maximum antenna gain, antenna tilt, and noise figure, are all extracted from the FCC's incumbent database \cite{Comsearch2018}. The subsequent results are averaged out over 1000 spatial realizations, where each one has a different deployment of UEs and different channel realizations. Additionally, unless otherwise stated, the results consider actual UEs direction, where gNB-UE association is performed first.
	
	\subsection{Validation of the 5G System}
\begin{figure}[t!]
		\centering
		\begin{subfigure}[t]{.4\textwidth}
			\centering
			\includegraphics[width=2.25in]{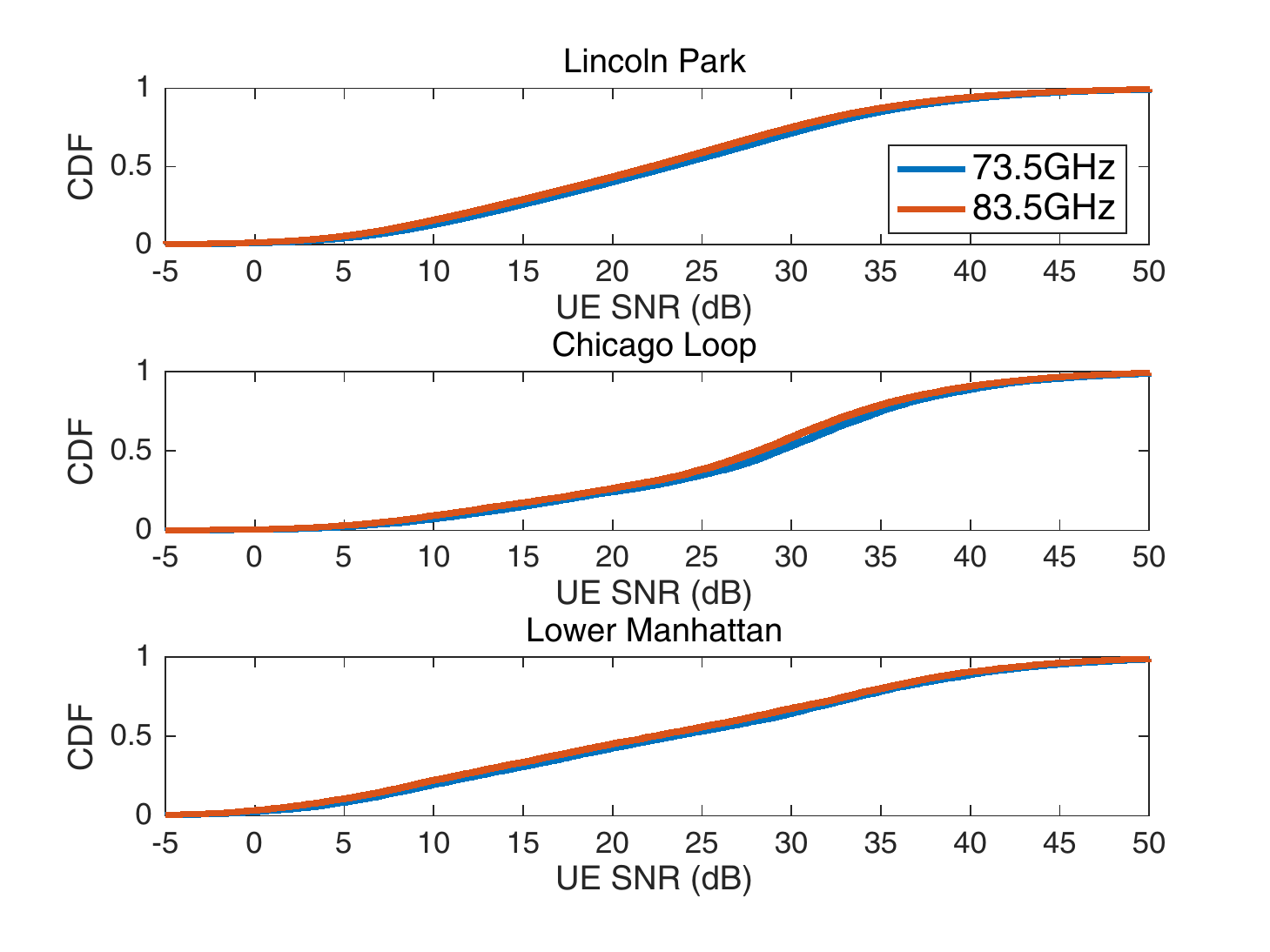}
			\caption{SNR CDF}
			\label{fig:SNR_cdf}
		\end{subfigure}~~
		\begin{subfigure}[t]{.4\textwidth}
			\centering
			\includegraphics[width=2.25in]{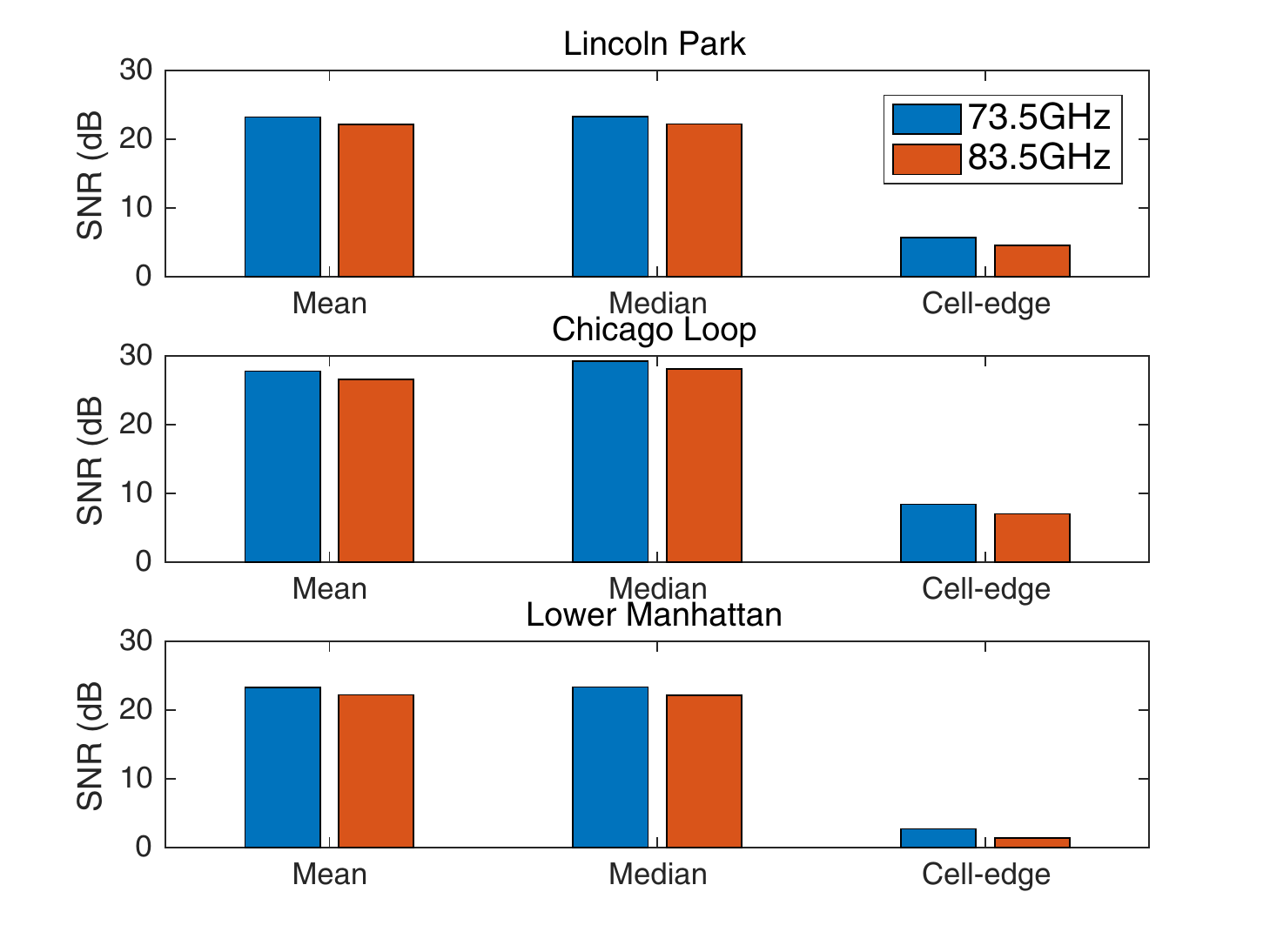}
			\caption{SNR statistics}
			\label{fig:SNR_statistics}
		\end{subfigure}
		\caption{SNR at 5G UE}
		\label{fig:SNR}
	\end{figure}

	We first verify that deployment of gNBs lead to reliable coverage for UEs. Fig. \ref{fig:SNR_cdf} shows the CDF of the signal-to-noise ratio (SNR) at the UE side after beam association, whereas Fig. \ref{fig:SNR_statistics} shows the main SNR statistics. Overall, it is shown that the deployment provides reliable coverage with positive cell-edge SNR values. Operating at 83.5GHz has slight SNR degradation due to higher path loss compared to operating at 73.5GHz.
	
\begin{figure}[t!]
		\centering
		\begin{subfigure}[t]{.3\textwidth}
			\centering
			\includegraphics[width=2.25in]{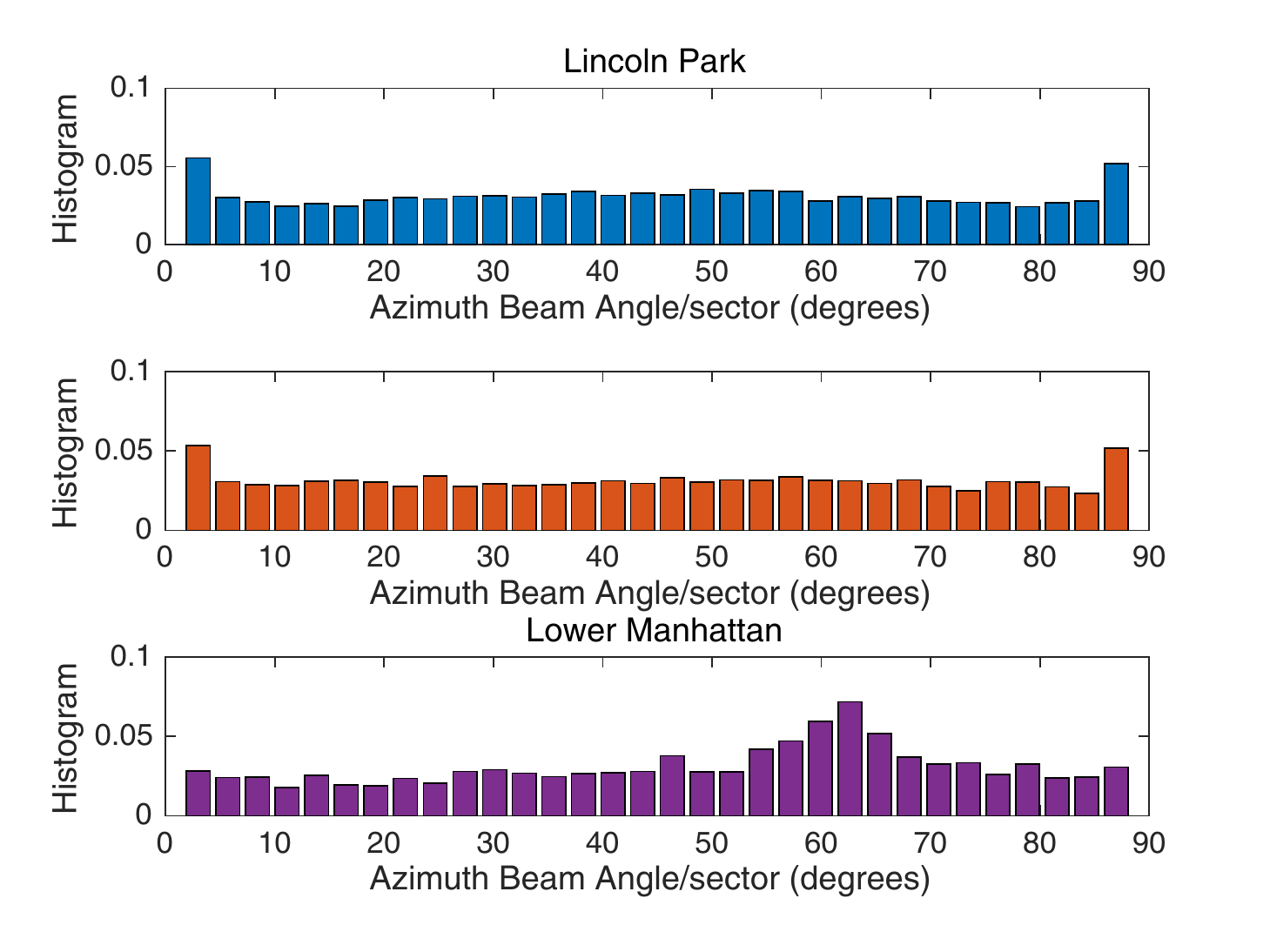}
			\caption{gNB azimuth beams}
			\label{fig:gNB_AzBeam}
		\end{subfigure}~~
		\begin{subfigure}[t]{.3\textwidth}
			\centering
			\includegraphics[width=2.25in]{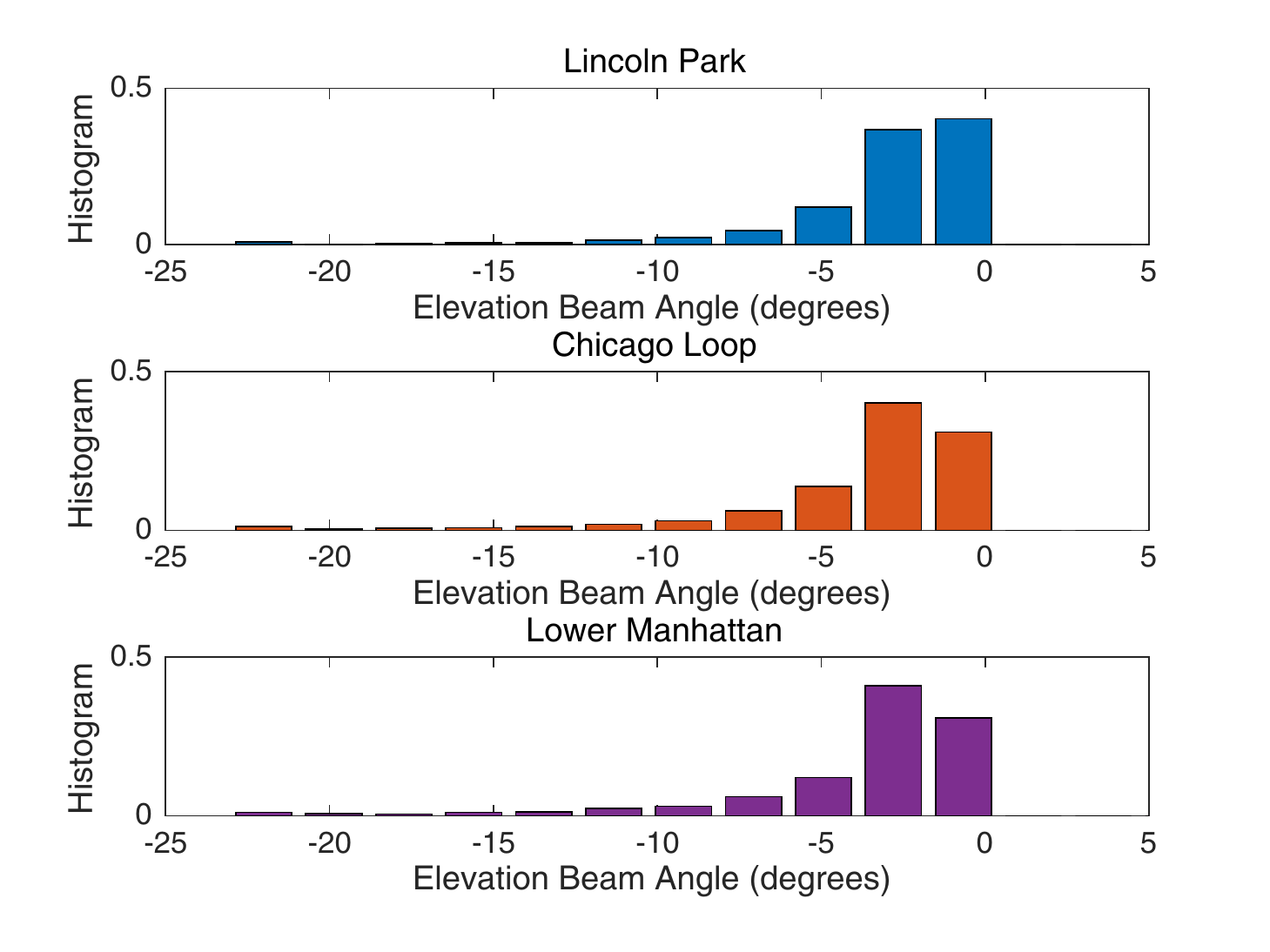}
			\caption{gNB elevation beams}
			\label{fig:gNB_ElBeam}
		\end{subfigure}~~
		\begin{subfigure}[t]{.3\textwidth}
			\centering
			\includegraphics[width=2.25in]{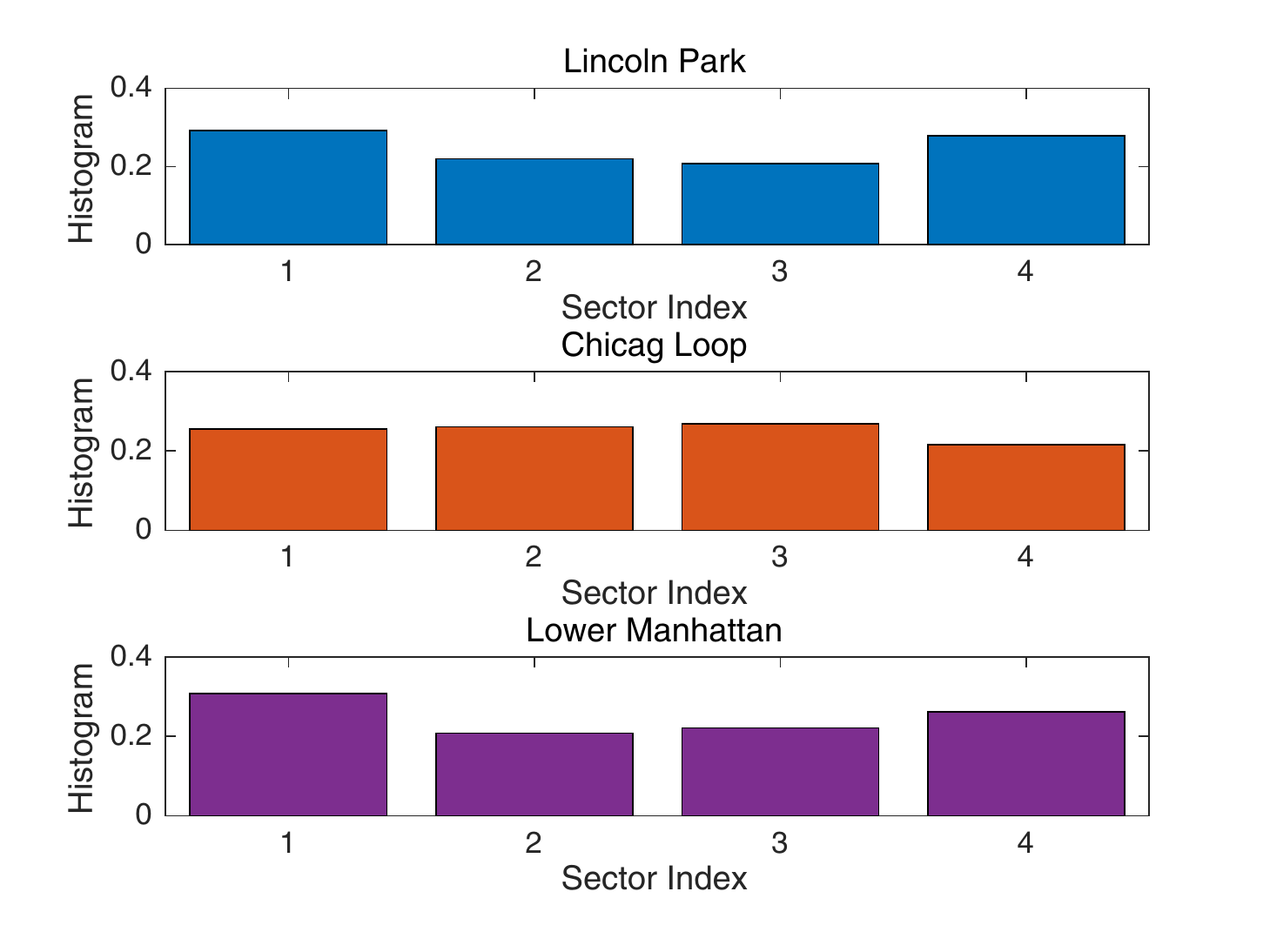}
			\caption{gNB sectors}
			\label{fig:gNB_Sector}
		\end{subfigure}\\
		\begin{subfigure}[t]{.3\textwidth}
			\centering
			\includegraphics[width=2.25in]{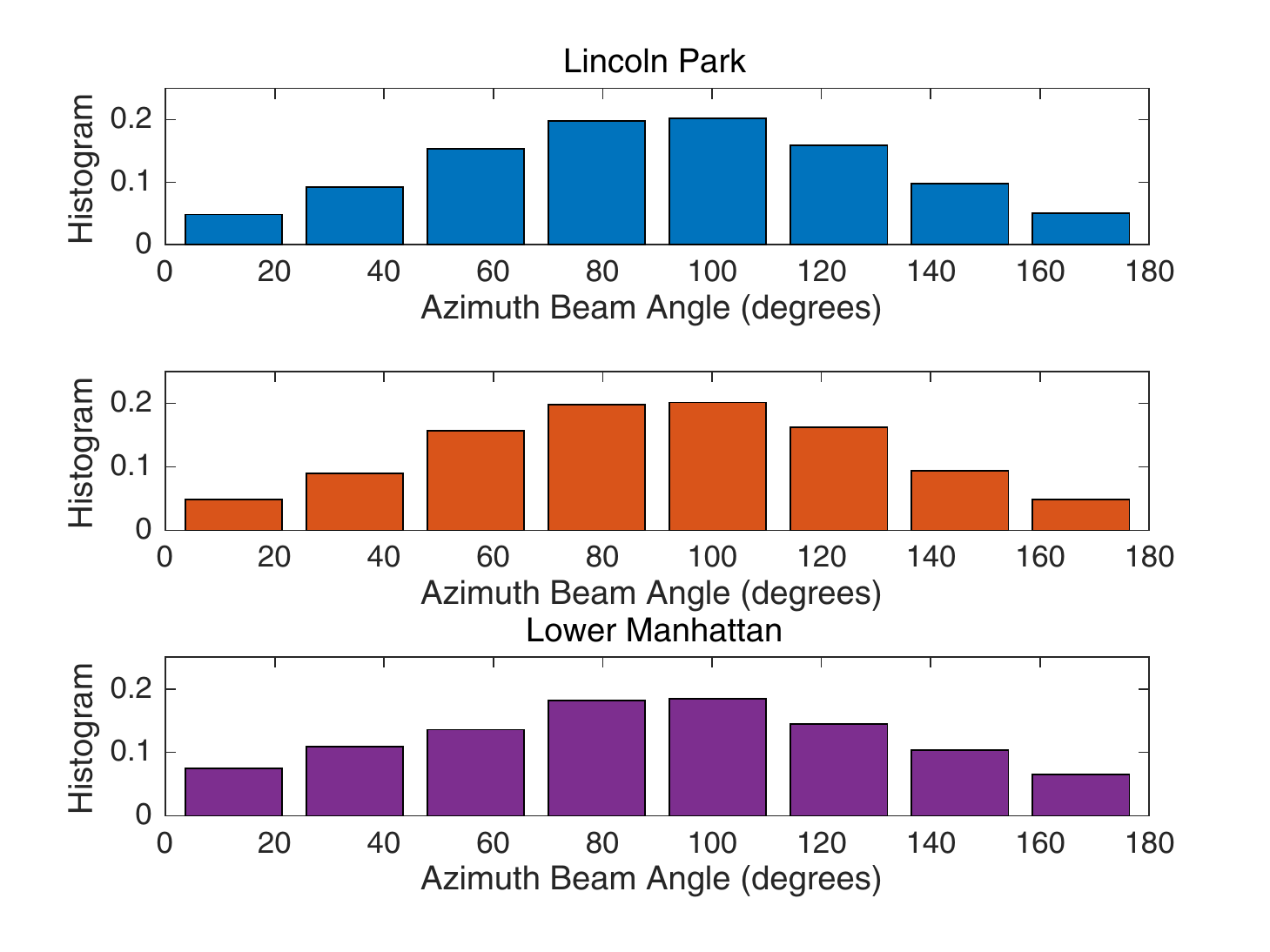}
			\caption{UE azimuth beams}
			\label{fig:UE_AzBeam}
		\end{subfigure}~~
		\begin{subfigure}[t]{.3\textwidth}
			\centering
			\includegraphics[width=2.25in]{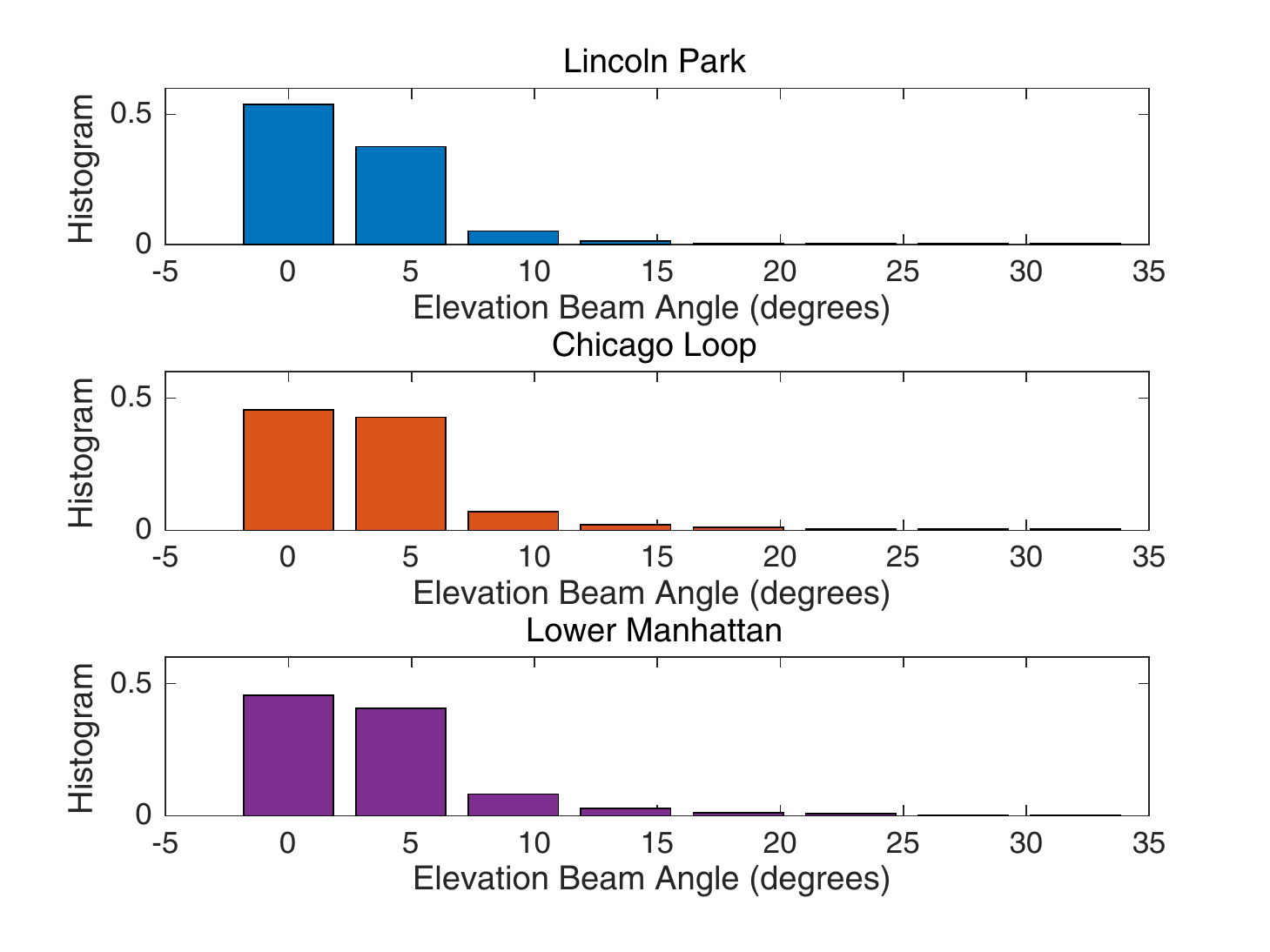}
			\caption{UE elevation beams}
			\label{fig:UE_ElBeam}
		\end{subfigure}	~~
		\begin{subfigure}[t]{.3\textwidth}
			\centering
			\includegraphics[width=2.25in]{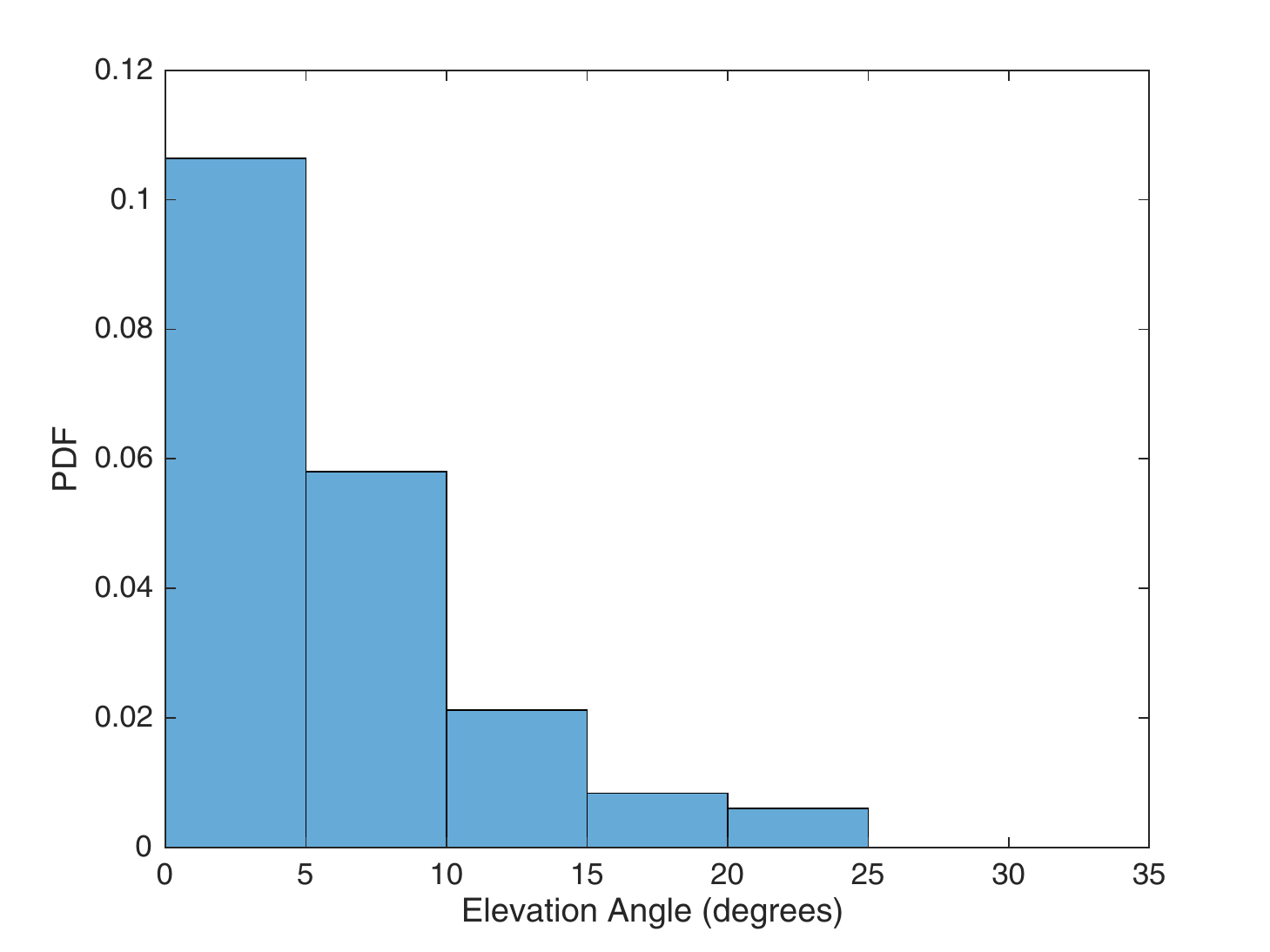}
			\caption{PDF of UE elevation angle under the random model}
			\label{fig:UE_ELBeamRandom}
		\end{subfigure}	
		\caption{Distribution of used beams and sectors}
		\label{fig:beams}
	\end{figure}

	Next, we look at the DL and UL beams used by gNBs and UEs, respectively, after user and beam association. This provides insights on which beams are likely to be used by the gNB and the UE for a realistic deployment scenario. In Fig. \ref{fig:beams}, we show the histograms of the beams used in azimuth and elevation by gNBs and UEs. It is shown that, overall, each azimuth gNB beam is equally likely to be used, with a similar observation regarding the gNB sectors. More importantly, only few elevation beams are active. This suggests that mobile operators should implement only a couple of elevation beams to serve outdoor users, which reduces the complexity of user association and codebook design. For the UE, only few elevation beams are used as well, with the majority of them being less than 10$^\circ$. Further, the UE uses the azimuth center beams more frequently because they have higher array gain. This suggests that at the UE side, only few candidate directions should be explored during user association, and particularly those that are centered around the physical orientation of the UE antenna boresight. This observation helps significantly reduce the beam search space in user association  \cite{Alkhateeb2017,Giordani2016}.
	
	\subsection{Distribution of INR}
\begin{figure}[t!]
		\centering
		\begin{subfigure}[t]{.3\textwidth}
			\centering
			\includegraphics[width=2.25in]{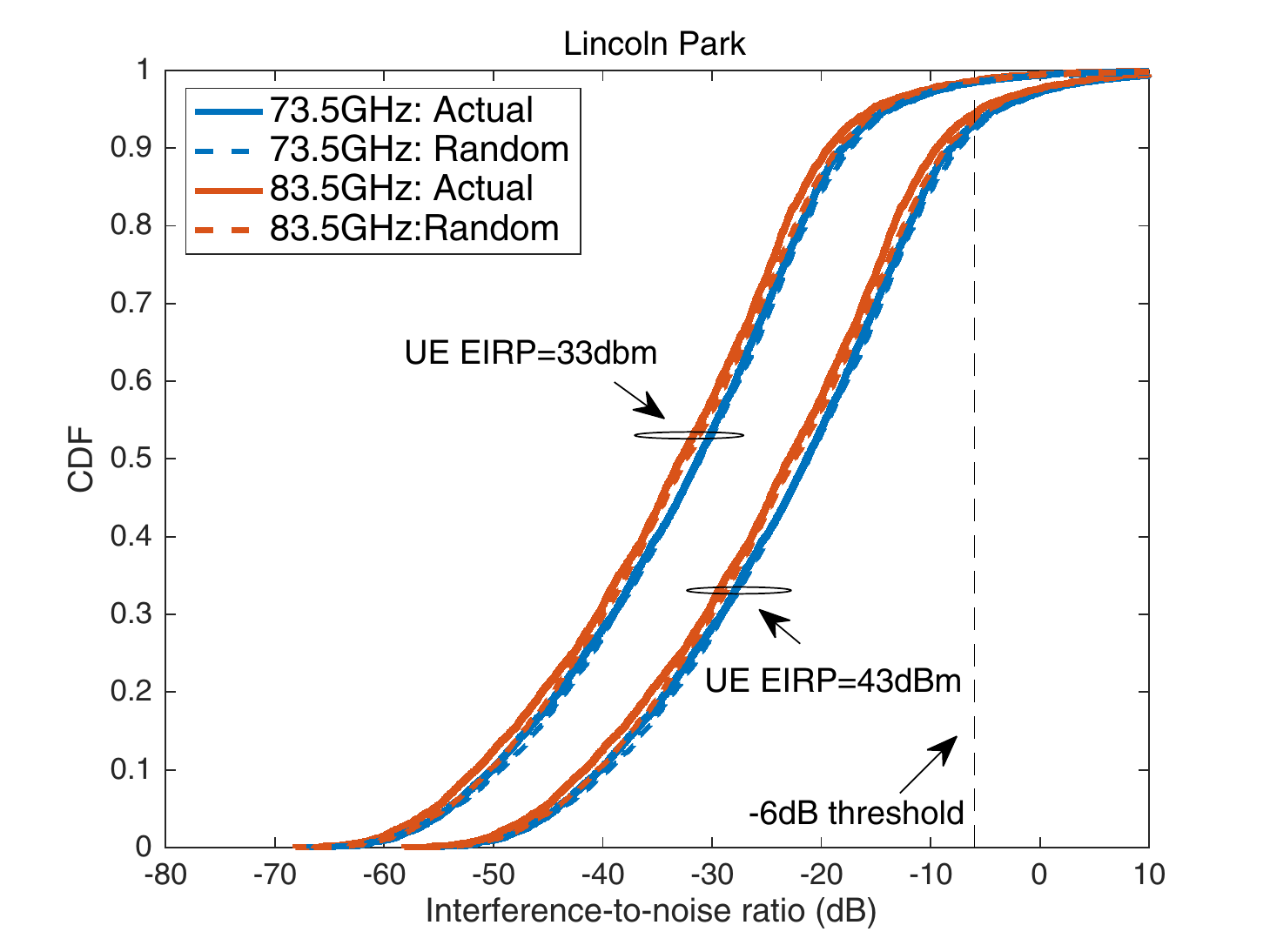}
			\caption{Lincoln Park}
			\label{fig:LP_INR_cdf}
		\end{subfigure}~
		\begin{subfigure}[t]{.3\textwidth}
			\centering
			\includegraphics[width=2.25in]{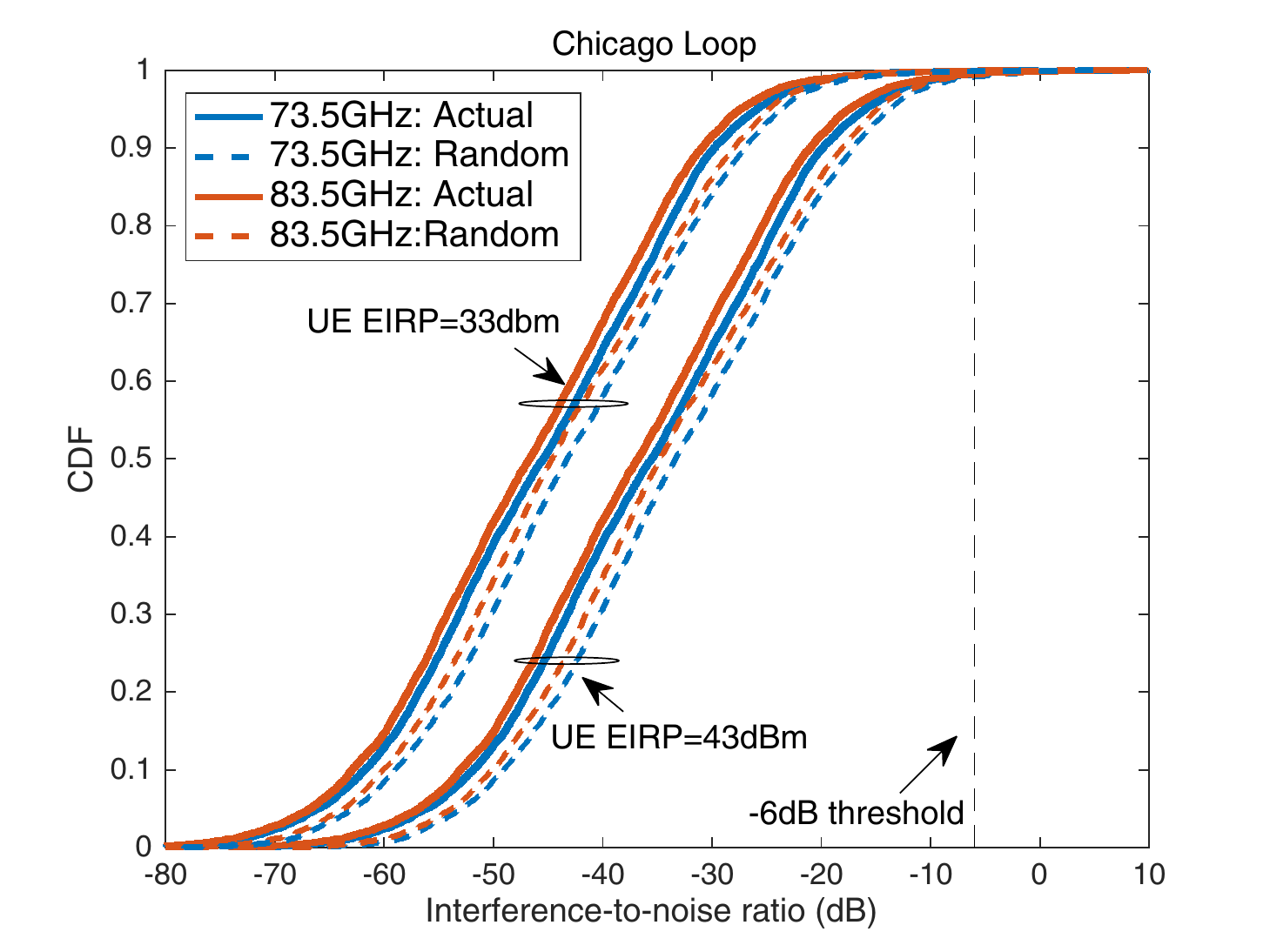}
			\caption{Chicago Loop}
			\label{fig:LOOP_INR_cdf}
		\end{subfigure}~
		\begin{subfigure}[t]{.3\textwidth}
			\centering
			\includegraphics[width=2.25in]{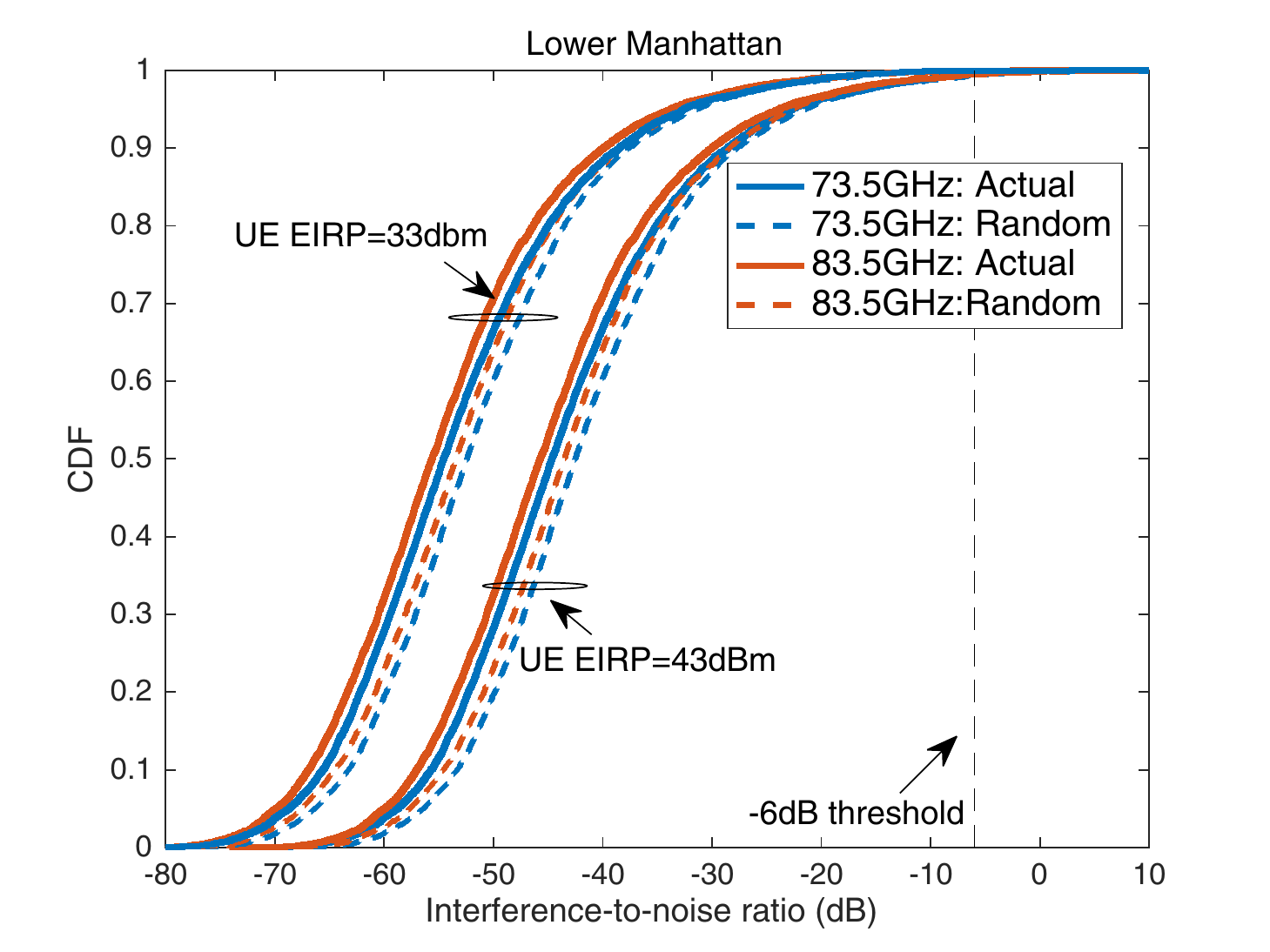}
			\caption{Lower Manhattan}
			\label{fig:MH_INR_cdf}
		\end{subfigure}
		\caption{CDF of INR}
		\label{fig:INRcdf}
	\end{figure}

	Fig. \ref{fig:INRcdf} shows the CDF of INR for the different case studies. We also show a reference INR threshold of $-6$dB, which corresponds to signal-to-noise-plus-interference ratio (SINR) degradation of 1dB, meeting the FCC's interference protection criterion \cite{FCC2017}. We have the following observations. First, using the random model, i.e., random UE azimuth and elevation directions, provides accurate results that match well with computing the actual pointing directions of the UE in the presence of gNBs. This follows because the deployment of gNBs is agnostic to the locations of FSs, and the distribution of used elevation directions (cf. Fig. \ref{fig:UE_ElBeam}) has a similar PDF to the one used in the random model (cf. (\ref{eq:randomEl}) and Fig. \ref{fig:UE_ELBeamRandom}). Second, the CDFs show that the INR is overall low, with the majority of FSs experiencing INR levels well below the noise floor. This follows due to the high attenuation at millimeter wave frequencies, i.e., the networks operate in a noise-limited regime, the stark height difference in deploying FSs and 5G systems, and the very low likelihood of UEs being aligned within 1$^\circ$ of the FS's beam. It is also shown that dense urban areas, e.g., downtown Manhattan, has lower INR due to the increased blockage resulted from the presence of high-rise buildings. Finally, the INR is slightly lower at 83.5GHz compared to 73.5GHz due to the higher path loss in the former. 
	
\begin{figure}[t!]
		\centering
		\begin{subfigure}[t]{.3\textwidth}
			\centering
			\includegraphics[width=2.25in]{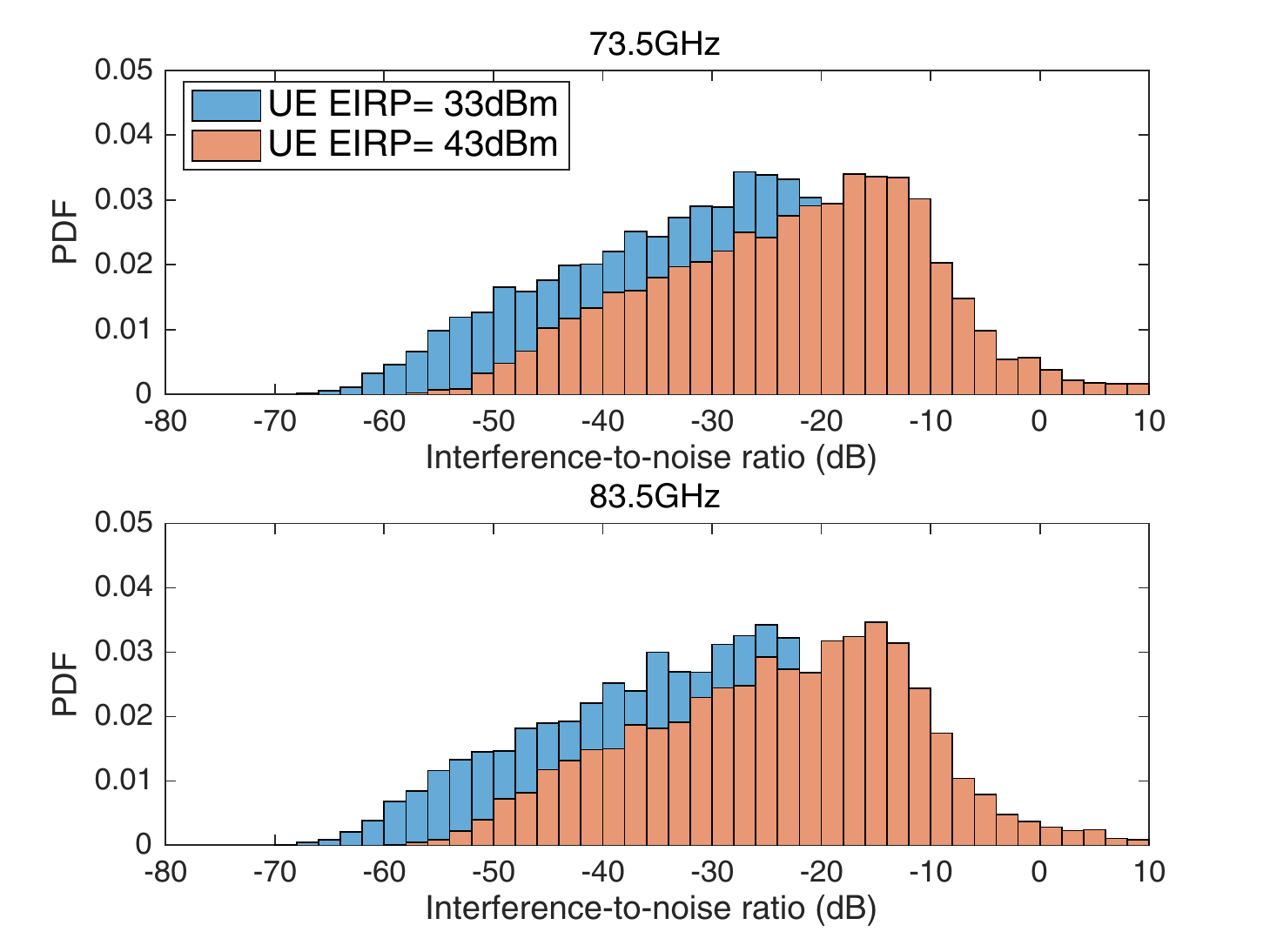}
			\caption{Lincoln Park}
			\label{fig:LP_INR_pdf}
		\end{subfigure}~~
		\begin{subfigure}[t]{.3\textwidth}
			\centering
			\includegraphics[width=2.25in]{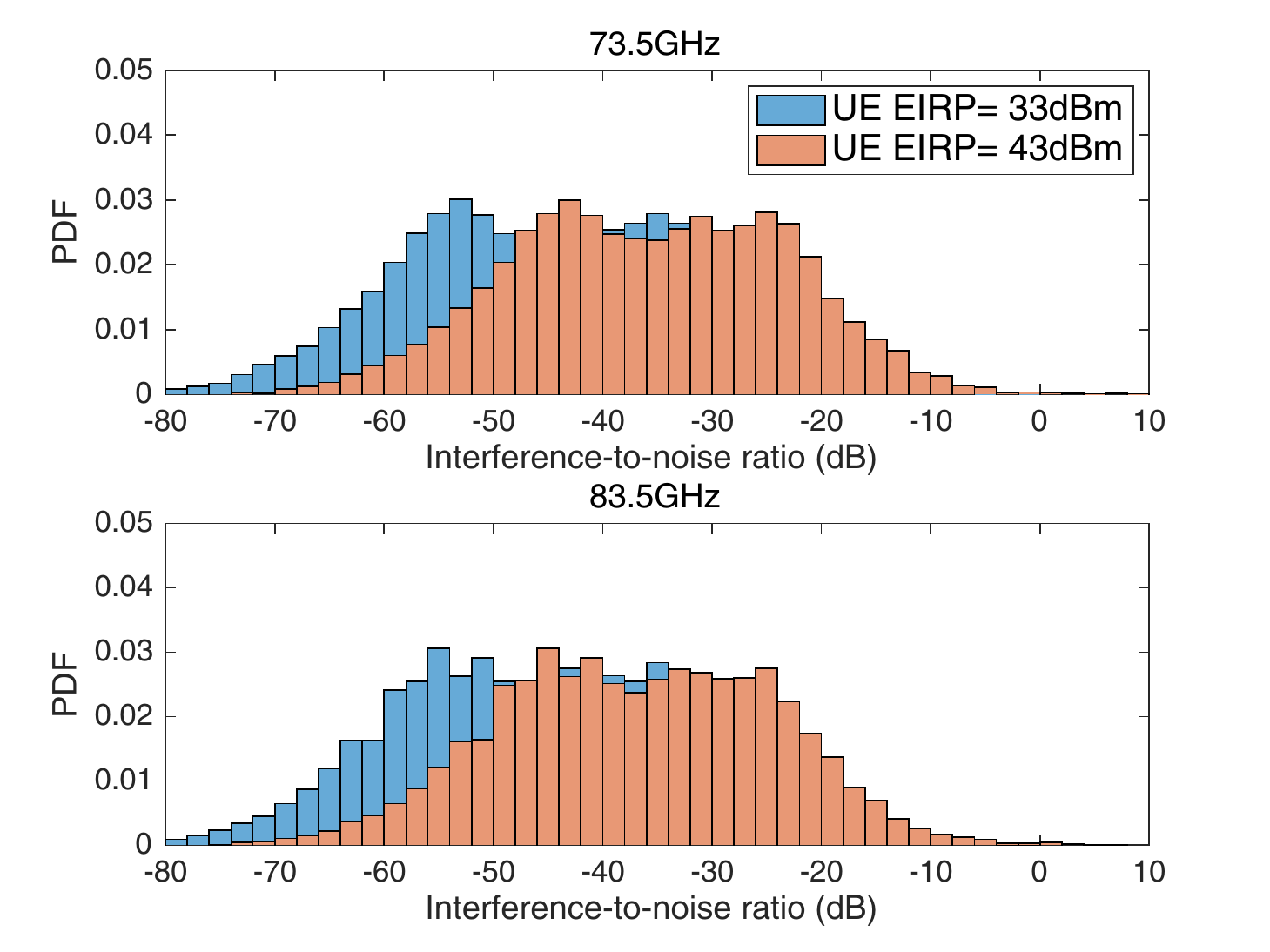}
			\caption{Chicago Loop}
			\label{fig:LOOP_INR_pdf}
		\end{subfigure}~~
		\begin{subfigure}[t]{.3\textwidth}
			\centering
			\includegraphics[width=2.25in]{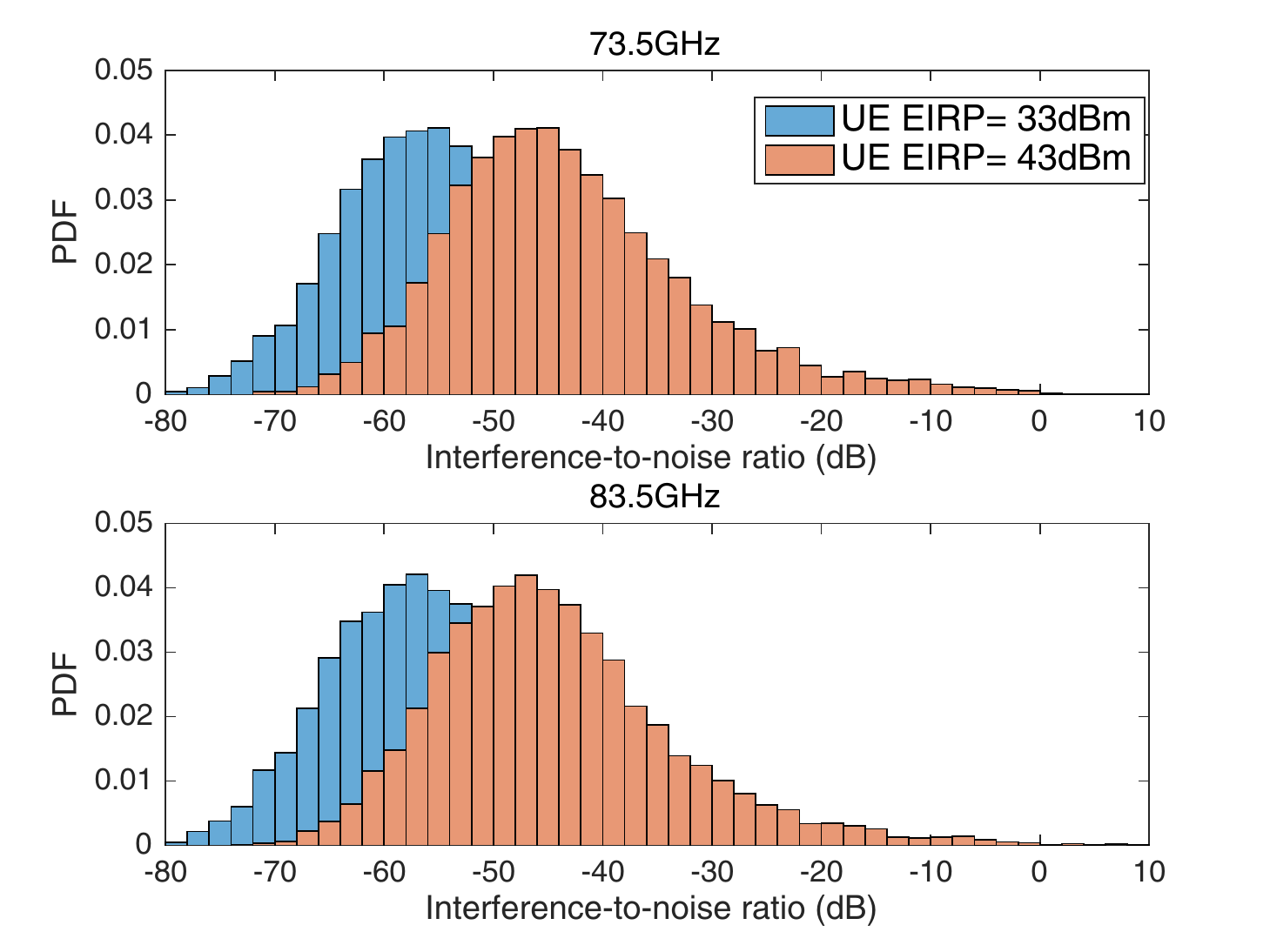}
			\caption{Lower Manhattan}
			\label{fig:MH_INR_pdf}
		\end{subfigure}
		\caption{PDF of INR}
		\label{fig:INRpdf}
	\end{figure}

	Fig. \ref{fig:INRpdf} shows the PDF of INR and its 95th percentile for the different case studies. As it can be seen, only very few FSs may experience high INR values, i.e., above the $-6$dB protection threshold, in Lincoln Park, whereas the rest are well protected. This motivates implementing the proposed mitigation techniques only to improve INR protection at those few FSs, simplifying the 5G coexistence. 
	
\begin{figure}[t!]
		\centering
		\begin{subfigure}[t]{.4\textwidth}
			\centering
			\includegraphics[width=2.5in]{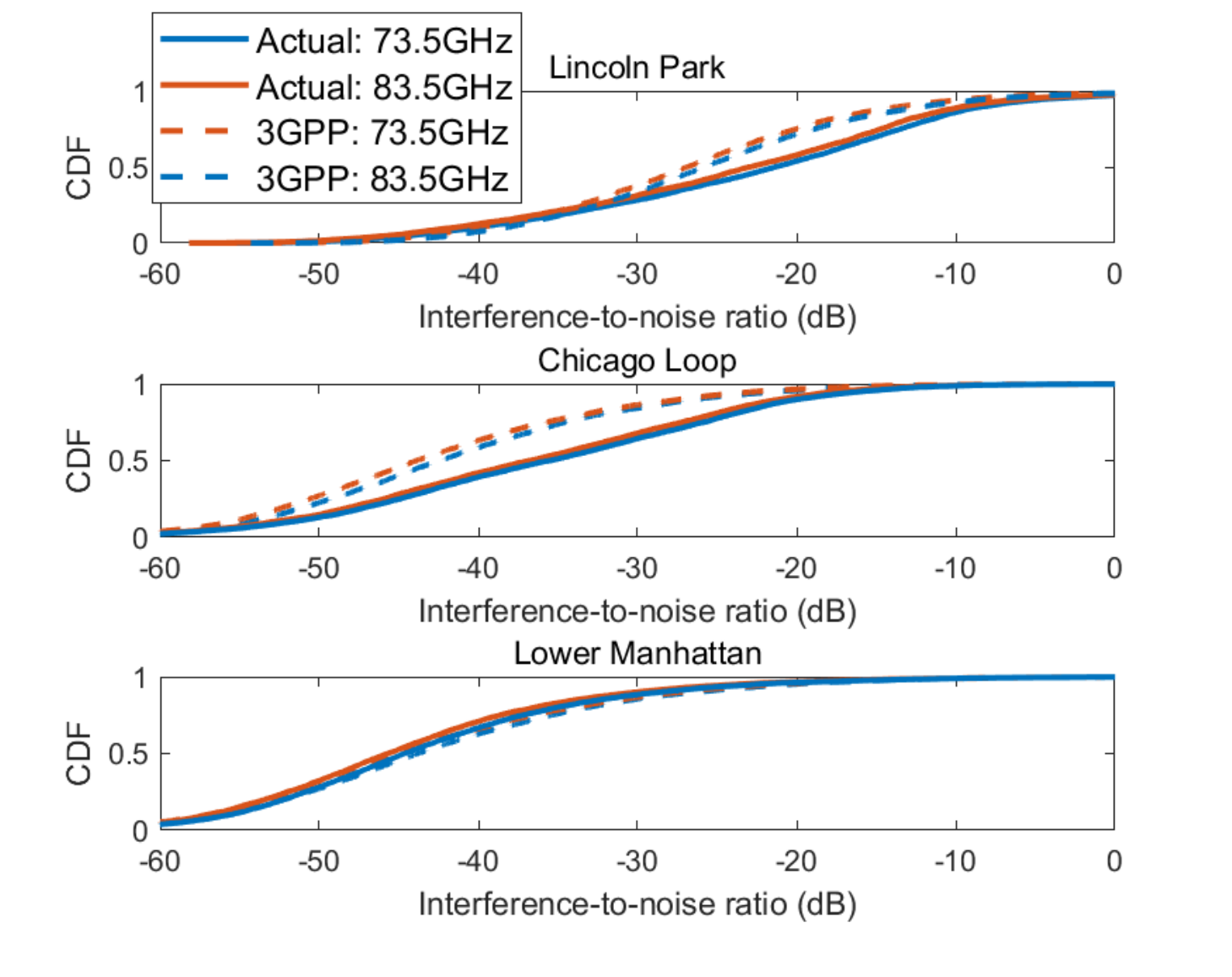}
			\caption{INR CDF}
			\label{fig:INRcdf_actualvs3gpp}
		\end{subfigure}~~
		\begin{subfigure}[t]{.4\textwidth}
			\centering
			\includegraphics[width=2.5in]{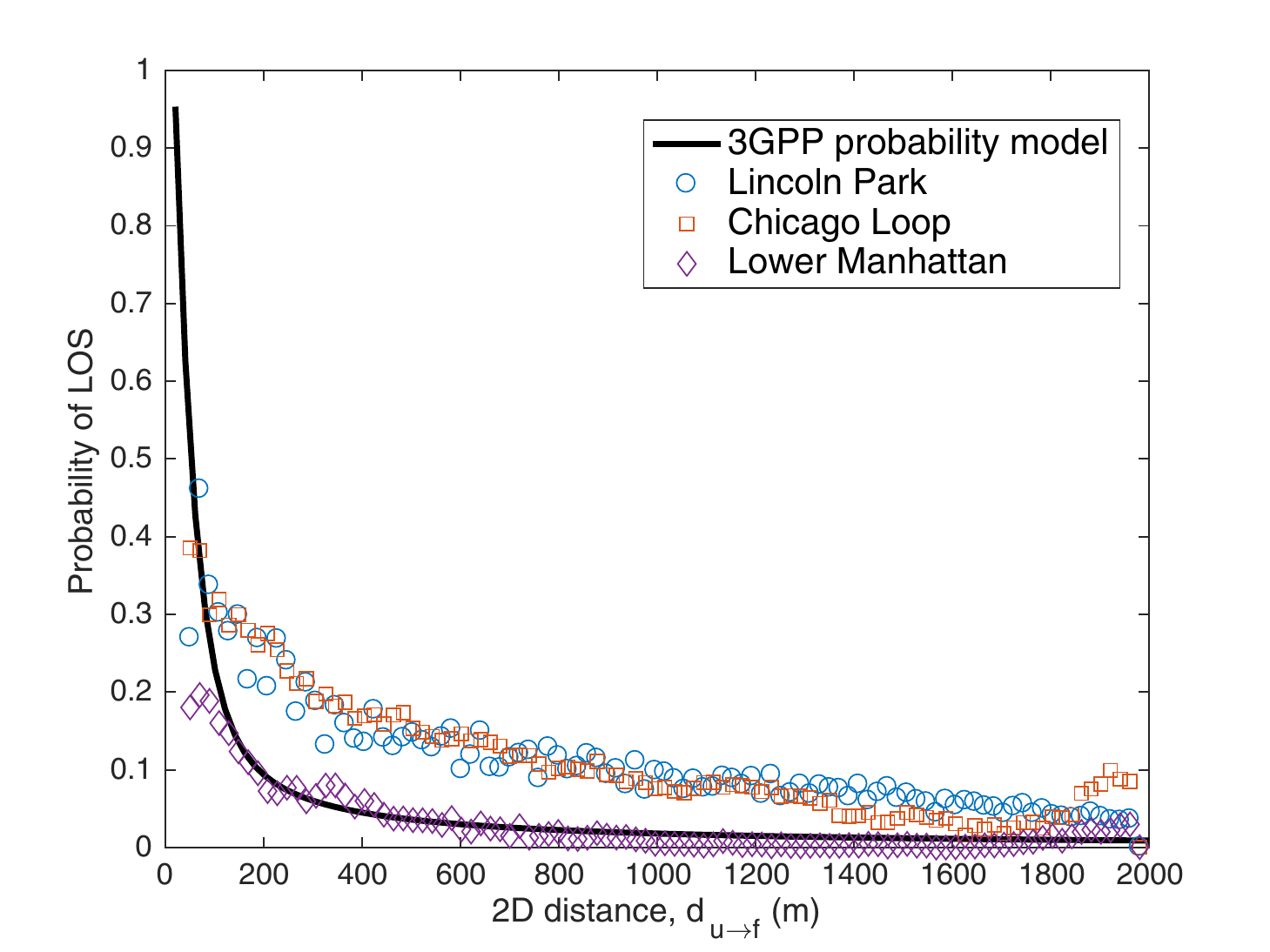}
			\caption{LOS probability}
			\label{fig:LOSprob}
		\end{subfigure}
		\caption{Comparison between the 3GPP LOS model and the actual LOS based on buildings layouts in terms of: (a) INR CDF and (b) LOS probability.}
		\label{fig:3gppVsActual}
	\end{figure}

	We then look at the INR performance when the 3GPP LOS model is used instead of the actual building layout. In Fig. \ref{fig:INRcdf_actualvs3gpp}, we show the CDF of the INR, where the UE EIRP is 43dBm. It is observed that for Lincoln Park and Chicago Loop, the INR using the 3GPP LOS model is lower than the INR using the actual building layout. This is because the 3GPP LOS model in (\ref{eq:3GPPLOS}) underestimates the LOS probability for larger distances in  these cities, as shown in Fig. \ref{fig:LOSprob}. This is not the case for Lower Manhattan due to the dense deployment of high-rise buildings, i.e., the 3GPP LOS model is shown to be more suitable for areas with denser high-rise buildings. We remark that we expect the LOS probability to be lower when blockage due to other objects is included, e.g., foliage, cars, etc., making FSs even better protected.

\begin{figure}[t!]
		\centering
		\begin{subfigure}[t]{.4\textwidth}
			\centering
			\includegraphics[width=2.5in]{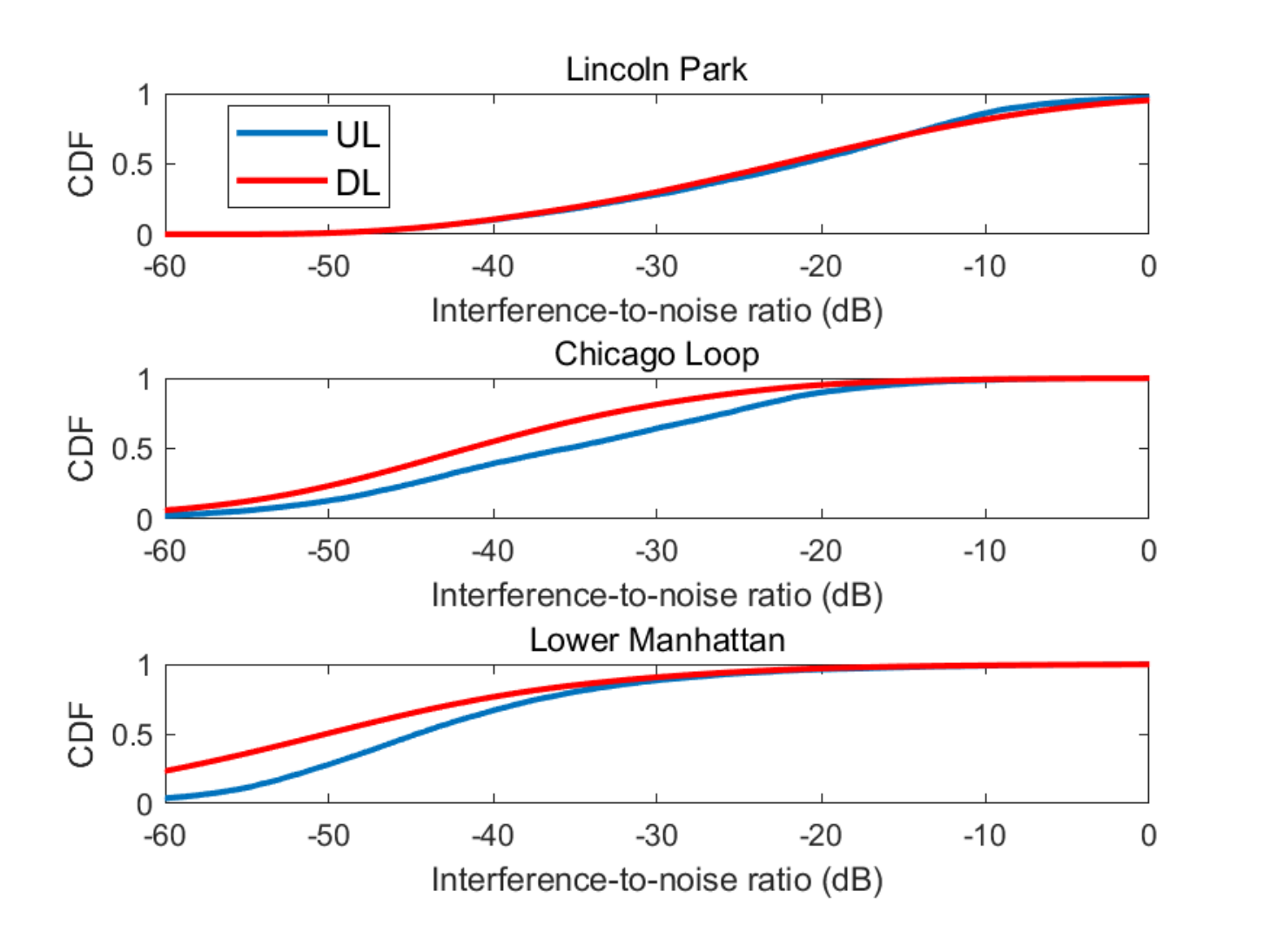}
			\caption{INR in the DL and the UL (73.5GHz)}
			\label{fig:INR_UL_DL}
		\end{subfigure}~~
		\begin{subfigure}[t]{.4\textwidth}
			\centering
			\includegraphics[width=2.5in]{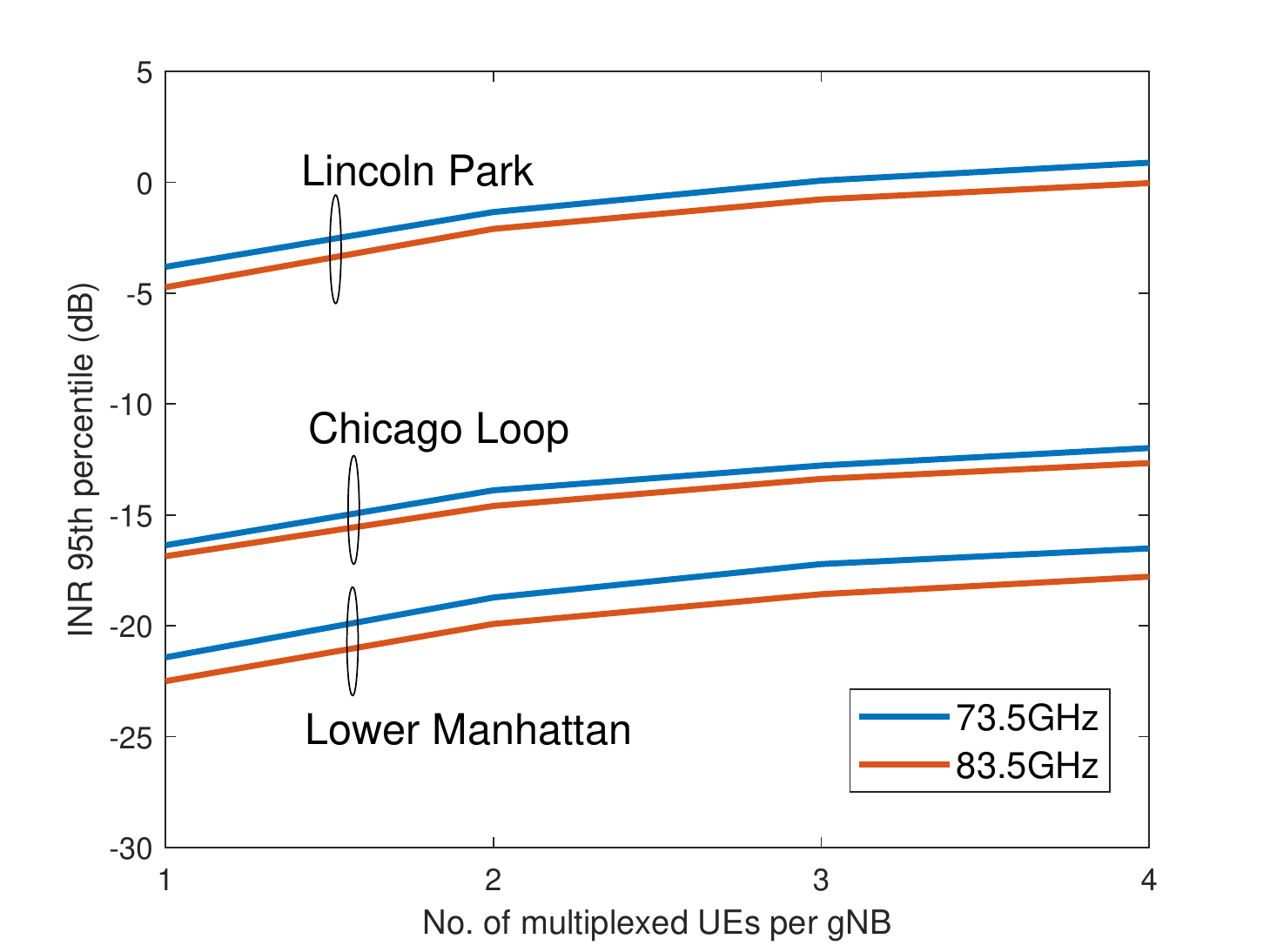}
			\caption{95th percentile for different number of multiplexed UEs}
			\label{fig:INR_vs_MUX}
		\end{subfigure}
		\caption{The INR performance under additional set-ups.}
		\label{fig:INRothers}
	\end{figure}

	We also study the INR performance for other set-ups, where we consider the UE EIRP to be 43dBm. For instance, in Fig. \ref{fig:INR_UL_DL}, we show the CDF of the INR in the DL, comparing it with that achieved in the UL. Here, the gNB EIRP is 57dBm. It is observed that although gNBs have higher EIRP, they do not incur higher interference, compared to UEs, primarily because their antenna tilts point to the ground. In Fig. \ref{fig:INR_vs_MUX}, we show the INR's 95th percentile for the multi-user case, where the number of spatially multiplexed UEs is varied from one to four. It shown, as expected, that increasing the number of multiplexed UEs increases the INR, yet it remains relatively low, particularly for the denser areas, e.g., Chicago Loop and Manhattan.
	
	\subsection{Impact of Sector-based and Beam-based Mitigation}
	We focus on a particular FS in Lincoln Park, which has relatively high INR in comparison with other FSs. The FS of interest is deployed at height of 34m with a wide open-space in its vicinity, making it more susceptible to interference from the 5G system. Here, we only consider operating at 73.5GHz with UE maximum radiated power of 43dBm, as this set-up leads to the highest interference. 
	
\begin{figure}[t!]
		\centering
		\begin{subfigure}[t]{.45\textwidth}
			\centering
			\includegraphics[width=2.5in]{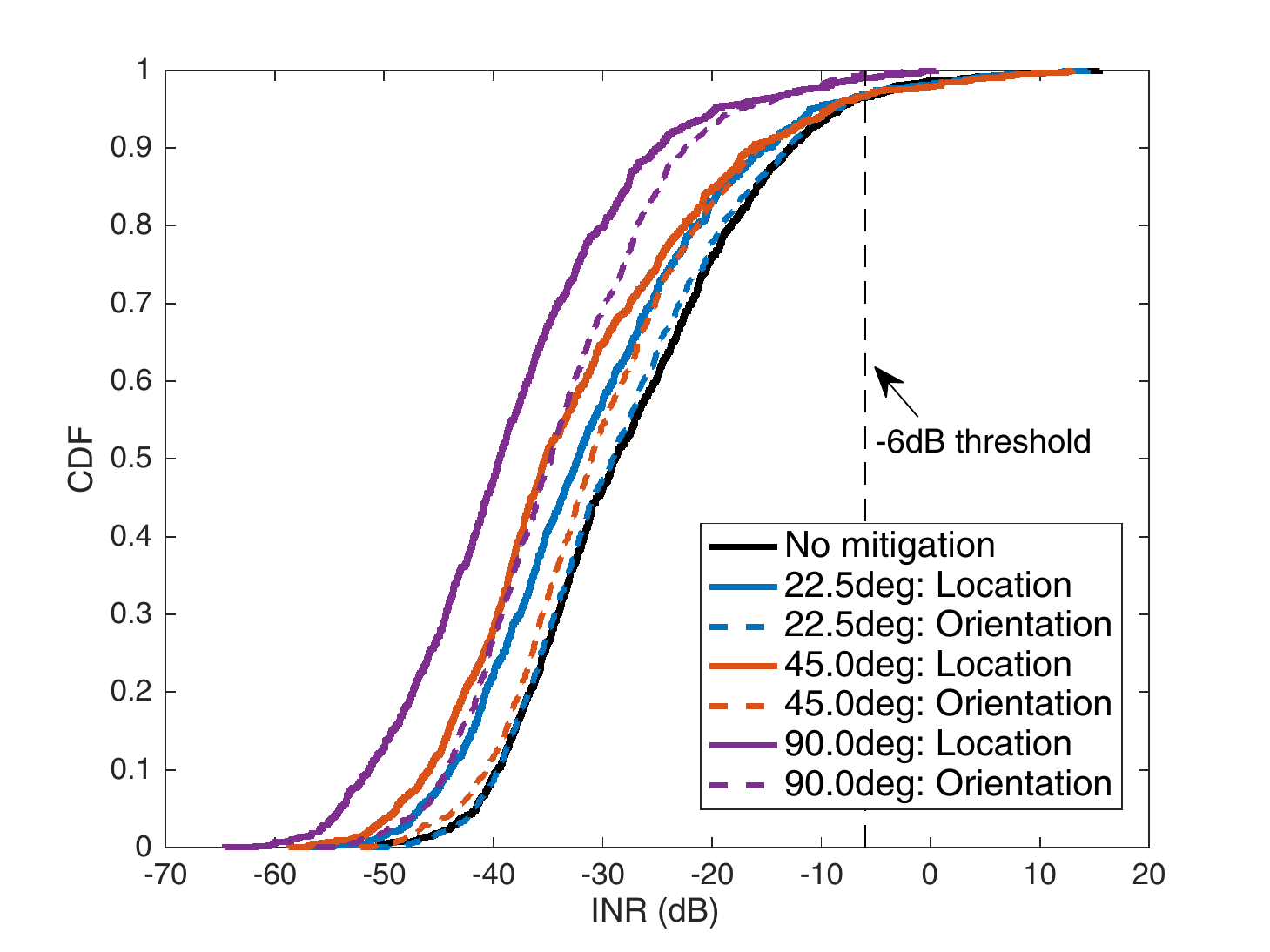}
			\caption{Sector-based}
			\label{fig:sectorMitigationCDF}
		\end{subfigure}~~
		\begin{subfigure}[t]{.45\textwidth}
			\centering
			\includegraphics[width=2.5in]{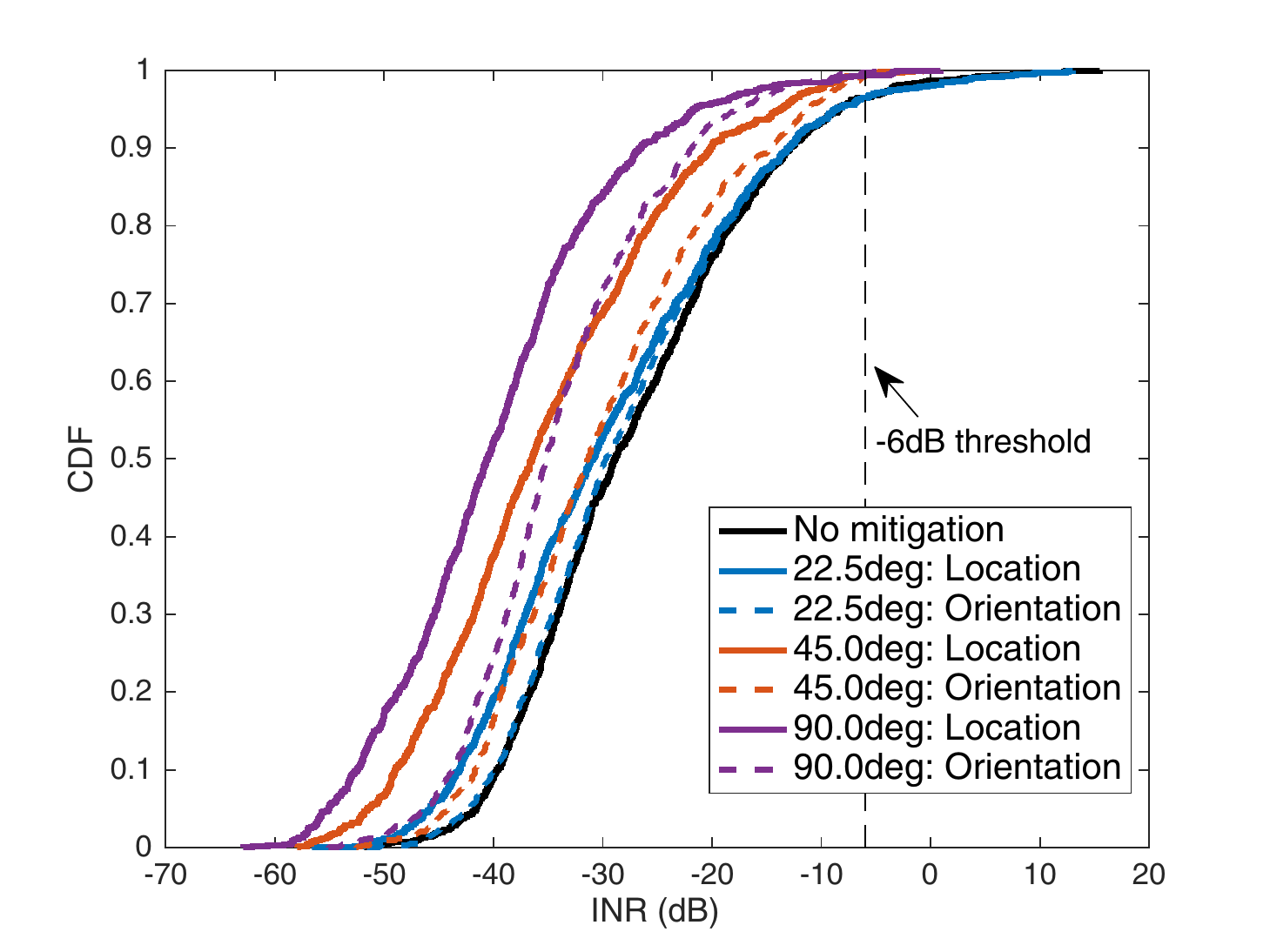}
			\caption{Beam-based}
			\label{fig:beamMitigationCDF}
		\end{subfigure}
		\caption{INR CDF in the presence and absence of passive mitigation}
		\label{fig:angularZones}
	\end{figure}

	Fig. \ref{fig:angularZones} shows the average INR on the FS in the presence of sector-based and beam-based mitigation. We have the following observations. First, using low protection thresholds, i.e., $\psi_s$ and $\psi_b$, does not result in tangible reduction in the 95th percentile of the INR. This follows because UEs tend to point randomly over space and even if the main lobe is not aligned, there is still a chance to have high interference from the side lobes. For this reason, larger thresholds provide much better protection. Second, it is shown that location-based protection is more reliable than orientation-based. This implies that to get very low INR, it is not enough to protect the boresight of the FS, i.e., signal attenuation due to FS pattern may not be sufficient if the UE effective radiated power is very high. Third, beam-based mitigation slightly outperforms sector-based mitigation, particularly for high thresholds. Equally important, the former also enables better 5G DL coverage, as it makes decisions at higher angular resolution compared to sector-based. The cost of using beam-based mitigation is the increased number of decisions needed to be made for each gNB in vicinity of the FS.
	
\begin{figure}[t!]
		\center
		\includegraphics[width=4in]{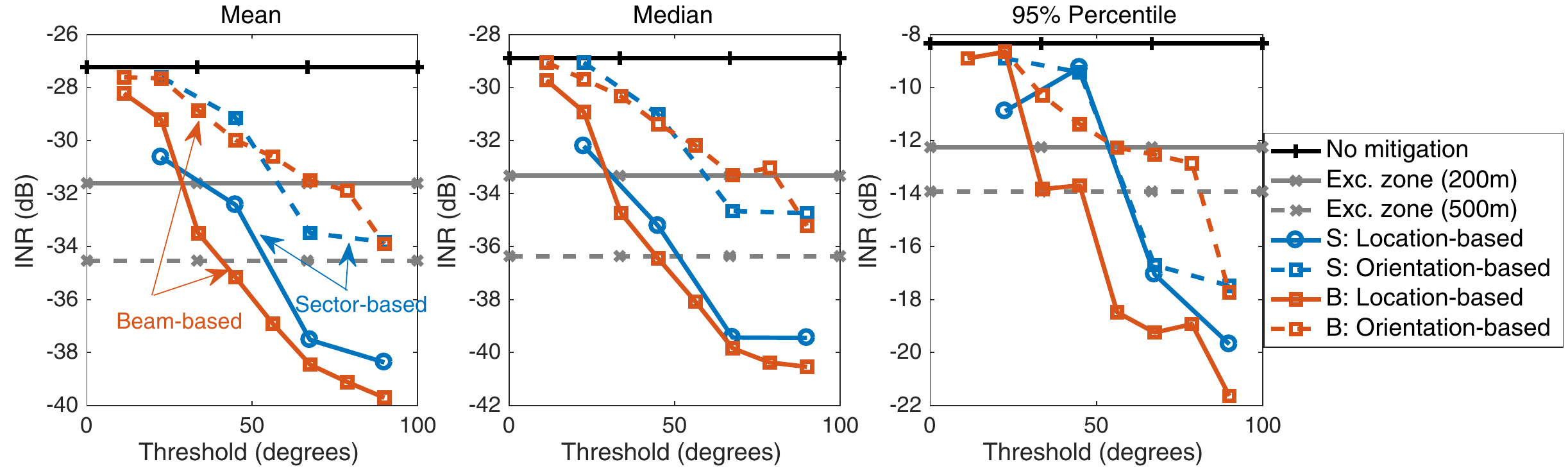} 
		\caption{INR statistics with variations of the angular protection threshold.}
		\label{fig:AngularMitigationStatistics}
	\end{figure}

	Fig. \ref{fig:AngularMitigationStatistics} shows the main INR statistics with variations of the angular protection threshold. We show the INR performance in the absence of mitigation for reference. We also show the INR in the presence of \emph{spatial} exclusion zones with radii 200m and 500, i.e., no gNBs are deployed inside these zones. As expected, the INR is significantly deceased for high protection thresholds. For instance, the 95th percentile decreases by approximately -5.5dB and -13dB when location-based beam mitigation is used with $\psi_b=45^\circ$ and $\psi_b=90^\circ$, respectively. Angular exclusion zones are more effective than spatial exclusions as the latter leads to coverage holes in the 5G system. This also emphasizes that the interference is not dominated by UEs that are close to the FS but rather by UEs that have beams directed towards the FS's boresight. 
	
\begin{figure}[t!]
		\centering
		\begin{subfigure}[t]{.5\textwidth}
			\centering
			\includegraphics[width=3.5in]{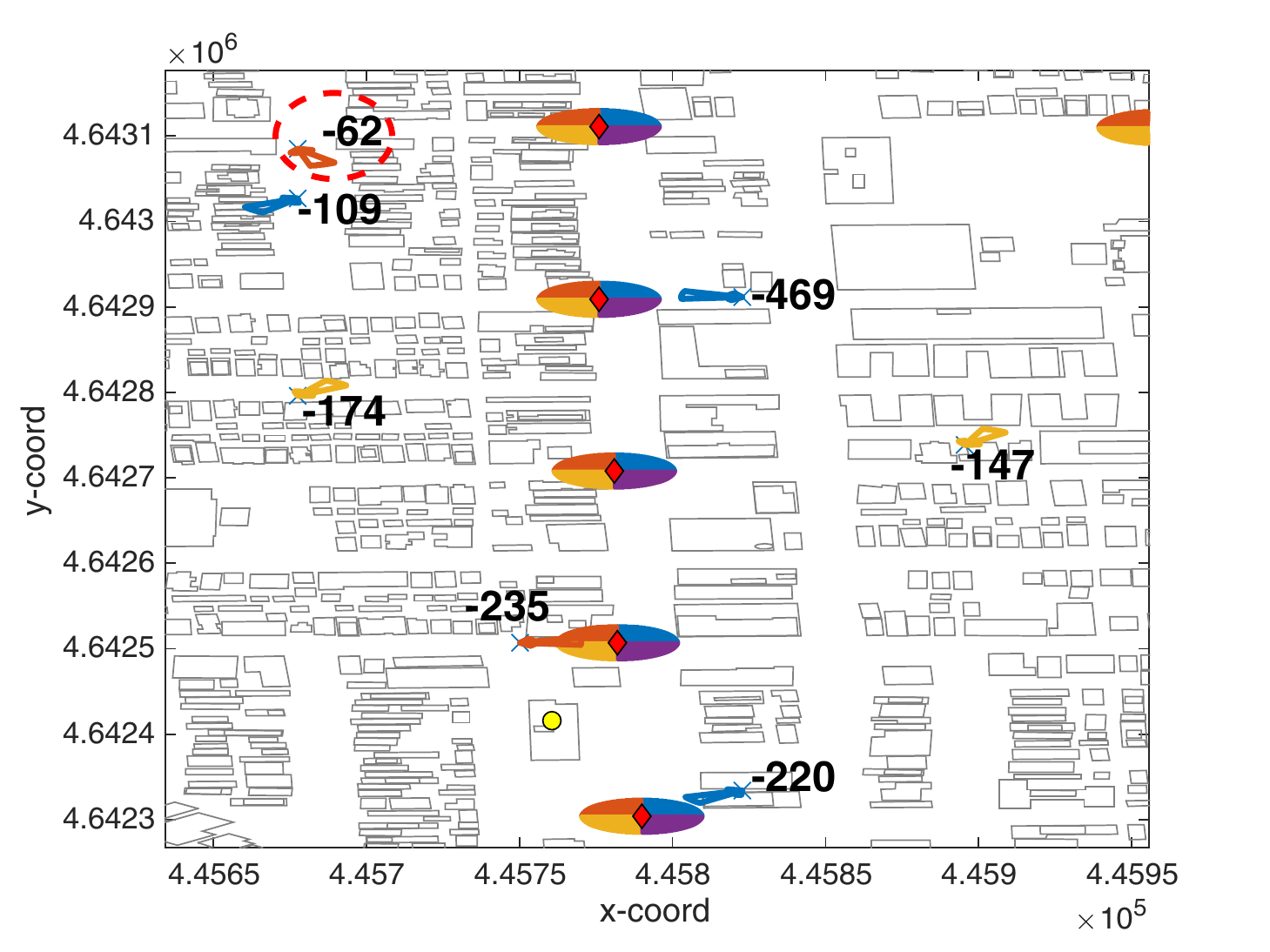}
			\caption{No mitigation}
			\label{fig:FS6NoMitg}
		\end{subfigure}~~
		\begin{subfigure}[t]{.5\textwidth}
			\centering
			\includegraphics[width=3.5in]{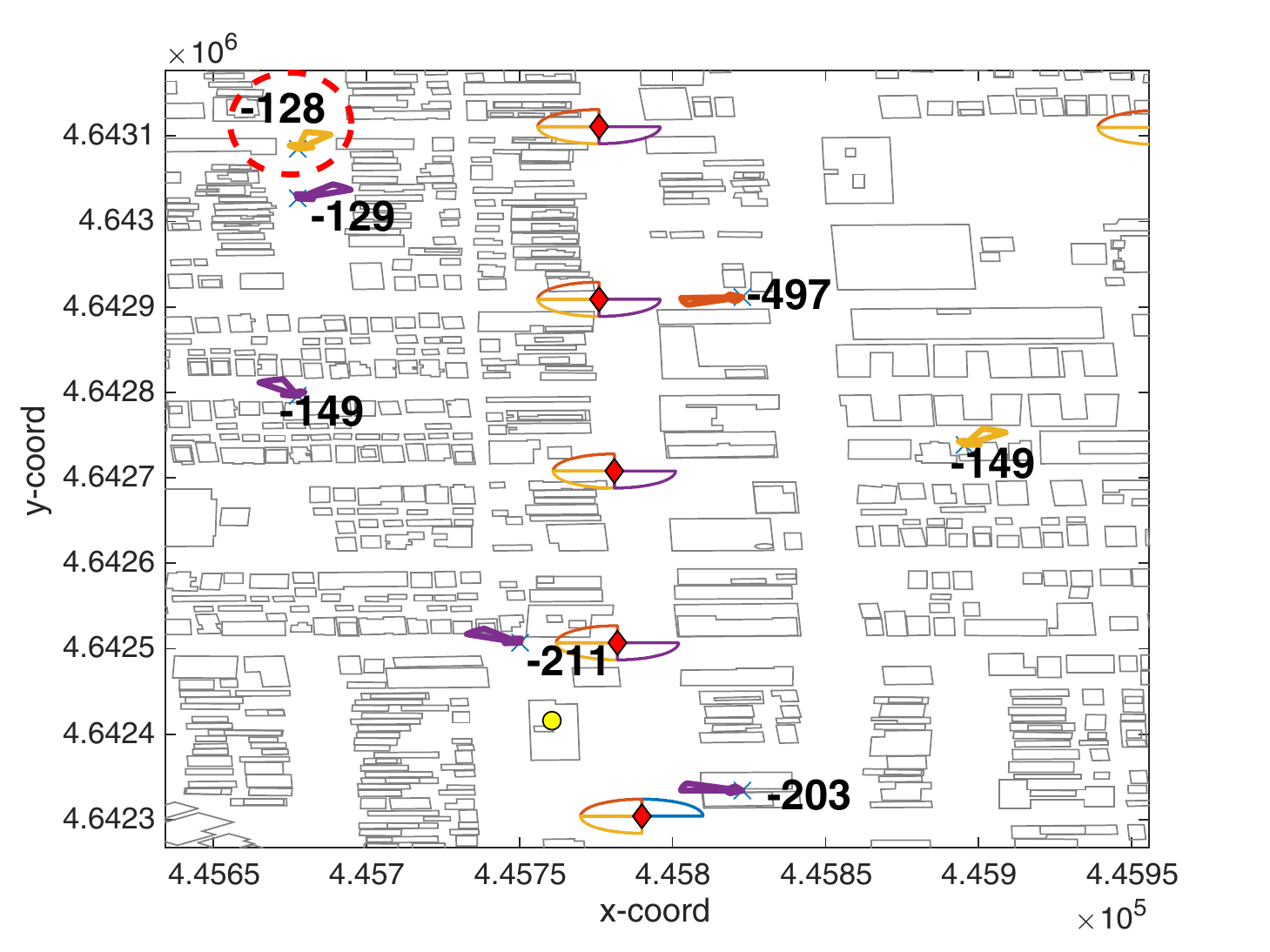}
			\caption{Sector-based}
			\label{fig:FS6SectorMitg}
		\end{subfigure}\\
		\begin{subfigure}[t]{.5\textwidth}
			\centering
			\includegraphics[width=3.5in]{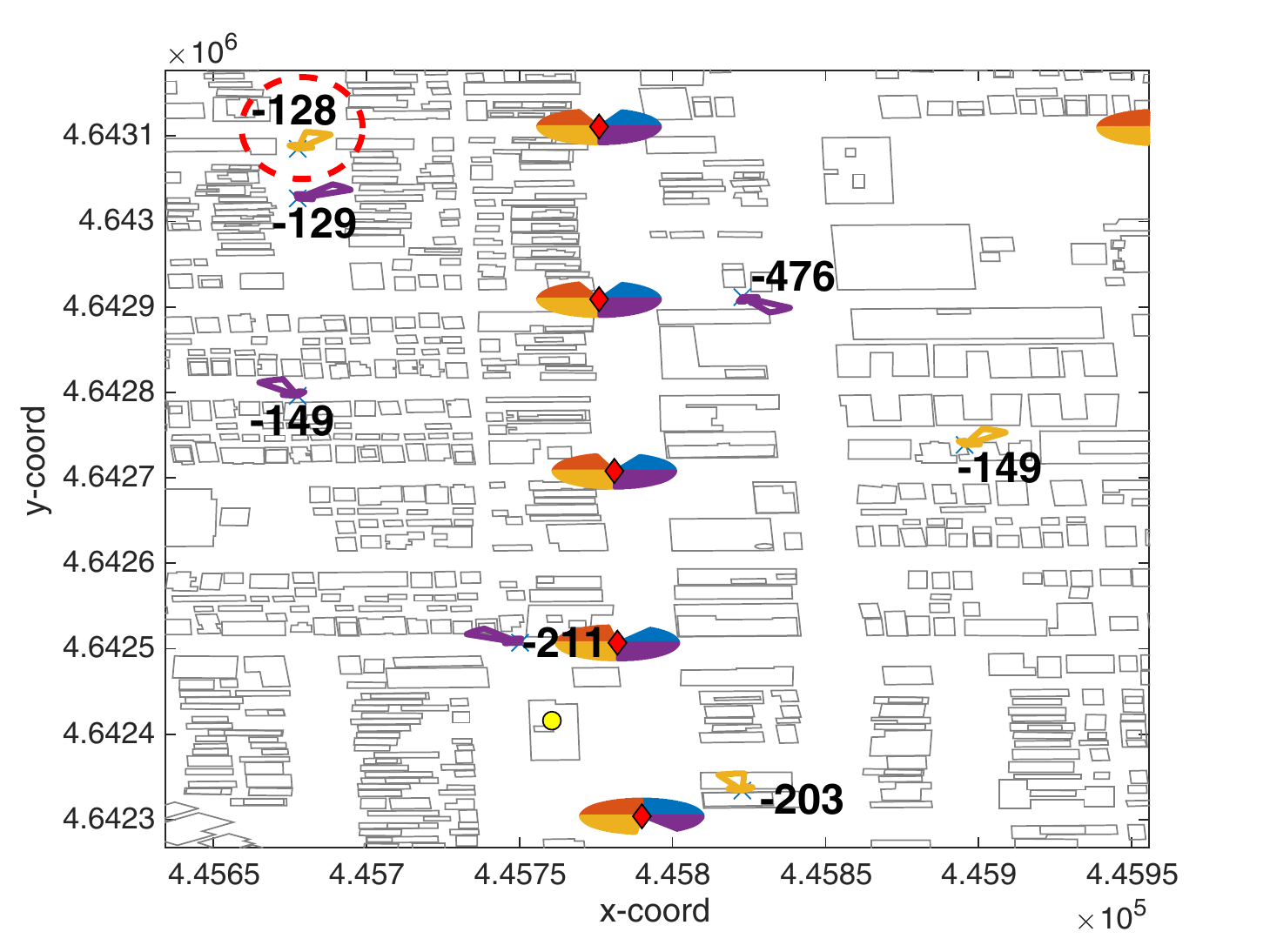}
			\caption{Beam-based}
			\label{fig:FS6BeamMitg}
		\end{subfigure}	
		\caption{One simulation snapshot of an FS that experiences high INR in the absence of mitigation. The beams used by UEs are shown, and the interference generated by each one into the FS is given in dBm: (a) No mitigation; (b) Sector-based; (c) Beam-based.}
		\label{fig:FS6}
	\end{figure}

	Fig. \ref{fig:FS6} shows one snapshot of the FS of interest and the 5G system in vicinity of the FS with and without the mitigation techniques. In the snapshot, we show the UE's beam used for data communication with its associated gNB as well as the interference generated from the UE into the FS (in dBm). In Fig. \ref{fig:FS6NoMitg}, the INR is high as it is dominated by a UE with an interference of $-62$dBm (the noise floor at the FS is approximately $-77$dBm). By using the location-based sector mitigation with $\psi_s=45^\circ$, it is shown in Fig. \ref{fig:FS6SectorMitg} that this particular UE switches to a different gNB, reducing its interference by 66dB! A similar observation is made for the beam-based approach, illustrated in Fig. \ref{fig:FS6BeamMitg}, where we use $\psi_b=22.5^\circ$, showing that angular exclusion zones at a finer scale are sufficient to protect the FS without compromising the 5G DL coverage.

	\subsection{Impact of spatial power control}
\begin{figure}[t!]
		\center
		\includegraphics[width=2.5in]{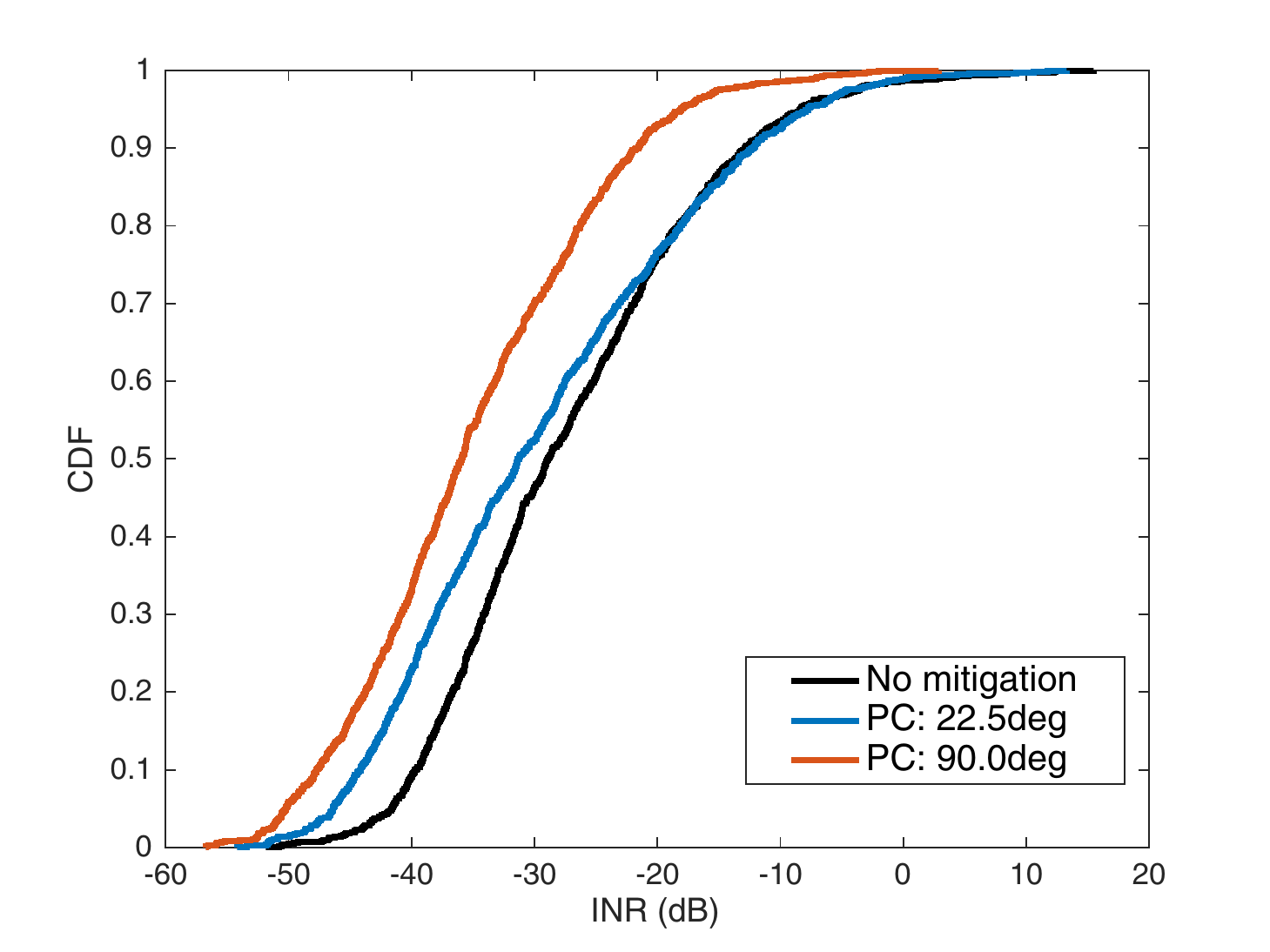} 
		\caption{INR CDF in the presence and absence of power control}
		\label{fig:PCMitigationCDF}	
	\end{figure}

\begin{figure}[t!]
		\center
		\includegraphics[width=4in]{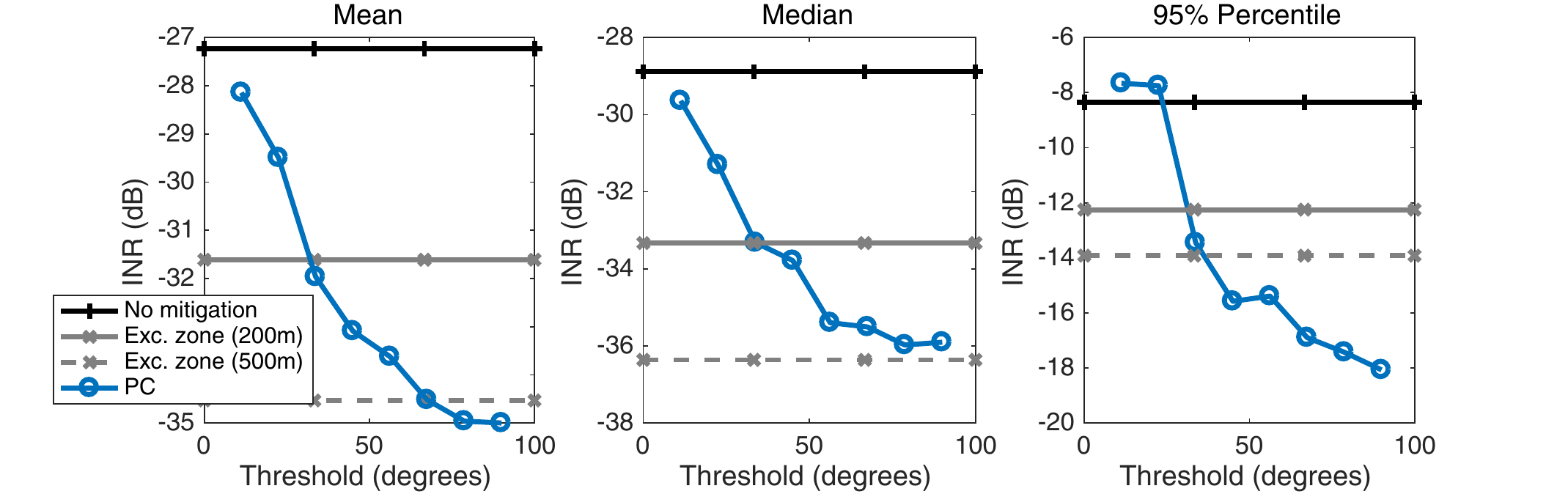} 
		\caption{INR statistics when spatial power control is used}
		\label{fig:PCMitigationStatistics}	
	\end{figure}

	We set $P_{\operatorname{lo}}$ and $P_{\operatorname{up}}$ such that the UE EIRP is 33dBm and 43dBm, respectively. Fig. \ref{fig:PCMitigationCDF} shows the INR's CDF at the FS when the power control (PC) in (\ref{eq:PC}) is used. It is evident that for higher protection thresholds, $\psi_b$, power control can be effective to reduce the INR. For instance, the 95th percentile reduces from -8dB to -15dB when $\psi_b=45^\circ$. Finally, Fig. \ref{fig:PCMitigationStatistics} shows the main INR statistics with variations of the protection threshold. It is shown that the 95th percentile can be reduced by approximately 10dB without the need to shut off any beams.  
	
	
	\subsection{Comparison of Mitigation Techniques}
\begin{figure}[t!]
		\center
		\includegraphics[width=4in]{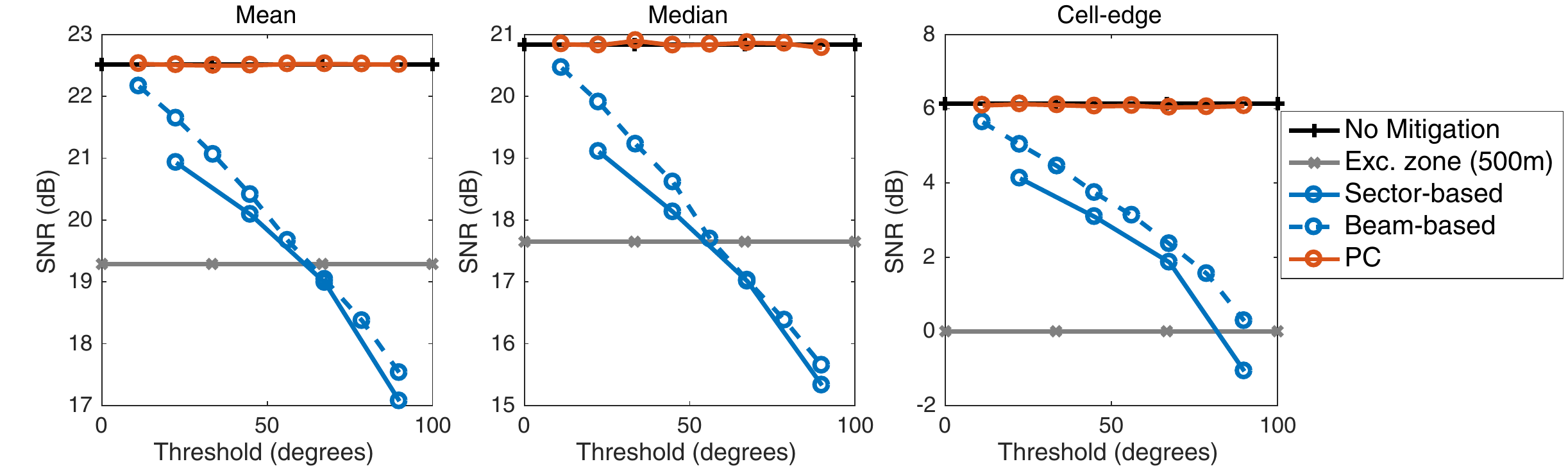} 
		\caption{SNR statistics when spatial power control is used}
		\label{fig:DLSNRcomparison}
	\end{figure}

	The aforementioned techniques have shown the effectiveness in mitigating interference at the FS. In this section, we compare them in terms of their impact on the DL coverage of the 5G system. Using the gNB antenna parameters, it can be shown that the maximum radiated power is 57dBm. Fig. \ref{fig:DLSNRcomparison} shows a comparison between the different techniques in terms of the DL coverage. We only consider location-based protection. Due to the angular exclusion zones created, using larger thresholds, i.e., $\psi_s$ and $\psi_b$, inevitably affect the DL coverage. This is not the case in spatial power control as all beams and sectors are active.  
	
\begin{figure}[t!]
		\centering
		\begin{subfigure}[t]{.45\textwidth}
			\centering
			\includegraphics[width=2.5in]{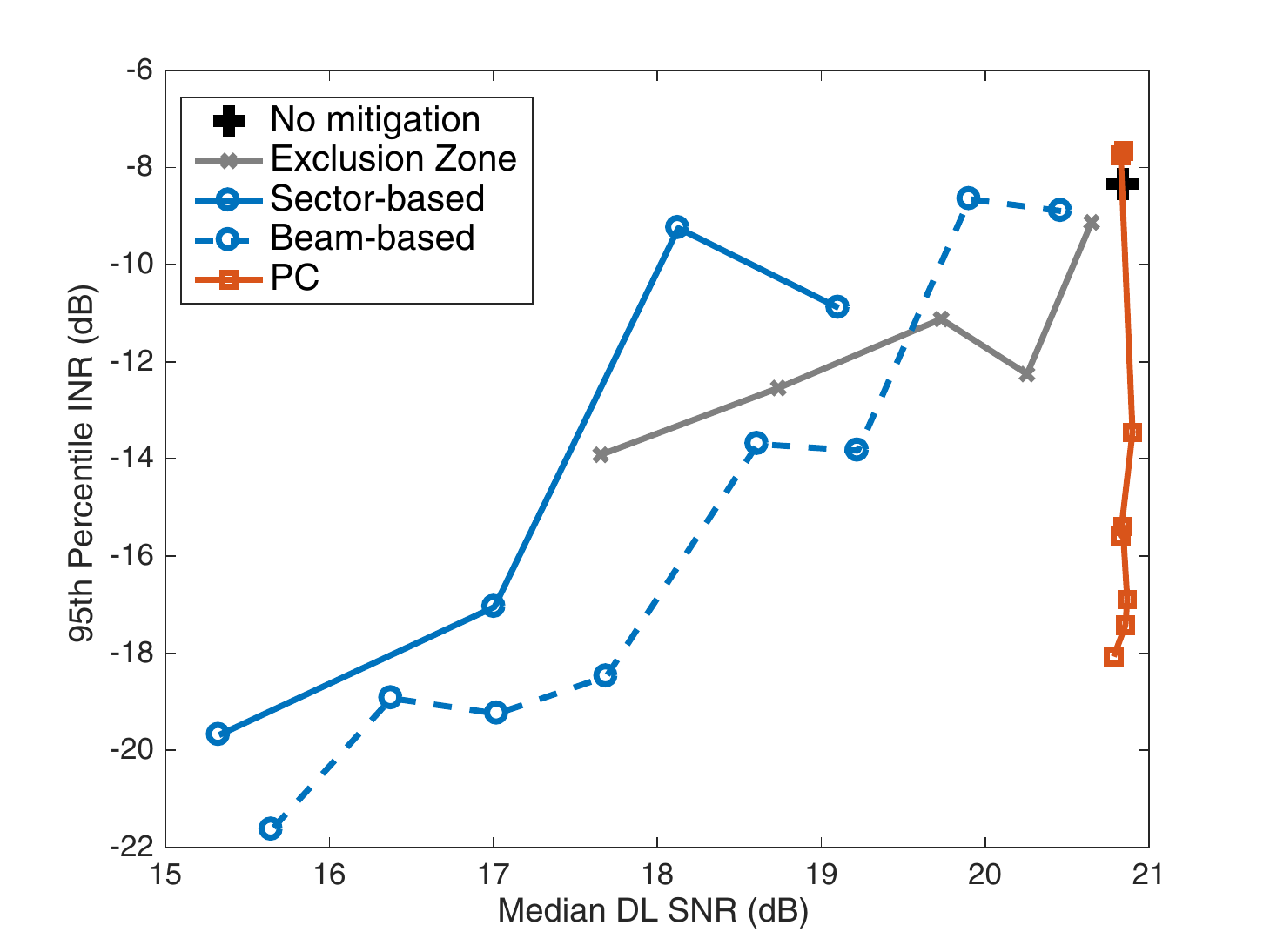}
			\caption{Median SNR vs 95th percentile INR}
			\label{fig:ComparisonMedian}
		\end{subfigure}~~
		\begin{subfigure}[t]{.45\textwidth}
			\centering
			\includegraphics[width=2.5in]{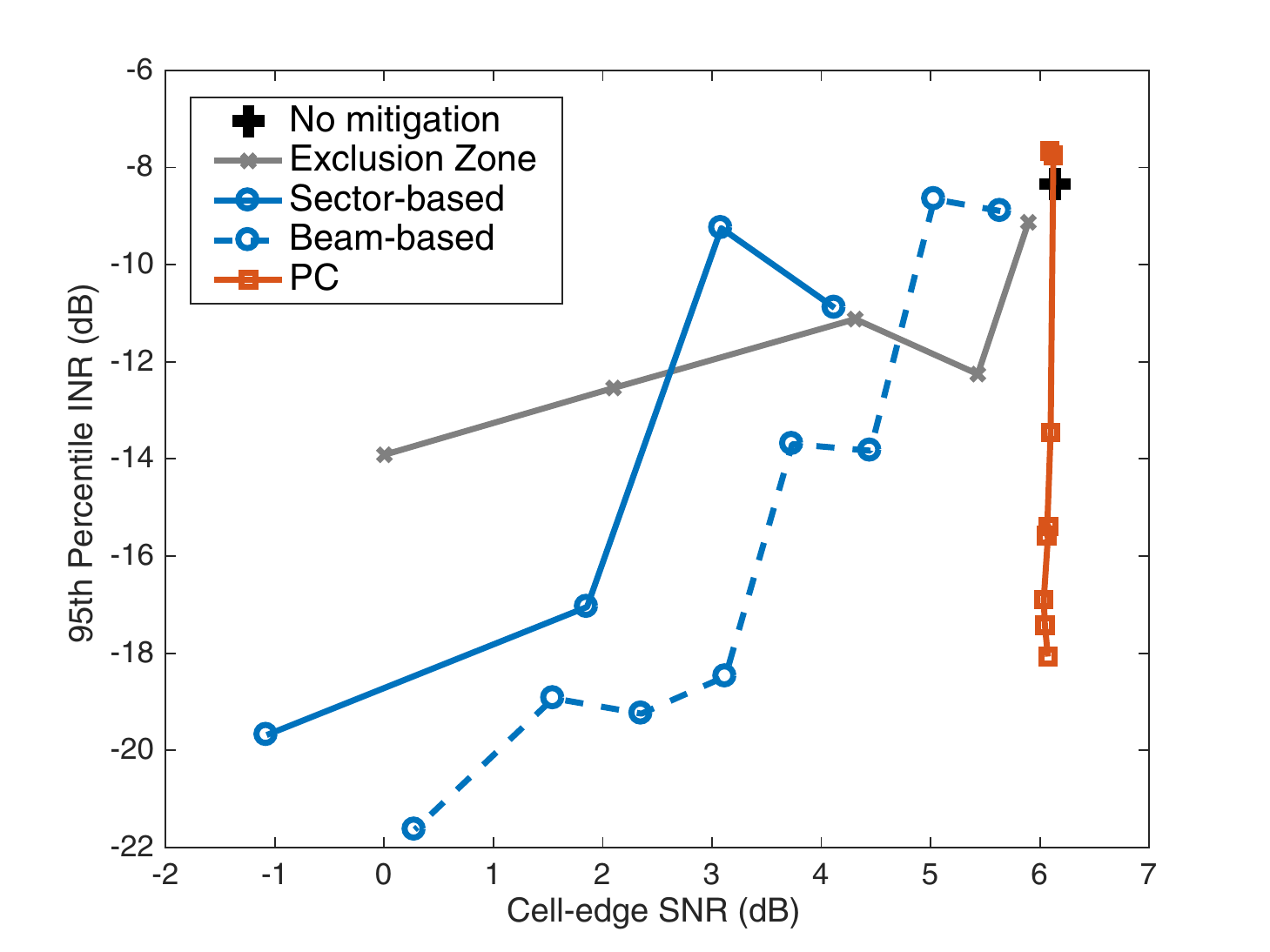}
			\caption{Cell-edge SNR vs 95th percentile INR}
			\label{fig:comparisonCellEdge}
		\end{subfigure}
		\caption{Comparison of SNR-INR curves of different mitigation techniques}
		\label{fig:compariosn} \vspace{-0.1in}
	\end{figure}

	Fig. \ref{fig:compariosn} shows the SNR-INR curves of the different mitigation techniques. The curves highlight the different possible operating points of the coexisting 5G and incumbent systems, i.e., the interference level expected on the incumbent for a target 5G DL coverage. We have the following observations. Comparing location-based beam angular zones with sector angular zones, it is evident that the former presents more operating points, making it more flexible and effective. However, both have an inevitable trade-off: higher INR protection incurs 5G coverage degradation. In addition, spatial exclusion zones are not as effective as angular exclusion zones. For instance, using a first-order approximation of the simulated curves, it can be shown that slopes for the beam-based and spatial exclusion zone techniques are approximately 3 and 1. In other words, by reducing the median SNR by 1dB, the INR reduces by 3dB when using beam-based angular exclusion zones and by 1dB when using spatial exclusion zones. Finally. spatial power control enables the reduction of INR with negligible coverage loss.

	\section{Conclusion}\label{sec:conclusion}
	The 10GHz of spectrum in the 70GHz and 80GHz bands have the potential to enable true mobile connectivity at gigabit speeds. A key obstacle to the 5G deployment of these bands is the presence of incumbent FS systems that require protection from harmful interference. To this end, we have thoroughly analyzed existing databases to understand the key features and properties of the incumbent system, including the spatial distribution and the antenna specifications. In addition, we have analyzed the aggregate interference from 5G UEs using realistic channel models and actual building layouts for accurate results. Our analysis and results have revealed that 5G coexistence beyond 70GHz is feasible thanks to the high propagation losses at millimeter wave frequencies, the high attenuation due misalignment with the FS antenna boresight, and the deployment geometry of FSs as they tend to be above 5G systems. 
	
	For FSs that are deployed at relatively low altitudes, we have proposed several passive mitigation techniques, including angular exclusion zones and spatial power control. Such techniques require minimal effort from mobile operators and do not require any coordination with the incumbents. We have shown that exclusion zones in the angular domain are more effective than spatial exclusion zones. However, the 5G DL coverage can be degraded due to the coverage holes introduced by exclusion zones. This can be overcome via power control, where the transmit power of the UE depends on the beam's direction with respect to the victim FS.

	\bibliographystyle{IEEEtran}
	\bibliography{C:/Users/ghait/Dropbox/References/IEEEabrv,C:/Users/ghait/Dropbox/References/References}
	
\end{document}